\documentclass[12pt,letterpaper]{article}
\usepackage{jheppub}
\usepackage{slashed}
\usepackage{amsmath}
\usepackage{amssymb}
\usepackage{epsfig}
\usepackage{subfig}

\usepackage{enumerate}
\usepackage[shortlabels]{enumitem}

\newcommand{\labell}[1]{\label{#1}}

\newcommand{\be}{\begin{equation}}
\newcommand{\ee}{\end{equation}}
\newcommand{\bea}{\begin{eqnarray}}
\newcommand{\eea}{\end{eqnarray}}
\newcommand{\ba}{\begin{eqnarray}}
\newcommand{\ea}{\end{eqnarray}}

\newcommand{\beq}{\begin{equation}}
\newcommand{\eeq}{\end{equation}}
\newcommand{\beqa}{\begin{eqnarray}}
\newcommand{\eeqa}{\end{eqnarray}}
\newcommand{\beqar}{\begin{eqnarray*}}
\newcommand{\eeqar}{\end{eqnarray*}}

\newcommand{\reef}[1]{(\ref{#1})}
\newcommand{\ssc}{\scriptscriptstyle}
\newcommand{\eg}{{\it e.g.,}\ }
\newcommand{\ie}{{\it i.e.,}\ }

\newcommand{\mt}[1]{\textrm{\tiny #1}}

\newcommand{\veps}{\varepsilon}

\newcommand{\ta}{{\tilde a}}
\newcommand{\tb}{{\tilde b}}
\newcommand{\tc}{{\tilde c}}

\newcommand{\te}{t_\mt{E}}

\newcommand{\myeq}[1]{\begin{equation} #1 \end{equation}}

\renewcommand{\c}{$c$}
\newcommand{\F}{$F$}
\newcommand{\cA}{{\cal A}}

\newcommand{\un}{{\hat{\bf n}}}
\newcommand{\um}{{\widehat{\bf m}}}
\newcommand{\ut}{{\hat{\bf t}}}
\newcommand{\uu}{{\hat{\bf u}}}

\newcommand{\ren}{R\'enyi\ }
\newcommand{\coss}{{c_0^{\ssc scalar}}}
\newcommand{\sg}{q_1} %{\sigma}
\newcommand{\roo}{q_2}
\newcommand{\GN}{G_\mt{N}}
\newcommand{\vepst}{\veps_{\textrm{topo}}} 

\arxivnumber{arXiv:1506.nnnnn}

\title{Mutual information and the \F-theorem}

\author[a]{Horacio Casini,}
\author[a]{Marina Huerta,}
\author[b]{Robert C. Myers}
\author[b,c]{and Alexandre Yale}

\affiliation[a]{Centro At\'omico Bariloche, %and Instituto Balseiro, 
8400-S.C. de Bariloche, R\'io Negro, Argentina}
\affiliation[b]{Perimeter Institute for Theoretical Physics, Waterloo, Ontario N2L 2Y5, Canada}
\affiliation[c]{Department of Physics \& Astronomy and Guelph-Waterloo Physics Institute,\\ University of Waterloo,
Waterloo, Ontario N2L 3G1, Canada}

\abstract{Mutual information is used as a purely geometrical regularization of
entanglement entropy applicable to any QFT. A coefficient in the mutual
information between concentric circular entangling surfaces gives a precise
universal prescription for the monotonous quantity in the \c-theorem for $d=3$.
This is in principle computable using any regularization for the entropy, and
in particular is a definition suitable for lattice models. We rederive the
proof of the \c-theorem for $d=3$ in terms of mutual information, and check our
arguments with holographic entanglement entropy, a free scalar field, and an
extensive mutual information model.}

\begin{document}
\maketitle

\section{Introduction}

A great part of the problem of understanding the structure of the space of
quantum field theories (QFT) reduces to the one of classifying possible
critical points. These critical points must be free of ultraviolet problems
which greatly constrains the consistent QFT's. In this picture, the theories
with mass scales are obtained by perturbing critical points and following the
corresponding renormalization group (RG) trajectories. However, interestingly,
not all critical points are joined to each other in such flows. For example, nontrivial 
constraints arise by anomaly matching conditions implying the ultraviolet (UV)
and infrared (IR) fixed points have the same chiral anomalies \cite{anom}.

A different class of constraints are provided by \c-theorems. Such theorems
state that a certain \c-function must decrease along any RG trajectory and so
they illustrate the irreversible character of RG flow. The possibility of such
a structure first emerged over twenty-five years ago with Zamolodchikov's proof
of the \c-theorem in two dimensions \cite{Zamo}. However, it is only recently
that a long sought proof was found for the corresponding $a$-theorem in four dimensions \cite{Komar1} --- see also \cite{luty}. Both the
$d=2$ and $d=4$ cases conform with a broader conjecture by Cardy \cite{cardy0}
that for any even number of spacetime dimensions, there exists a \c-theorem
where at the fixed points, the \c-function coincides with the coefficient of
the Euler term in the trace anomaly, the so-called A-type anomaly term
\cite{deser}. Unfortunately, progress towards proving such \c-theorems in
higher dimensions (\eg $d=6$ \cite{hdim}) has been limited. Of course, Cardy's
proposal cannot be extended to odd dimensions where the trace anomaly vanishes.

Recently, a new perspective on \c-theorems was proposed in \cite{myers1}, in particular,
one which incorporates both {\em odd} and even dimensions. Motivated by holographic studies, they
suggested that the \c-function could be identified as the universal coefficient
appearing in the entanglement entropy of a region enclosed by a sphere. In
fact, in even dimensions, this universal
coefficient yields precisely the central charge in the A-type trace anomaly at fixed points of the RG flows 
\cite{myers1,CHM} --- see also \cite{sphere1}. Hence this new proposal
provides a new perspective on Cardy's conjecture for even dimensions. However, more
importantly, entanglement entropy provides a framework to consider \c-theorems for quantum
field theories in an odd number of dimensions. Recently, entanglement entropy was also discussed as a
framework to extend the two-dimensional $g$-theorem \cite{bound} to higher dimensional boundary \c-theorems \cite{defect} --- see also \cite{davide}. 

Much of the subsequent work focused on the case of $d=3$. In particular, an
independent formulation of the three-dimensional \c-theorem was proposed in \cite{Jaff1}, where
the \c-function was identified at conformal fixed points with the constant term
in the free energy evaluated on a three-sphere. A great deal
of evidence for this \F-theorem was gathered from a wide variety of field theory examples,
\eg \cite{igor1,evident}.\footnote{See \cite{tarum} for an interesting application of the
\F-theorem to derive nonperturbative results for strongly interacting quantum field theories.} However, it was also known that the \F-theorem
coincides with the previous proposal using entanglement entropy \cite{CHM}. In
particular, for any odd $d$, it was shown that the free energy of a $d$-sphere
and the entanglement entropy of a $(d-2)$-sphere can be mapped to each other in
a CFT \cite{CHM,Dowk0}. Next, a proof was found to show that the
constant term in the entanglement entropy for a circle decreases monotonically
under RG flow in $d=3$ QFT's \cite{proof}. This proof of the three-dimensional
\F-theorem relies only on unitarity and Lorentz invariance of the underlying
QFT, as well as strong subadditivity of entanglement entropy \cite{strong}.
Unfortunately, this proof does not (easily) extend to higher
dimensions due to the appearance of more singular terms in the entanglement
entropy.

In the context of QFT, an essential feature of entanglement entropy is that it cannot be defined without 
reference to a UV regulator because the result is dominated by short-distance correlations in the vicinity of the entangling surface.
Considering a $d$-dimensional QFT, if we choose a spherical entangling surface
of radius $R$ in flat space,\footnote{Implicitly, we also assume that the
theory is in its ground-state throughout the following.} the entanglement
entropy generically has an expansion
 \beqa
S &=&  c_{d-2} \left( \frac{R}{\delta} \right)^{d-2}+ c_{d-3} \left(
\frac{R}{\delta} \right)^{d-3} + c_{d-4} \left(
\frac{R}{\delta} \right)^{d-4} + \cdots
\labell{unis}\\
&&\qquad\qquad\qquad\qquad\qquad\qquad+ \left\lbrace
\begin{matrix}
(-)^{\frac{d}{2}-1}\,4\,c_0 \log(R/\delta)+\cdots&\quad&{\rm even\ }d\,,\\
  (-)^{\frac{d-1}{2}}\,2\pi\,c_0 \ +\ \cdots\ \ \ \ \ \ \ \ \
  &\quad&{\rm odd\ }d\,.
\end{matrix}\right.
\nonumber 
 \eeqa
where $\delta$ is the short-distance UV cut-off. For a CFT, the coefficients
$c_k$ are pure numbers but for a generic QFT, they will depend on
other mass scales $\mu_i$ through the dimensionless combination $\mu_i\delta$
\cite{liu,m1,mass2}. Of course, the first term in the above expansion is the celebrated `area law' term \cite{area}. 

The coefficients of the power law divergent terms depend on the details of the UV regulator and only $c_0$ can be universal. In an even number of spacetime dimensions, this universality is guaranteed by the logarithmic character of the associated singluarity in eq.~\reef{unis}. However,  for odd
dimensions where $c_0$ appears in the finite contribution, there are a variety of potential ambiguities which may ruin the universality of this constant. For example, the regulator may limit the radius $R$ from being determined with a resolution of better than the UV cut-off $\delta$. However, any shift $R\to R'=R+\alpha\,\delta$ will result in mixing $c_0$ with all of the higher order coefficients in the expansion \reef{unis}. We will expand on such ambiguities further in section \ref{prob}.

Important progress in this regard was made in \cite{liu,gente}\footnote{See also \cite{zz33,zomis} for discussion from the perspective of the free energy.} where a class of regulators was identified which yield an unambiguous definition of $c_0$. The key feature was that the regulator must be covariant or `geometric,' and as a result, the UV divergent terms in the entanglement entropy all have a geometric character. That is, these contributions can all be expressed as integrals of local geometrical quantities over the entangling surface.\footnote{Note that with such
a covariant regulator, the expansion \reef{unis} will be restricted to contributions involving odd (even) powers of $R/\delta$ for odd (even) $d$. That is, all of the coefficients $c_{d-3-2k}$ will vanish.} However, identifying various properties to define a `nice regulator' leaves
the discussion somewhat incomplete, in that one might be left to conclude
that $c_0$  is only an artifact of the choice of a geometric regulator. If this constant is a physical quantity, we should be able to use any regularization scheme which
defines the continuum theory of the underlying QFT. 

%\newpage

Given these issues, we introduce a list of requirements needed to properly establish a \c-theorem:
\it
%\begin{enumerate}[(a),topsep=2pt]
\begin{enumerate}[topsep=2pt] 
\itemsep-1pt
\item The \c-function must be a dimensionless, well-defined quantity, which
    is independent of the regularization scheme. This means that it must be
    a renormalized  physical quantity, which can in principle be computed
    with any regulator.
\item The \c-function must be a quantity which is intrinsic to the fixed point of interest.
 In particular, for an RG flow running between two fixed points, the \c-function must be independent of the details of the RG flow at the infrared fixed point.
\item Finally, for the \c-theorem to hold, the \c-function must decrease
    monotonically along any RG flow connecting a UV fixed point to an IR
    fixed point.
\end{enumerate}
\rm

Our proposal is to use the concept of mutual information to define the \c-function 
and we will demonstrate that the corresponding \c-theorem in three dimensions
satisfies all three of the above conditions. The
mutual information is another measure of information between two
systems\footnote{For example, this quantity can be used to bound the connected correlators of observables in two regions \cite{core4}. See \cite{proofd2,mut,split,swingle,cardymutual} for various studies of general properties of mutual information in QFT.} and for non-intersecting
regions $A$ and $B$, it may be defined as
\begin{equation}
I(A,B)=S(A)+S(B)-S(A\cup B)\,.\labell{mutualdef}
\end{equation}
Unlike entanglement entropy, mutual information is finite and independent of
the UV cut-off of the theory, \ie the divergences in $S(A)$ and $S(B)$ are
canceled by those in $S(A \cup B)$. We note that, in fact, mutual
information can be defined without any reference to entanglement entropy. The latter definition uses the Araki formula for the
relative entropy of two different states in an operator algebra \cite{araki}
--- see also \cite{narn}. Unfortunately, this approach seems unsuited for
practical calculations and so our discussion will only consider the definition above in eq.~\reef{mutualdef}.

Beginning with eq.~\reef{mutualdef}, we provide a new definition of $c_0$ where the mutual
information serves as a `universal' geometric regulator of the circle entanglement
entropy, which can be applied for any QFT with any UV regulator.\footnote{Similar ideas were discussed previously in \cite{split,black,newcas}.} Since the
mutual information is finite and independent of the UV regulator, the constant
$c_0$ defined in this way automatically satisfies our first criterion above. With
further care in formulating our prescription, we are able to show that the second
criterion is also satisfied. Hence with the mutual information, we are able
to identify $c_0$ as a renormalized physical quantity, which can be calculated
using any regulator. Finally, we will also show the proof of the \c-theorem can
be formulated in terms of our mutual information prescription. The latter
clarifies the validity of certain assumptions in \cite{proof}, as well as
establishing that our final requirement is satisfied within this new
framework.

An overview of the paper is as follows: We begin the next section with a
discussion of some ambiguities which arise in evaluating $c_0$ directly the entanglement entropy \reef{unis}. In section \ref{newdef}, we provide a cleaner definition of
a similar entanglement `central charge' using the mutual information. Then in
section \ref{critb}, we establish that the new definition protects the coefficient $\tc_0$ from
the high energy details of the theory. In section \ref{entropy}, we argue that in fact $\tc_0=c_0$ with a suitable choice of regulator. In particular, we examine
with which regulators, the charge defined by the mutual information can be
obtained directly from that in the entanglement entropy of a circle. 
Further, with a detailed examination of the
example of a free massless scalar, we highlight some ambiguities in defining $c_0$, even when
using some geometrical regularizations, and we show that they can be eliminated
using our mutual information prescription. In section \ref{proof}, we prove the
monotonic decrease of $c_0$ in terms of the mutual information. We conclude
with a brief discussion of the results in section \ref{discuss}. In appendix
\ref{holographic}, we calculate the mutual information for a circle within a
holographic framework. Appendix \ref{extensive} presents calculations of mutual
information in a particular model where the mutual information satisfies a
certain extensivity property \cite{split}. Finally, in appendix \ref{scalar},
we calculate the entanglement entropy of a circle for a free massless scalar,
first by mapping the problem to the hyperbolic cylinder $R\times H^2$ and then
mapping to de Sitter space.

The discussion in sections \ref{mutual} and \ref{entropy}, as well as the
examples in the appendices, can be easily generalized to any number of
spacetime dimensions. However, the proof of monotonicity is section \ref{proof}
is only valid for $d=3$. For conciseness, the material in sections \ref{mutual}
and \ref{entropy}, as well as in the appendices will be limited to considering $d=3$.

\section{Mutual information as a `geometric regulator'} \labell{mutual}

A central issue raised in the introduction was that entanglement entropy in QFT cannot be defined without explicit reference to a UV regulator. Further, without restricting the class of allowed regulators, this introduces various potential ambiguities in trying to define $c_0$ directly from eq.~\reef{unis} for odd dimensions, in particular, for $d=3$. In section \ref{prob} below, we discuss these ambiguities in more detail and, in particular, we explicitly show that evaluating $c_0$ with a lattice regulator is problematic. Next in section \ref{newdef}, we consider using mutual information as a geometric regulator for entanglement entropy of a circle and this allows us to define a related coefficient $\tc_0$. Since mutual information is a well defined (finite) quantity, $\tc_0$ can be regarded as a physical quantity, which is independent of the details of the regulator. However, our prescription still has a certain freedom which introduces ambiguities in this constant which are somewhat analogous to those found using the entanglement entropy. Hence we refine the definition of $\tc_0$ in section \ref{critb} to ensure that at RG fixed points, it is a quantity intrinsic to the underlying CFT, \ie that it is independent of any higher UV scales. With this refined definition, we also explicitly show that $\tc_0$ can be computed using a lattice regulator. With our new mutual information approach, $\tc_0$ is a candidate \c-function which satisfies both the first and second criteria which required for a legitimate \c-theorem. Of course, it will still remain to show that this new \c-function actually decreases monotonically along RG flows, our third criterion.  We will return to this point in section \ref{proof}.

\subsection{Ambiguities in $c_0$} \labell{prob}

The \F-theorem in three dimensions was originally proposed in terms of the
constant $c_0$ appearing in the entanglement entropy \reef{unis} of a circle \cite{myers1}. At a
conformal fixed point, we have
\begin{equation}
S(R)=2\pi R\left(\frac{a}{\delta}+b\right)-2\pi\,c_0\,,\labell{enti}
\end{equation}
where $R$  is the radius of the circle and $\delta$ is a short distance cut-off.
As noted above, with a covariant regulator, it can be shown that $c_0$
always decreases along RG flows \cite{proof}. However, without further 
information on the regularization procedure, this dimensionless constant is ambiguous.

The latter ambiguity is readily demonstrated with a square lattice regulator \cite{review}. The results of an explicit calculation for a free scalar on a square lattice are shown in figure \ref{entro-circ}. In this figure, $c_0$ is determined by evaluating the entanglement entropy of a circle as a function of radius and then subtracting the term proportional to $R$ in a linear fit to the data. According to eq.~\reef{enti}, the result should correspond to $-2\pi c_0$, however, rather than converging to a specific
value, the result fluctuates apparently forever as a function of the circle radius.
It is straightforward to understand the size of the fluctuations as follows: A typical
error in the radius of the circle will be of order of the lattice
spacing $\delta$. Given the form of eq.~\reef{enti}, this error will then feed into $c_0$ with errors of order
$a$. In the example shown here, from the coefficient of the linear fit, one finds $2\pi a\sim
0.5$ which then matches the approximate size of the noise in figure
\ref{entro-circ}. We emphasize that this situation will \emph{not} be improved by simply carrying out
the lattice calculations with greater resolution.  Attempting to describe a circle on a square lattice simply fails at the scale of the lattice spacing $\delta$. 
That is, we simply cannot be sure what the precise geometry of the entangling surface is with a
resolution of better than $\delta$ and so the length of this surface cannot be determined with an accuracy of better than $\delta$. Hence,
as described above, in a direct calculation of the entanglement entropy, the result for $c_0$ will always `polluted' by UV data because of this intrinsic
error.  Therefore one must expect the fluctuations in figure \ref{entro-circ} will never diminish even if the calculations were extended out to arbitrarily large values of $R$. 
Hence we must be able to find a better definition of $c_0$ if this constant represents a true physical quantity.
\begin{figure}[t]
  \centering
   \includegraphics[width=0.4\textwidth]{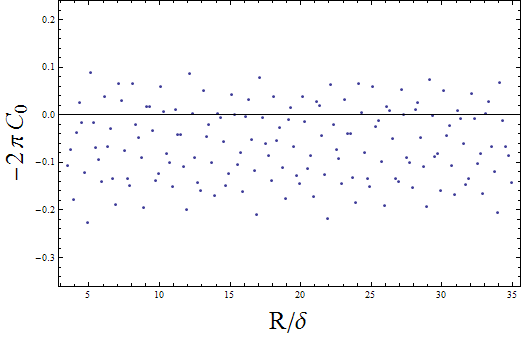}
  \caption{
The entropy of the set of lattice points inside circles of radius $R$ for a free scalar in a square lattice. The radius is taken from $3.5$ to $35$, with equal spacing of $0.2$. The values of
$-2 \pi c_0$ shown are obtained by computing $S(R)$ and then subtracting the term proportional to $R$ in a linear fit to
the data.} \labell{entro-circ}
\end{figure}

The above issue actually arises more broadly. Even if the underlying theory is already taken in the continuum
limit, it is generally a subtle problem how to define $c_0$ unambiguously
through eq.~\reef{enti} because evaluating the entanglement entropy explicitly requires a cut-off.
In particular, it may be that the latter sets a limit $\delta$ on the resolution of distances for the entire theory
and so there is a lack of precision in defining $R$ in eq.~\reef{enti}.
Now again if we shift $R\rightarrow R+\alpha\, \delta$, $c_0$ at the IR fixed
point gets mixed with $a$ and hence is polluted by contributions from much higher
UV scales. We illustrate this issue further with calculations for a free scalar field in section \ref{entropy}.

In addition, gauge fields introduce further ambiguities in the entanglement entropy that spoil the universality of the constant term appearing in eq.~\reef{enti}. In particular, for theories with gauge fields, the entropy depends on the exact prescription for assigning a gauge-invariant operator algebra to a given region \cite{gauge}. Prescriptions that eliminate superselection sectors for the local algebras (in order to remove the classical Shannon contribution to the entanglement entropy arising from the  probability distribution assigned to these sectors) are related to certain gauge fixings. Hence one finds an alarming gauge dependence in entanglement entropy.

Returning to eq.~\reef{enti}, let us consider the two constants, $a$ and $b$, appearing in the area law term. Of course,  the dimensionless constant $a$ appearing as the coefficient of the linear divergence depends on the details of the UV regulator. The appearance of the finite constant $b$ is a slightly more subtle issue. Recall that eq.~\reef{enti} describes the entanglement entropy at a conformal fixed point. However, this expression also clearly requires that $b$ have the dimensions of mass and so
%rob
na\"ively, it would seem that a finite contribution of this form simply cannot appear in eq.~\reef{enti}. However, we are assuming here that the fixed point theory may
arise in the deep infrared of an RG flow from some UV fixed point. In such a scenario, the $b$ term will typically appear with mass scales from this RG flow, \eg \cite{frank,frank2,mass2,calc2}.\footnote{Further explicit examples of this $b$ term appear in our calculations with
the extensive model in the appendix \ref{extensive}. Let us also note that similar terms will appear in our regularization using the mutual information.} Hence, we wish to emphasize that even in the far infrared where the theory flows to a CFT, the `area law' term
in eq.~\reef{enti} contains data external to the CFT, \eg masses which appear at higher scales. So, interestingly this
contribution to the entanglement entropy in the infrared is not an
intrinsically infrared quantity but rather it depends on the whole running of
the theory.  

With an appropriate regulator, the fixed point theory will be well-defined in the continuum and implicitly, the meaning of distances between spacetime points is defined with infinite
precision. For example, generally, a CFT can be thought of as defined by the
spectrum of its operators and their correlation functions at finite separation. In such a framework, the above ambiguities can be avoided following the proposal of \cite{liu} --- see also \cite{gente}. This approach implicitly
assumes that the regulator respects the property $S(A)=S(\bar{A})$
for global pure states, where $\bar{A}$ is the region complementary to $A$.
Further as discussed above, the regulator must be covariant or `geometric.' As a result, all of the divergent
terms in eq.~\reef{unis} can be expressed as integrals of local geometrical quantities over the entangling
surface. For the circle in three dimensions, the latter only includes the area law term but in higher dimensions, a rich array of such terms can appear \cite{liu,m1,mass2,calc2,klop}. Further for `scalable' geometries, these geometric
contributions can be eliminated to isolate the $c_0$ coefficient in eq.~\reef{unis} by combining various derivatives of the entanglement entropy \cite{liu}, \ie for the circle in three dimensions, $c_0=\frac1{2\pi}\left[R\,\partial_R S(R) - S(R)\right]$. This construction has the virtue that it removes the entire area law contribution and so eliminates the finite $b$ term, as well as the divergent term proportional to $a/\delta$. 

As noted in the introduction, we feel the preceding approach is somewhat
unsatisfactory. In particular, we should be able to identify $c_0$ as a
physical quantity of the underlying quantum field theory. That is, it should
be defined with any regularization scheme which defines the continuum theory.
Without such a definition, one cannot be sure that this constant is not an
artifact of the `nice choice' of a geometric regulator. We believe that the
mutual information provides precisely such a physical definition.

\subsection{Mutual information} \labell{newdef}

As commented in the introduction, we propose to define the appropriate \c-function with mutual information. The mutual information $I(A,B)$ between two
non-intersecting regions $A$ and $B$ is given by
\begin{equation}
I(A,B)=S(A)+S(B)-S(A\cup B)\,.\labell{mutualdef2}
\end{equation}
Let us reiterate that mutual information can be defined without any reference to entanglement entropy \cite{araki,narn}. However, we only consider the above definition since it is better suited for our calculations. The advantage of working with this quantity is that the mutual information is a well-defined
finite quantity.  In particular, it is independent of any regularization scheme used to define the underlying QFT. In
eq.~\reef{mutualdef2}, the divergences in $S(A)$ and $S(B)$ are canceled by
those in $S(A \cup B)$.  Moreover, mutual information in the continuum theory does not suffer any ambiguities in gauge theories \cite{gauge}.  
Hence in defining the \c-function with mutual information, we will have automatically
satisfied the first criterion which we listed in the introduction to establish a legitimate \c-theorem. 
\begin{figure}
  \centering
   \includegraphics[width=.80\textwidth]{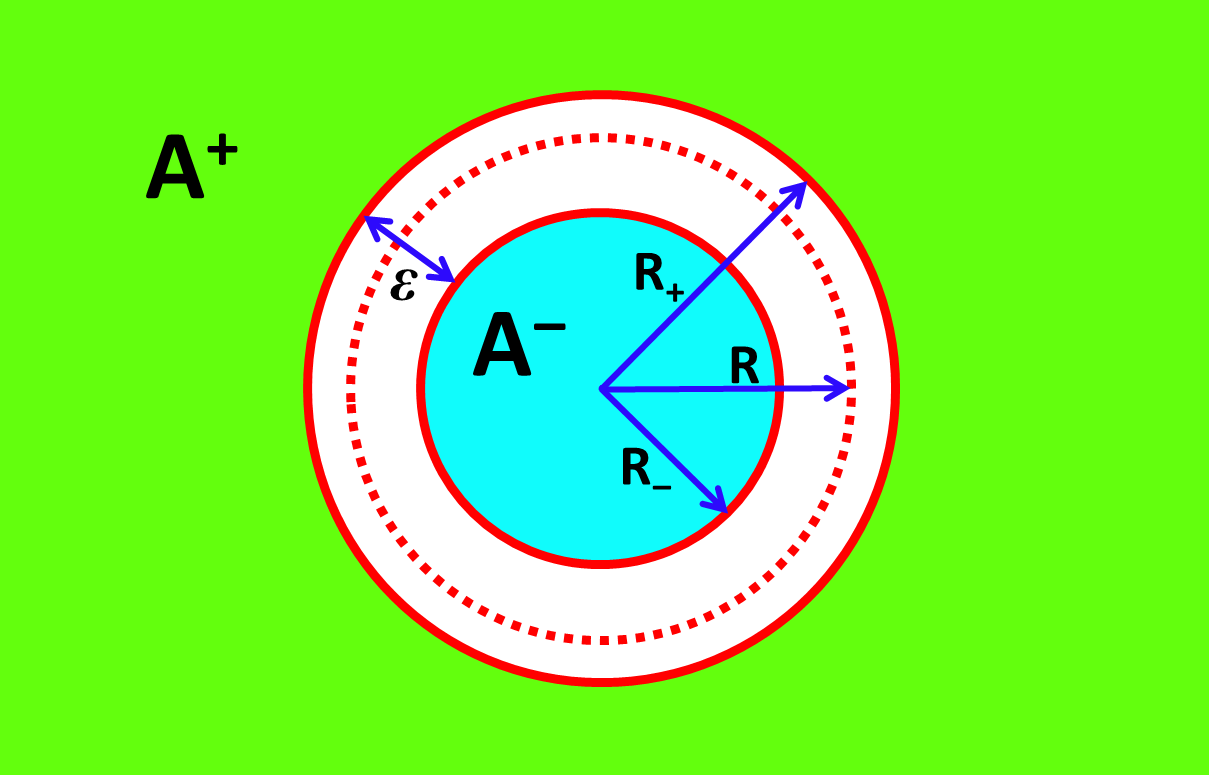}
  \caption{(Colour online)
We compute the mutual information between the interior of the circle of radius
$R_-$ and the exterior of a concentric circle of radius $R_+$ with $R_+>
R_-$.} \labell{fig:Mutual0}
\end{figure}

Now given a disk $A$ of radius $R$, the mutual information can be used to construct a `regulated' form of the entanglement entropy \reef{enti} as follows:\footnote{A similar strategy for regulating calculations of entanglement entropy was recently discussed in \cite{newcas}.}  First we replace the original entangling surface by two new (concentric) circles --- see figure \ref{fig:Mutual0}.  The first has a slightly smaller radius $R_-$ enclosing a smaller disk $A^-$ and the second with a slightly larger radius $R_+$ which is taken as the boundary of $A^+$, the exterior of this circle. The quantity of interest is then the mutual information between $A^+$ and $A^-$, \ie
\begin{equation}
I(A^+,A^-)=S(A^+)+S(A^-)-S(A^+\cup A^-)\,.\labell{mutualdef30}
\end{equation}
Let us denote the separation of the two radii as $\veps\equiv R_+-R_-$. We will always be thinking of the two radii and their separation as macroscopic scales much larger than any UV cut-off implicit in defining the underlying quantum field theory, \ie $R_+, R_-, \veps \gg \delta$. 

In the following, we focus on evaluating this mutual information in
the groundstate of the underlying quantum field theory. Hence given that we are working with a pure quantum state, we can also expect that the entanglement entropies of a region and its complement will be the same, \ie $S(A)=S(\bar{A})$. Hence in eq.~\reef{mutualdef30}, $S(A^+)$ can replaced by $S(\bar{A}^+)$, the entanglement entropy of the disk enclosed by the second circle of radius $R_+$.
Similarly, $S(A^+\cup A^-)$ can be replaced by $S(\overline{A^+\cup A^-})$, the entanglement entropy 
of the annular strip between the two circles. Then the mutual information \reef{mutualdef30} becomes
\begin{equation}
I(A^+,A^-)=S_{disk}(A^-)+S_{disk}(\bar{A}^+)-S_{strip}(\overline{A^+\cup A^-})\,.\labell{mutualdef3}
\end{equation}
However, we must emphasize that in general, care must be taken with these substitutions, as will be discussed in section \ref{entropy}.

Now we wish to consider the mutual information $I(A^+,A^-)$ when the separation of the circles is small, \ie in the limit $\veps/R\rightarrow 0$.\footnote{There is an implicit order of limits here which still maintains $\veps\gg\delta$.} However, we must first designate the precise relation between the original radius $R$ and the two new radii $R_\pm$:
  \beq
R_-=R-\left(\frac12-\alpha\right)\veps\,, \qquad
 R_+=R+\left(\frac12+\alpha\right)\veps\,.
 \labell{eps0}
 \eeq 
Alternatively, we have
  \be
R=\frac{R_++R_-}{2} - \alpha\,\veps\,.
 \labell{radiusR}
 \ee 
We allow the dimensionless parameter $\alpha$ to take some value in $-1/2\le
\alpha\le 1/2$ so that the original radius lies in the range $R_-\le R\le
R_+$. However, ultimately we will fix $\alpha=0$ below.

If we take the limit $\veps\rightarrow 0$ at a conformal fixed point, the
mutual information $I(A^+,A^-)$ will diverge in a form reminiscent of
eq.~\reef{enti} with $\veps$ playing the role of the cut-off $\delta$. That is,
we will have an expansion of the form
\begin{equation}
I(A^+,A^-)= 2\pi R\left(\frac{\ta}{\veps}+\tb\right)-4\pi\,\tc_0+O(\veps)\,,
 \labell{mi1}
\end{equation}
where $\tilde{a}$, $\tilde{b}$ and $\tilde{c}_0$ are new constants. In fact,
since $I(A^+,A^-)$ is a well defined quantity, $\tilde{a}$, $\tilde{b}$ and
$\tilde{c}_0$ can all be regarded as physical quantities, \ie all of these coefficients are independent of
the regularization scheme. Our previous discussion of the entanglement entropy \reef{enti} provides a useful intuitive understanding of the nature of these parameters. Heuristically, we can say $\tilde{a}$ contains information about physics at high scales of order $1/\veps$ while $\tilde b$ 
tells us about the RG flow down to the infrared scale $1/R$.\footnote{Let us note here that in principle if the RG flow is driven by operators with sufficiently large anomalous dimensions, one may find additional `area law' terms with diverge in the limit of vanishing $\veps$ --- see further discussion in appendix \ref{kuprate}. \labell{foot88}} Finally, we can expect that the constant $\tc_0$ characterizes the fixed point theory, when defined with suitable precision -- as will be discussed in the next section. In comparing the above expression with eq.~\reef{enti}, one may observe 
that we have introduced an additional factor of 2 in the constant term in eq.~\reef{mi1} 
for convenience, \ie we will argue in section \ref{entropy} that $\tc_0=c_0$ under the appropriate conditions. Heuristically, one can think that this factor arises because in the $\veps\rightarrow 0$ limit, the mutual information yields twice the entanglement entropy of a single
circle, \ie $I(A^+,A^-)\simeq 2S(A)$ \cite{mut,split} --- of course, this na\"ive expectation
is spoiled by divergences.

Our proposal will be that $\tc_0$ will serve as the \c-function in our approach --- see discussion below. However, first we note that while $\tilde{c}_0$ is a well-defined 
regulator-independent quantity, it depends on our choice for the intermediate radius $R$, or equivalently on $\alpha$.
If we change $\alpha\rightarrow \alpha^\prime$, the expansion \reef{mi1} becomes
\begin{equation}
I(A^+,A^-)=2\pi R'\left(\frac{\ta}{\veps}+\tb\right)-4\pi\,\tc'_0+O(\veps)\,,\labell{expec}
\end{equation}
with
\begin{equation}
\tilde{c}_0^\prime=\tilde{c}_0+\frac{\alpha-\alpha^\prime}{2}\, \tilde{a}\, . \labell{changes}
\end{equation}
Hence depending on our choice of $\alpha$, the value of $\tc_0$ is contaminated by the coefficient
$\ta$, which depends on the details of the underlying QFT at high scales. That is, for generic values of $\alpha$, $\tc_0$ does
not correspond to an intrinsic quantity of the IR theory alone and so further attention is required in choosing $\alpha$ if a legitimate \c-theorem is to be defined here. However, we observe that if the constant term in the mutual information is decreasing from
UV to IR for some choice of $\alpha$, it is also decreasing for any other $\alpha$ since $\ta$ and hence
the difference in eq.~\reef{changes} remains constant. In closing, we note that the choice of $\alpha$ is creating an ambiguity here which is similar to that discussed for the entanglement entropy in the previous section, where the resolution in determining $R$ was limited by the UV regulator.\\

\noindent{\bf Strategy:} Let us take a moment to outline our overall strategy in using the mutual information to examine RG flows. As noted above, the mutual information is a finite regulator-independent quantity and in our geometric construction above, $R$ and $\veps$ are both macroscopic distance scales much larger that the UV cut-off. In other words, we are calculating the mutual information $I(A^+,A^-)$ for the renormalized quantum field theory in its continuum limit. Now we wish to examine an RG flow between two fixed points in this continuum theory and in such a situation, there must be some mass scales $m_i$ which are introduced in perturbing the UV fixed point to generate the flow. We will choose $1/\veps$ to be much larger than the largest of these scales, \ie $\veps\ll 1/m_i$. While $\veps$ is held fixed, the radius $R$ is our probe scale and we will vary the radius from $R\ll 1/m_i$ (still with $\veps\ll R$) where the mutual information is (only) probing the correlations in the UV CFT to $R\gg1/m_i$ where the IR physics of the second fixed point is being probed. In particular then, we expect to find
\beqa
\veps\ll R\ll 1/m_i&:&\qquad I(A^+,A^-)= 2\pi R\,\frac{\ta}{\veps}-4\pi\,\tc^{\ssc UV}_0+O(\veps)\,,\\
\veps\ll 1/m_i\ll R&:&\qquad I(A^+,A^-)= 2\pi R\left(\frac{\ta}{\veps}+\tb\right)-4\pi\,\tc^{\ssc IR}_0+O(\veps)\,.
\labell{forrest}
\eeqa
As we will discuss below, the area term appears with the same coefficient $\ta$ in both expressions since this is determined by the UV physics at the scale $1/\veps$. However, the RG flow will generally also induce a finite contribution proportional to the area, \ie $2\pi R\,\tb$, which appears in the mutual information at IR scales. Now the central object in our analysis is $\tc_0$, which is proposed to serve as the \c-function in this approach. More precisely, a \c-function will be constructed using the mutual information such that it matches $\tc_0$ at the conformal fixed points. In analogy to the discussion of the entropic \c-function \cite{proof,liu}, we may define
\beq
C(R) = \lim_{\veps\to0}\frac1{4\pi}\left(R\,\frac{\partial  I(A^+,A^-)}{\partial\,R}-I(A^+,A^-)\right)
\labell{guff}
\eeq
so that $C(R)=\tc_0$ for fixed points of the RG flows. In fact, a slightly refined version \reef{funix} of this expression will appear in proving the \F-theorem in section \ref{proof}.

\subsection{UV independence of $\tilde{c}_0$}
\labell{critb}

As described above, the mutual information \reef{mi1} provides a regulator-independent means of defining a candidate \c-function and so this approach automatically satisfies our first criterion for a proper \c-theorem. However, at an RG fixed point, the \c-function yields the parameter $\tc_0$, which fails to satisfy our second prerequisite for general $\alpha$ because it depends on physics at higher UV scales. Hence the question is if for some specific $\alpha$, $\tc_0$ is a quantity which is intrinsic to the fixed point of interest.

\begin{figure}[t]
\centering
   \subfloat[]{\includegraphics[width=.48\textwidth]{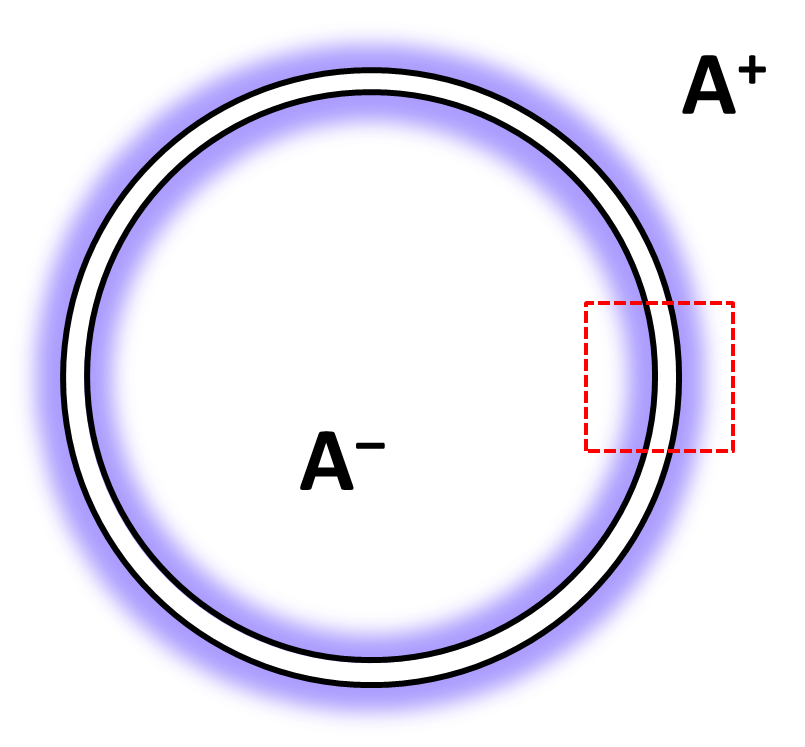}}
   \hspace{2.6cm}
   \subfloat[]{\includegraphics[width=.3\textwidth]{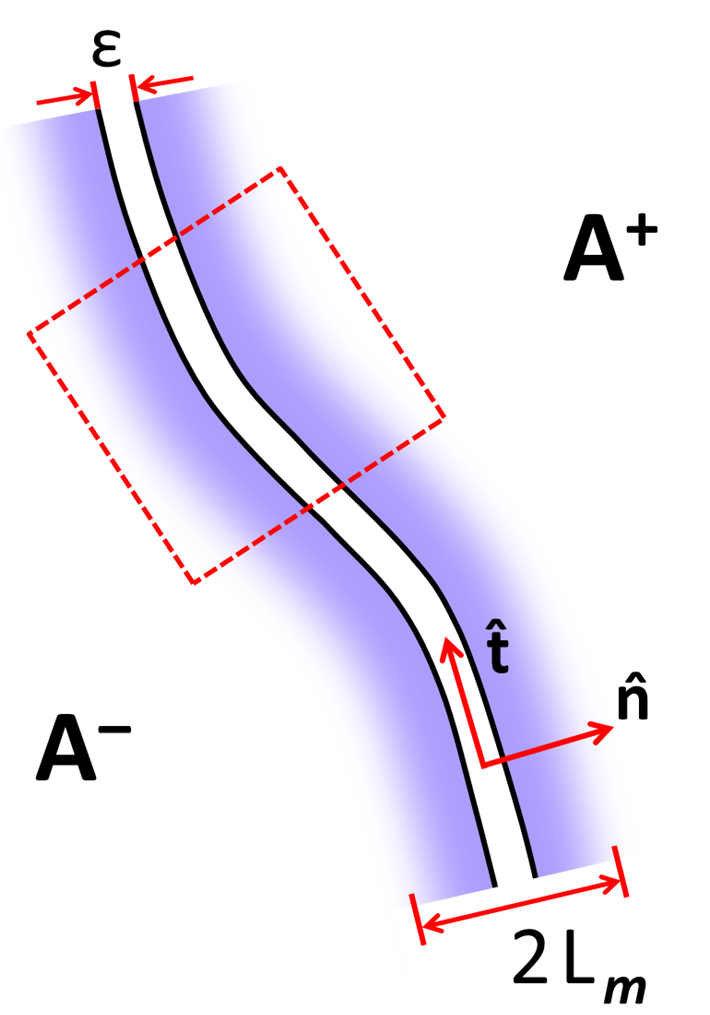}}
\bigskip
\caption{(Colour online) In both figures, the shaded areas represent regions on both sides of the boundaries where the correlations between $A^+$ and $A^-$ contributing to the mutual information are sensitive to higher UV scales. As indicated in panel (b), this band has a width of roughly $2L_m$ in the regime where $\veps\ll L_m$. These (nonconformal) correlations give local contributions calculable from a small region, such as that enclosed by the dashed (red) box.
In panel (a), these local regions tend to become flat in the limit of large $R$ and so the local factor can be written as a series in inverse powers of $R$, as in eq.~\reef{stinkpot}. Panel (b) illustrates the more general case where the local factor can be expressed in terms of local geometric quantities in terms of the unit normal and tangent vectors, $\un$ and $\ut$, as in eq.~\reef{inte}. } \labell{dashed}
\end{figure}
Let us consider evaluating the mutual information $I(A^+,A^-)$ at some RG fixed point. As prescribed above, the geometry is constructed with $R\gg\veps$. However, we also take the radius to be much larger than the distance scale set by the lowest mass $m$ appearing in the definition of the fixed point, \ie $R\gg L_m \sim 1/m$. As we are interested in the limit $\veps\to0$, we also suppose that $\veps\ll L_m$.
Now in this regime, the region where the correlations contributing to the mutual information are sensitive to $m$ is restricted to a narrow band near the annulus separating $A^+$ and $A^-$, as shown in figure \ref{dashed}. Beyond this region, the correlations between $A^+$ and $A^-$ are the purely conformal ones of the fixed point theory. Hence in the regime where $\veps\ll L_m \ll R$, the contribution to the mutual information, which is sensitive to $m$ and higher scales, will be local and extensive on the annulus. That is,  this high energy contribution will take the form
\beq
I(A^+,A^-)_{HE}=2\pi R\left(\sigma_0 + \frac{\sigma_1}{R} +  \frac{\sigma_{2}}{R^2}+\cdots\right)
\labell{stinkpot}
\eeq
where the factor in parentheses is some local contribution calculable from any small region, such as that enclosed by the dashed box in figure \ref{dashed}. If we increase the radius $R$ while keeping the size of the box fixed, %(along with $\veps$ and $L_m$), 
this region grows flatter and eventually becomes a straight strip of width $\veps$ in the limit $R\to\infty$. This is why the local factor can be arranged in the series in inverse powers of $R$ shown in eq.~\reef{stinkpot}. The first term $\sigma_0$ in this series corresponds to the local contribution for the straight strip.  Now we see that if $\sigma_1$ is nonvanishing, the second term in the local factor produces a constant contribution in $I(A^+,A^-)_{HE}$ which is independent of $R$. This constant $2\pi\sigma_1$ would then be included as a part of $-4\pi \tc_0$  in eq.~\reef{mi1} and hence our candidate \c-function would be sensitive to higher UV scales. Therefore our objective is to find a choice of $\alpha$ in eq.~\reef{eps0} which ensures that $\sigma_1$ vanishes.

To proceed, let us think about more general situations where the two boundaries defining the `interior' region $A^-$ and the `exterior' region $A^+$ are two parallel curves defining some curved strip of width $\veps$. The only restriction\footnote{We are also assuming that the strip is not self-intersecting.}  on the strip geometry will be that everywhere along the curves, the scale of their curvatures is much larger than both $\veps$ and $L_m$. Now as above in this regime, the high energy contribution to the mutual information will be local and extensive on this narrow ribbon. Hence, we will be able to write $I(A^+,A^-)_{HE}$ here in terms of local geometric quantities integrated along the strip. In particular, we can write 
\begin{equation}
I(A^+,A^-)_{HE}=\int ds\, \left(\sigma_0-\sigma_1 \, \un\cdot\partial_s\ut-\sigma_2\,\ut\cdot\partial_s^2\ut+\cdots\right)\,, \labell{inte}
\end{equation}
where $\un$ and $\ut$ are the unit normal and unit tangent vectors, respectively, while $s$ is the coordinate running along the length of the strip.\footnote{A variety of geometric identities have been used to simplify eq.~\reef{inte}. This expression contains all of the distinct geometric terms up to second order in $s$ derivatives, up to total derivatives.} In the case of a circular strip of radius $R$, this expression reduces to eq.~\reef{stinkpot}. Again, $\sigma_0$ corresponds to the local contribution for a straight strip of width $\veps$, while $\sigma_1$ gives the problematic contribution which would make $\tc_0$ sensitive to higher UV scales in our construction above.

At this point, we turn to the choice of $\alpha$ and in particular, we choose $\alpha=0$. The latter places the curve along which
 the integral is performed precisely half way between $A^-$ and $A^+$. Hence this choice is distinguished because it establishes a symmetry between $A^-$ and $A^+$ where $I_{HE}(A^+,A^-)=I_{HE}(A^-,A^+)$ and this symmetry will then be reflected in the expansion in eq.~\reef{inte}. Specifically, it requires that the local contribution must be unchanged when the direction of the normal vector $\un$ is reversed, \ie under the replacement $\un\to-\un$. Therefore, with the choice $\alpha=0$, the term proportional to $\un\cdot\partial_s\ut$ must be absent in eq.~\reef{inte} and there is no constant contribution in $I(A^+,A^-)_{HE}$. Hence $\tilde{c}_0$ is protected from the details of the UV physics with $\alpha=0$. Let us add that in the next section, we will show $\tc_0=c_0$, \ie the constant terms agree using
the mutual information and the entanglement entropy (with a covariant regulator). 
Now we have verified these results with various explicit calculations, however, before describing those, it is useful to examine eq.~\reef{inte} in more detail. 

First, recall that $\sigma_0$ corresponds to the local contribution for a straight strip of width $\veps$.  Further, we observe that the term proportional to $\sigma_0$ in eq.~\reef{inte} will make a contribution proportional to $2\pi R$ in the mutual information and in particular, this term will produce the contribution proportional to $\ta\,R/\veps$ in eq.~\reef{mi1}. Now considering eq.~(\ref{mutualdef}) for the mutual information, we see that the latter contribution must come from $S(A^+\cup A^-)$ since the entanglement entropies for the individual circular boundaries will be independent of $\veps$. Therefore combining these observations, we may identify the coefficient $\ta$ as that appearing in the universal contribution to the entanglement entropy of a straight strip. That is, in a three-dimensional quantum field theory, the latter takes the form, \eg see \cite{review,m1}:
\beq
S_{\ssc strip} = c_1\, \frac{L}\delta - \ta\,\frac{L}\veps +\cdots
\labell{strip}
\eeq
for a straight strip of length $L$ and width $\veps$ with $L\gg\veps$ and where the ellipsis denotes terms that vanish in the limit that $\delta,\veps\to0$.\footnote{Additional `area law' terms, which diverge in this limit, may appear in eq.~\reef{strip} if the RG flow is driven by nearly marginal operators --- see  appendix \ref{kuprate}.}
The universal coefficient $\ta$ is known for free scalar and fermion fields \cite{review}, and has been studied holographically \cite{m1}. In the present context, it is clear that this universal term is determined by the physics of high scales of order $1/\veps$ and hence in RG flows, $\tilde{a}$ depends entirely on the UV fixed point.

To confirm the previous arguments, we have computed mutual information in a holographic framework in appendix \ref{holographic}. As expected, we find that only
the case $\alpha=0$ gives place to a constant term which is independent of UV terms and in fact, the constant term appearing in the mutual information is precisely twice that
for a circle. That is, comparing eqs.~\reef{enti} and \reef{mi1}, we have $\tc_0=c_0$ in the holographic model. For $\alpha\neq 0$, the constant term is contaminated with the linear coefficient $\ta$, which we noted above is equivalent to the coefficient in the entanglement entropy of a straight strip.  We also computed the mutual information in a model with extensive mutual information in appendix \ref{extensive}. In this example, we can follow explicitly the renormalization group flow of the constant term and check that it decreases as $R$ increases and that in the infrared, it is independent of the details of the RG flow provided $\alpha=0$. 

Finally, this new approach should allow us to compute $\tc_0$ using a lattice regulator.
Figure \ref{mutu-1-5} shows the results of a preliminary calculation for a free massless scalar using a (square) lattice regulator.
In particular, figure \ref{mutu-1-5}(a) shows the mutual information $I(A,B)$ as a
function of the radius in lattice units $R/\delta$. Here we have set $\alpha=0$ and so $R=(R_++R_-)/2$. We approach the continuum limit by increasing $R$ in terms of the lattice spacing,
while holding the ratio $\veps/R=2/11$ fixed (or equivalently $R_+/R_-=6/5$ held fixed). As shown in the figure, the data does indeed seem to converge quite well. As in the entanglement entropy calculation shown in figure \ref{entro-circ}, the noise arises because the circumference of the entangling surfaces is uncertain at a resolution of the order of the lattice spacing. That is, there is an uncertainty of order the lattice spacing $\delta$ in the radius $R$ in the formula \reef{mi1} for the mutual information. Hence we can expect fluctuations of the order of $2\pi \tilde{a}\,\delta/\veps$ but these will vanish in the continuum limit where $\delta/\veps\rightarrow 0$. Recall from eq.~\reef{strip} that $\ta$ appears as the coefficient of the universal term in the entanglement entropy for a strip. Using this connection, the value of this coefficient can be computed analytically for a massless scalar \cite{review}:\footnote{It can also be obtained efficiently by directly evaluating the strip entropy on the square lattice.} $\tilde{a}\simeq0.0397$. Combining this value with the ratio $\veps/R=2/11$, our estimate for the expected fluctuations becomes $1.37\,\delta/R$, which seems to match quite well with the order of magnitude of the noise in the figure \ref{mutu-1-5}.
\begin{figure}[t]
  \centering
  %\subfloat[]
  \subfloat[]{\includegraphics[width=.39\textwidth]{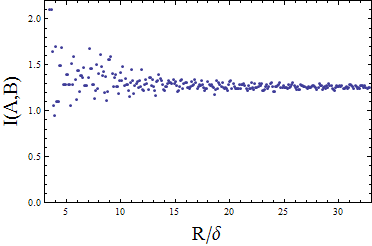}}
   \subfloat[]{\includegraphics[width=.4\textwidth]{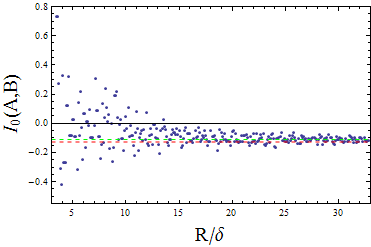}}
   % \hspace{1.cm}
  \caption{(Colour online) Panel (a) shows the mutual information $I(A^+,A^-)$ calculated on a square lattice for a free scalar. The mutual information is plotted as a function of $R/\delta$ while holding the ratio $\veps/R$ fixed with $\veps/R=2/11$. Panel (b) shows the same data as in panel (a) but now on the
vertical axis, we have plotted the quantity $I_0(A)=I(A^+,A^-)-2\pi R\,\ta/\veps$. For the limit of small $\veps/R$, this
quantity should converge to the universal value of the constant term $-4\pi \tc_0$. The green dashed line shows
the approximate value of $-4\pi \tc_0$ obtained by averaging over the values for $R\in (25,33)$. The exact value
shown with the red dashed line is slightly smaller.} \labell{mutu-1-5}
\end{figure}

In figure \ref{mutu-1-5}(b), we plot $I_0(A)=I(A^+,A^-)-2\pi R\,\ta/\veps$ as a function of $R/\delta$. Comparing with eq.~\reef{mi1}, we see that this
quantity should converge to the universal constant term $-4\pi \tc_0$ in the limit $\veps/R\to0$.\footnote{Note that $\tb=0$ in the present case since we are not considering an RG flow.} Averaging over the values in the figure for $R/\delta\in (25,33)$, our lattice calculations yield an approximate value of $4\pi \tc_0\simeq 0.110$.
Comparing this result with the exact value \cite{igor1,Klebfree},
\begin{equation}
 4\pi \coss \equiv \frac{1}{4}\left(\log2-\frac{3\zeta(3)}{2\pi^2}\right)\simeq 0.12761410956\,,\labell{plat}
\end{equation}
we find that our preliminary lattice calculation has already produced a good approximation. In fact, the difference of roughly $15\%$ between the exact and lattice values
is as good as we could expect since the lattice calculation is limited to $\veps/R=2/11\simeq0.18$, as well as $R/\delta\approx 30$. Of course, the lattice result can be improved by making $\veps/R$ smaller which in turn, requires considering circles of larger radii. 

Recall that the choice of $\alpha=0$ was essential to protect the constant term in the mutual information from the details of the UV physics.
That is, with a fixed large $R$ and $\alpha=0$, changing the
UV behavior in an RG flow to a given IR conformal fixed point will result in a renormalization of the `area' term in eq.~\reef{mi1} but
the constant term remains unchanged. Notice with eq.~\reef{inte}, we have framed our argument in a very general context and so
this property also holds for the mutual information across any smooth shape\footnote{Similarly, it also extends to smooth shapes in any odd
number of dimensions.} as we scale the size of the region to infinity. The key point is that we must adopt
the prescription of evaluating the area along the curve which is precisely half way between the boundaries of $A^+$ and $A^-$. 
However, for general shapes other than a circle, it is not expected that the constant term should decrease monotonically as the size of the region increases.

Finally, we note that our general argument for the UV independence of $\tilde{c}_0$ is very similar to those made for extracting a universal number for the constant term of the circle entropy \cite{liu}. However, those arguments require some specific properties for the regularization used in evaluating the entanglement entropy.
In particular, as noted above, the regulator must be covariant which ensures that the UV contributions to the entropy have a geometric expansion analogous to that shown in eq.~\reef{inte}. Further, the regulator must respect the symmetry $S(A)=S(\bar{A})$, where $\bar A$ is the complement of $A$. The latter again requires that the entanglement entropy will be invariant under the replacement $\un\to-\un$. This invariance then ensures that there can not be a term linear in the curvature of the entangling surface, \ie the second term in the expansion \reef{inte} must again vanish, and hence the constant term is protected from UV contributions.\\

At this point, we have established that the mutual information for the geometry illustrated in figure \ref{fig:Mutual0} can be used to construct a candidate `central charge' $\tc_0$ for three-dimensional conformal fixed points. In particular, this new charge satisfies the first two requirements which we set out in the introduction as necessary to properly establish a \c-theorem. Of course, it remains to be shown that this quantity can be extended to a \c-function $C(R)$ satisfying $C(R)=\tc_0$ at conformal fixed points and also satisfying the third criterion, \ie that the \c-function actually decreases monotonically along RG flows. In the next section, we will argue that generally we expect $\tc_0=c_0$, \ie our new universal coefficient will coincide with that extracted from the entanglement entropy (using a covariant regulator). The latter is already very suggestive with regards to the monotonic flow of $C(R)$ but we will return to this point in section \ref{proof}.

%%%%%%%%%%%%%%%%%%%%%%%%%%%%%%%%%%%%%%%%%%%%%%%%%%%%%%%%%%%%%%%%%%
%%%%%%%%%%%%%%%%%%%%%%%%%%%%%%%%%%%%%%%%%%%%%%%%%%%%%%%%%%%%%%%%%%
%%%%%%%%%%%%%%%%%%%%%%%%%%%%%%%%%%%%%%%%%%%%%%%%%%%%%%%%%%%%%%%%%%
%%%%%%%%%%%%%%%%%%%%%%%%%%%%%%%%%%%%%%%%%%%%%%%%%%%%%%%%%%%%%%%%%%

\section{Matching mutual information and entanglement entropy} \labell{entropy}

In the previous section, we focused on the expansion \reef{mi1} of the mutual information at a conformal fixed point. In particular, we showed that with $\alpha=0$, \ie with $R=(R_++R_-)/2$, the constant term is free of contamination from high energy scales. In this section, we discuss under which conditions the constant $\tc_0$ appearing there will equal $c_0$ in eq.~\reef{enti}, the constant appearing in the analogous expansion of the entanglement entropy of a circle. 

Recall that the purity of the vacuum state yields the identity $S(A)=S(\bar A)$ for the entanglement entropies of a region $A$ and its complement $\bar A$.\footnote{However, as we show below, the regularization can spoil this equality for subleading terms.} This identity then allows us to rewrite the original expression \reef{mutualdef30} for the mutual information $I(A^+,A^-)$ in terms of the entanglement entropies of two disks and of a thin annular strip, as in eq.~\reef{mutualdef3}. Now the assertion above is that under suitable conditions, the constant term $4\pi\tc_0$ appearing in eq.~\reef{mi1} is equivalent to twice $2\pi c_0$ appearing in eq.~\reef{enti} for a single disk. Given eq.~\reef{mutualdef3}, this match can only hold if the constant term in the entropy of the annular strip vanishes. Hence next we turn to understanding when the latter constant will vanish, \eg what conditions are required for the regularization. 

However, first let us remark that much of the following analysis still applies without applying $S(A)=S(\bar A)$. Hence while the discussion is facilitated by considering the entanglement entropy of the annular strip, the same analysis establishes that the constant term vanishes in the entropy of the complementary region, $A^+\cup A^-$.
The robust result is then that if the constant term vanishes in the entropy of $A^+\cup A^-$, then the constant term in the mutual information is given by the sum of those in $S_-(R)$ and $S_+(R)$, the entropies of the interior and exterior of a circle of radius $R$, respectively. Since our discussion is implicitly considering a conformal fixed point, the radius $R$ of the circle in either of these entropies can be set freely.

The entanglement entropy of the narrow annular strip is a purely UV contribution, \ie it probes correlations at the (length) scale $\veps$. This is because contributions of the correlation functions across distances much larger than the width $\veps$ cancel due to the opposite contributions coming from the two sides of the strip. This cancellation becomes more transparent when thinking in terms of the \ren entropies. For the \ren entropies, there are two twist operators (\eg see \cite{calcar1}) running along each of the boundaries of the strip but with opposite orientations. That is, opposite `twists' will be induced in a local field operator if it is carried in the same direction around these line operators of either edge of the strip. Because of their opposite orientations, the connected correlators of the twist operators with a distant operator should go to zero with
$\veps$. In other words, we may think that the first term in the effective OPE (\eg see \cite{twist}) of the parallel twist
operators will be proportional to the identity operator. 

The example of a free scalar field is useful to illustrate this cancellation. For a free scalar, the Renyi entropies for a region $V$ can be expressed as \cite{review},
\begin{equation}
 S_n(V)=\frac1{1-n} \sum_{k=1}^{n-1}\ \log Z\!\left[e^{2 \pi i\frac{k}{n}}\right]
\end{equation}
where $Z[e^{i a}]$ is the partition function where a phase $e^{i a}$ is imposed as a boundary condition on the field in approaching $V$ from above and below. This boundary condition can be enforced by coupling a (complex) scalar field to an external gauge field $A_\mu$, which is pure gauge everywhere except at the boundary of $V$. To be concrete consider an infinite straight strip running along the direction $\ut$ and with the edges separated by $\vec{\veps}$. One choice of the gauge field is
\begin{equation}
 A_\mu(x)=a\, \veps_{\mu\nu\sigma}\, {\hat t}^\nu \left(\frac{x^\sigma}{|x|^2} -\frac{(x+\veps)^\sigma}{|x+\veps|^2}\right)\,.
\end{equation}
Hence the magnitude of this gauge field vanishes as $|A|\sim \veps/|x|^2$ for small $\veps$ and fixed $x$, and the effect of the two boundaries
on distant regions vanish at least linearly in $\veps$.

The UV character of the entanglement entropy of the thin annular strip is also evident in holographic models. As discussed in appendix \ref{holographic}, the minimal surface determining the holographic entanglement entropy in this case is an annular region in the bulk connecting the boundaries of the strip and it resides entirely near the AdS boundary --- see figure \ref{fig:Mutual2}. A similar cancellation of large distance contributions from the two edges of the thin strip is also seen in the extensive mutual information model considered in appendix \ref{extensive}. The contributions of the two boundaries have different signs due to the factor $\ut_x\cdot\ut_y$ in eq.~(\ref{yy1}), where $\ut_x$ and $\ut_y$ are the tangent vectors of the two boundaries.

Because of the UV character of the entanglement entropy of a narrow strip, it is natural to think that it should be local and extensive, \ie it can be written as an integral of local quantities along the length of the strip. One can argue that deviations from extensivity will go to zero with some positive power of the width $\veps$ as follows: Consider the
mutual information between small portions of the strip with a linear size $\veps$ and which are separated by some distance $|x|$ along the strip. By the definition \reef{mutualdef}, the mutual information measures precisely the degree of nonextensivity of the entropy between two regions. This mutual information should go to
zero with a positive power of $\veps/|x|$ due to clustering of correlation functions and conformal invariance. This is known to hold for free fields \cite{review}
and was recently proved for distant spherical regions in any conformal field theory (in any number of dimensions) \cite{cardymutual}. In these cases, mutual information between distant portions on the strip would decays as $(\veps/|x|)^{\sigma}$ where the power satisfies $4\Delta\le\sigma\le 6$ where $\Delta$ is the lowest dimension in the spectrum of CFT (\ie $\Delta=1/2$ for the conformal scalar in three dimensions) or $\sigma=6$ if $\Delta>3/2$.

As noted above, the local and extensive character of the entropy of the annular strip means it is given by an integral of local objects along the length of the strip. Therefore the same argument given in section \ref{critb} for the vanishing of the constant term in high energy contribution to the mutual information serves here to show the vanishing of the constant term in the entanglement entropy of the strip. The only complication is that rather than being purely geometric, the local quantities appearing in the integral are introduced by the UV regulator.\footnote{This complication does not arise for the mutual information since the latter
is independent of the regulator.} According to the discussion in section \ref{critb}, a minimal requirement to ensure that the regulator does not spoil our argument is that the regularization respect the symmetry interchanging $A^+$ and $A^-$, which inverts the curvature of the strip. This is achieved, for example, if the divergences of the two edges of the strip are regulated with the same geometric prescription because the curvatures of the two boundaries have opposite signs (taking the normal to
the boundary pointing outwards from the strip). In contrast, this symmetry is not respected with a square lattice since the details of the regularization depend strongly on
the precise position of the boundaries. With the symmetry requirement, the annular strip's entropy is given by an expansion as in eq.~(\ref{inte}) and the parity argument eliminates the term proportional to the curvature of the strip, \ie $\un\cdot\partial_s\ut$. As a consequence, there is no constant contribution independent of the radius $R$ and from eq.~\reef{mutualdef3}, the constant term in the mutual information coincides with twice that in the entanglement entropy of a disk --- or the sum of the constant
terms in $S_+(R)$ and $S_-(R)$.

The vanishing of the constant term in the entanglement entropy of the annular strip was verified for holographic entanglement entropy in appendix \ref{holographic}. Below we confirm our arguments for a free scalar field with heat kernel and lattice regularizations.

However, before proceeding, we must say that there is one situation which seems to evade these arguments, namely topologically ordered phases. Here we are referring to phases which despite being gapped %(\ie having zero correlation length in the IR)
 still have some kind of long range entanglement, \eg see \cite{topogap}. In this situation, the edges of the strip decouple but the entanglement entropy of the annulus will still have a topological contribution which is constant, \ie independent of $R$. This contribution is clearly nonlocal and depends on the organization of entanglement all along the closed edges of the annular strip. However, we expect that when $\veps$ is smaller than the correlation length of the topological sector, the annulus contribution becomes local and as a result, the corresponding constant term vanishes. We return to this point in section \ref{discuss}.

\subsection{Scalar field} \labell{scall}

In a conformal field theory, the constant term in the entanglement entropy \reef{enti} across a circle is equal to the constant term in the free energy of a three-dimensional Euclidean sphere \cite{CHM},\footnote{Similar regularization ambiguities arise in identifying the universal
coefficient using the free energy of a $d$-sphere in odd dimensions.} \ie $F=-\log Z[S^3]$. This provides a relatively simple approach to evaluate $c_0$ for free fields. In this case, one of
the best methods to calculate the free energy is using zeta-function regularization, which eliminates all of the divergent terms and leaves simply the constant term. We already quoted the result for a conformally coupled free scalar \cite{igor1,Klebfree} in eq.~\reef{plat} when comparing to the results from our lattice calculations using the mutual information approach.
%\begin{equation}
% c_0^{\textrm{scalar}}\equiv \frac{1}{2^4}(2\log2-\frac{3\zeta(3)}{\pi^2})\approx 0.06380705478\,.\labell{platx}
%\end{equation}
Let us add that this number coincides with that found in \cite{Dowk0} using a slightly different calculation. Using a radial lattice,  the same value for $\coss$
was also obtained numerically from the circle entropy in \cite{liu} --- see also \cite{Klebanov2012}. However, we note that in this numerical calculation, the authors had to choose a particular definition $R=(n+1/2)\,\delta$ of the circle radius in terms of the lattice spacing $\delta$ to reproduce the `correct' value of $\coss$. The prescription $R=(n+1/2)\,\delta$ also gives the correct constant for a free fermion \cite{Safdi2012}. We discuss the reason for the success of this choice in the radial lattice below.

We now show that the constant in the mutual information \reef{mi1} coincides with $4\pi\coss$ for a free scalar, as expected. According to previous discussion, this agreement will follow provided we impose the same cut-off for both, $S_+(R)$ and $S_-(R)$. Of course,
the same cut-off, which is imposed to regulate the divergent contributions of the two circles, must also be used for the inner and outer
boundaries of the strip.

In appendix \ref{scalar}, we review the calculation of the entanglement entropy for the circle by conformally mapping the circle causal domain to the space $H^2\times R$, where $H^2$ is the two-dimensional hyperbolic space \cite{CHM}. The calculation reduces to the integration of the local entropy density at a fixed temperature over the hyperbolic space. One must then regulate the volume of $H^2$ to get a finite entropy. This can be done, for example, by integrating the entropy density in $H^2$ over a domain which corresponds to the region $|x|<R-\delta$ on the disk enclosed by the original circle, where $\delta$ is the short-distance cut-off in the CFT. This calculation yields
\begin{equation}
c_0=\frac{3}{2}\,\coss\,\labell{vv11}
\end{equation}
and hence the result seems to be off by a factor of $3/2$.  We might generalize this scheme by choosing the domain in $H^2$ to correspond to the disk extending out to $r_{\textrm{max}}=R-\hat\delta$ where
\begin{equation}
\hat\delta=\delta\,f(\delta/R) \quad {\rm with}\ \ f(\delta/R) = f(0) + f'(0) \frac{\delta}{R} + f''(0) \frac{\delta^2}{R^2}+\cdots\labell{expa11}\,.
\end{equation}
We are assuming that $f(\delta/R)$ has a well behaved Taylor expansion around $\delta/R=0$. 
In this case, if we express the resulting entanglement entropy with an expansion in terms of $\delta/R$, the constant becomes
\begin{equation}
c_0=\left(\frac{3}{2}+\frac{f'(0)}{f(0)}\right)\coss\,.\labell{fitoz}
\end{equation}
This result is a manifestation of the regulator ambiguities in the constant term in the entropy.

Above we noted that a robust result would be that that if the constant term vanishes in the entropy of the strip, then the constant term in the mutual information is given by the sum of those in $S_-(R)$ and $S_+(R)$, the entropies of the interior and exterior of a circle of radius $R$. Further, we are guaranteed that the constant contribution vanishes for the annular strip if we use the same geometric regulator at both boundaries. Now as above, the calculation of the entanglement entropy of this exterior region can again be related to a calculation of thermal entropy in the same space $H^2\times R$  \cite{twist,Renyi}. In addition, the calculation of $S_-(R)$ was regulated by choosing a region  $|x|=R-\delta$ and so we must impose the same short-distance cut-off $\delta$ in calculating $S_+(R)$.  That is, we should again only consider contributions up to a distance $\delta$ from the boundary and so we consider the region $|x|\ge R+\delta$ in calculating $S_+(R)$. While the latter region can be mapped directly to a domain on the hyperbolic plane, it is more informative to first map this region to the interior of the circle using the inversion 
\begin{equation}
 \hat x^\mu= \frac{R^2}{|x|^2}\,x^\mu\,. \labell{infert}
\end{equation}
With this conformal transformation the cut-off $\delta$ for the exterior region is mapped to the cut-off
\begin{equation} 
\hat\delta=R-\frac{R^2}{R+\delta}=\delta\,\left(1-\frac{\delta}{R}+\frac{\delta^2}{R^2}+\cdots\right) 
\labell{wacka}
\end{equation}
for the inner disk. Comparing the above with eq.~(\ref{expa11}), we find with eq.~\reef{fitoz} that the constant term for $S_+(R)$ is
\begin{equation}
 c_0=\frac{1}{2}\,\coss\,.\labell{vv21}
\end{equation}
Now the constant appearing in the mutual information \reef{mi1} is just the sum of those appearing in eqs.~(\ref{vv11}) and (\ref{vv21}) and so we find
\begin{equation}
\tc_0=\frac{c_0^++c_0^-}{2}=\coss\,. \labell{sdf}
\end{equation}
Hence despite the ambiguities in defining the individual entanglement entropies, the mutual information has reproduced precisely coefficient expected from the calculation of the free energy.

It is instructive to test the robustness of this calculation using the general scheme outlined in eq.~(\ref{expa11}). Na\"ively, averaging the contributions of
the inner and outer regions with the same $\hat\delta$ only gives the desired constant as in eq.~(\ref{sdf}) if $f'(0)=0$.\footnote{Analogous calculations in higher odd
dimensions would require that all of the odd derivatives of $f(\delta/R)$ vanish.} This apparent contradiction is easily understood and corrected by realizing that our specification was that the regularization scheme must have a geometric character, \eg we should be able to apply it for general boundaries. In making the short-distance cut-off
$\hat\delta$ explicitly depend on $R$ in eq.~(\ref{expa11}), we are implicitly constructing a regularization procedure specific to the circle and the particular calculation of interest. However, we can restore the geometric nature of the cut-off by interpreting the $R$-dependence in eq.~(\ref{expa11}) as dependence on the curvature of the boundary (in which case, it could be applied with a generic boundary geometry). Then, since the curvature, \ie $\un\cdot\partial_s\ut$, of the circle is reversed when it is the boundary of the exterior region from when it is the boundary of the inner disk, we must choose
\begin{equation}
 \delta_-=\delta\,f(\delta/R)\quad{\rm and}\qquad \delta_+=\delta\,f(-\delta/R)\,.\labell{epp}
\end{equation}
With this additional sign reversal, it is straightforward to verify that this geometric cut-off again reproduces the correct result as in eq.~\reef{sdf}.

Applying an analogous geometric cut-off in higher dimensions yields a result for $S_+(R)+S_-(R)$ with specific parity
because of the symmetry of the conformal transformation \reef{infert}. That is, this sum only contains even powers of $\delta/R$ for even
dimensions and odd powers for odd dimensions, except for the universal terms, \ie the constant term in odd dimensions and the $\log(\delta)$ term in even dimensions. Hence these universal contributions are independent of any of details in the choice of the function defining the cut-off in eq.~\reef{epp}. Similarly, the entanglement entropy of the annular strip cannot generate any contributions with the wrong parity and hence the universal coefficient in the mutual information is robust against the freedom in defining the (geometric) cut-off.\footnote{We might note that the cut-off used in \cite{Klebfree} in a similar
calculation of the constant term of the entanglement entropy --- see the expansion in eq.~(\ref{parti}) ---
gives the correct result without summing the contributions of the inner and outer regions. This calculation succeeds because this particular cut-off happens to be invariant under inversion \reef{infert}. That is, in this case
\begin{equation}
 \hat\delta[\delta,R]=R-\frac{R^2}{R+\hat\delta[\delta,-R]}\,.\nonumber
\end{equation}
Therefore in this case, the contributions of $S_+(R)$ and $S_-(R)$ are equal and it is sufficient to compute only $S_-(R)$. This choice of cut-off was motivated by comparing with the cut-off appearing the corresponding holographic calculations, which respects the conformal inversion invariance.}

Exactly the same approach solves the mismatch found for the calculation of $c_0$ with a mapping to de Sitter space, as described in appendix \ref{scalar}. In the mapping to de
Sitter space, the constant term is given directly by the corresponding constant in the free energy. This also explains why the results coincide with those
obtained from the zeta-function regularization since this latter is clearly symmetric under inversions.

\subsubsection*{Lattice calculations}

With these new insights, we can also understand why the numerical results using a radial lattice yield the correct constant provided that the circle radius is defined as
$R=(n+1/2)\,\delta$ with $n$ being the number of lattice points inside the circle, and $\delta$, the lattice spacing \cite{liu,Klebanov2012,Safdi2012}. The entropies
of the disk including the first $n$ points and the complementary outer region, extending from the point $n+1$ to infinity, are equal on the lattice due to the purity of the vacuum state. The prescription $R=(n+1/2)\,\delta$ simply ensures that these contributions give the same constant term once one subtracts the contribution linear in the variable $R$ from the entropy. The lattice representation of the annular strip includes the radial points from
$n+1$ to $m$, in the limit $n\rightarrow \infty$ with $m-n\ll n$. With the same cut-off prescription as above for the circle entropy, this is interpreted as an annulus extending from $r=R-\veps/2=(n+1/2)\,\delta$ to $r=R+\veps/2=(m+1/2)\,\delta$. The general argument indicates that the corresponding entanglement entropy has a vanishing
constant contribution in the small $\veps$ expansion. Thus, the constant term in the mutual information must coincide with twice the numerical result for the
circle with this prescription.

In order to verify this reasoning, we calculated the mutual information for the concentric circles numerically on a radial lattice. We computed mutual information for concentric circles with a radius $R/\delta$ from $30$ to $100$, and with $\veps/\delta$ from $4$ to $10$. We used an infrared cut-off for the total lattice size from $500$ to $1000$ lattice points and fitted the entropies with the infrared cut-off first --- the details of this procedure can be found in \cite{loha}. Then we fitted the mutual information for fixed $\veps$ as $x_1 (R/\delta)+x_2 +x_3 (R/\delta)^{-1}$. The constant term $x_2$ was subsequently fitted for different $\veps$ as $y_1 +y_2 (\delta/\veps)$. In this way, we found that the constant term $y_1\simeq 0.1320 $, which is compatible with the analytical result for the constant term on the mutual information of a scalar $4\pi \coss\simeq 0.1276$, as given in eq.~(\ref{plat}), within numerical errors, \ie roughly $3\%$ for our simulations.

%rob
Let us now turn to regulating the scalar theory with a square lattice. As illustrated in figure \ref{mutu-1-5}, the expected value for $c_0$ for a scalar can be extracted from an explicit numerical computation of the mutual information of concentric circles on a square lattice. However, the constant term cannot be obtained directly from the circle entropy alone, as illustrated in figure \ref{entro-circ}. These results seems to be at odds with the arguments in the above discussion. However, we believe that the resolution is that the square lattice regularization does not allow the contribution of the entropy of the strip to be written as a geometric integral. More precisely, even if this contribution is local in sizes of the order $\veps$ along the strip, the quantity that is integrated will break rotational symmetry and also is not invariant under translations. The contribution will fluctuate according to the precise position of the strip. Of course, the same holds for the circle entropies but there is a cancellation when these quantities are combined in the mutual information.

Still this might suggest that the circle entropies can still be used to get the correct result for $c_0$ with a square lattice, if we  follow a slightly more ellaborate prescription: For a circle of radius $R$ in the plane, let $S(R)$ be the entropy of the lattice points inside the circle. The entropy has an area term $2 \pi a R/\delta$ proportional to $R$, where the latter can be determined with increasing precision in the lattice for large $R$. Then let us subtract this area term (as was done in figure \ref{entro-circ} for a circle) but then average the result over different $R$ to eliminate the fluctuations. For a discretization of the radial variable, $R_i=R_0+i \Delta R$ with $i=0,1,\cdots$  with some small interval $\Delta R$, we can construct the following function
\be
\bar{S}(R_j)=\frac{1}{j}\sum_{i<j} (S(R_i)-2\pi a \frac{R_i}{\delta})\,.
\ee
The idea would be that $\bar{S}(R_j)$ could converge to $-2\pi c_0$ for large $j$. 
This procedure does indeed give results which are near the actual value of $c_0$. For the points shown in figure \ref{entro-circ} involving circles with radii up to $35$  (and where we
took equally spaced radii\footnote{Our procedure was to define the disk for which we evaluate $S(R)$ by draw an imaginary circle on the square lattice and including all points within the circle as comprising the disk. This allows the circles to have an arbitrary radius and allows the step size to be less than the actual lattice spacing.} with steps of $0.2$), we found $2\pi c_0=0.0622$. The correct value is $0.0638$, and so the error is only roughly $2.5\%$. However, it seems that the ``noise'' in figure \ref{entro-circ} is not as random as one could have expected. In particular, after some critical radius, further averaging does not seem to improve the result --- see figure \ref{random}. Further, changing the step $\Delta R$ in averaging the radius,  we obtain results that do not seem to coverge to the same value (and it is again unclear whether the sequence converges at all) ---  see figure \ref{random}. Changing the point where the circle is centered also changes the result, pointing to the lack of translational invariance in this averaging procedure. 
\begin{figure}[t]
  \centering
  %\subfloat[]
 {\includegraphics[width=.49\textwidth]{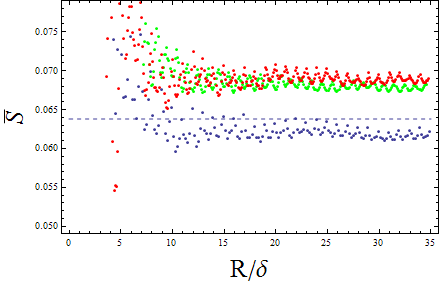}}
   % \hspace{1.cm}
  \caption{(Colour online) The averaged constant term in the entropy of circles $\bar{S}(R)$ in a square lattice as a function of the radius in lattice units, $R/\delta$. The blue points are for a radius step $\Delta R=0.2$ (as in figure \ref{entro-circ}). The red points are for $\Delta R=(3\pi)^{-1}\sim 0.106$. The green points are for the same $\Delta R$ but the circles are centered at the point $(1/3,1/2)$ in lattice units, rather than at the origin. The correct value of $2\pi c_0$ is shown with the dashed line.} \labell{random}
\end{figure}
   
\bigskip

Summarizing our discussion in this section, we have shown the universal constant $\tc_0$ appearing in the expansion of the mutual information \reef{mi1} matches $c_0$ appearing in the entanglement entropy of a circle \reef{enti} under appropriate conditions. A more robust result was that $\tc_0$ can be computed by evaluating the constant term in the sum of the entropies of a disk and of the complementary exterior region. However, the latter equality still requires a geometric regulator that is applied in the same way in calculating both entropies. The latter ensures that no contribution to the constant comes from the annular region. Being able to evaluate the universal constant from the entanglement entropy of disk with an appropriate regulator is very useful  from a practical point of view since the calculation of the mutual information function is generally much more difficult. In the conformal case, summing the two entropies also ensures that the resulting coefficient is invariant under conformal inversions. In this case, the inversion symmetry guarantees that the calculation of $\tc_0$ (or more generally, the universal contribution in higher dimensions) respects the na\"ive symmetry $S(V)\leftrightarrow S(\bar{V})$ of the entanglement entropy. However, for general lattice models, there is certainly no simple procedure which allows us to extract the correct universal coefficient $c_0$ from data for the circle entropies alone. Rather, it seems that an actual computation of mutual information is mandatory to produce the correct result for the constant term. 

\section{Mutual information and proof of the \F-theorem}
\labell{proof}

The proof \cite{proof} of the \F-theorem in $d=3$  uses strong subadditivity (SSA) of the
entanglement entropy of boosted circles located on a common null cone, as
depicted in figure \ref{cone}(a). Strong subadditivity is the
inequality
\begin{equation}
S(A)+S(B)\ge S(A\cap B)+S(A \cup B)\,, \labell{ssaa}
\end{equation}
for any spatial regions $A$ and $B$ which lie in a common Cauchy surface. Using many circles rotated around the axis of the light cone, one arrives at
an inequality involving ``wiggly circles'', as shown in figure \ref{cone}(b). To be more precise, the proof is based on the following assumptions:
\it
\begin{enumerate}[(a),topsep=2pt] 
\itemsep-1pt
\item There is a Lorentz invariant regularization of entanglement entropy which is defined for regions with smooth boundaries except for a finite number of corners lying on null planes.
\item This regulated entanglement entropy satisfies strong subadditivity at least for sets which cut each other producing new corners on null planes --- such as the circles in
figure \ref{cone}(a).
\item The wiggly circles have an entanglement entropy which approaches that of a circle with the same perimeter, as the number $N$ of null corners goes to infinity.
\end{enumerate}
\rm
\begin{figure}[t]
  \centering
  \subfloat[]{\includegraphics[width=.4\textwidth]{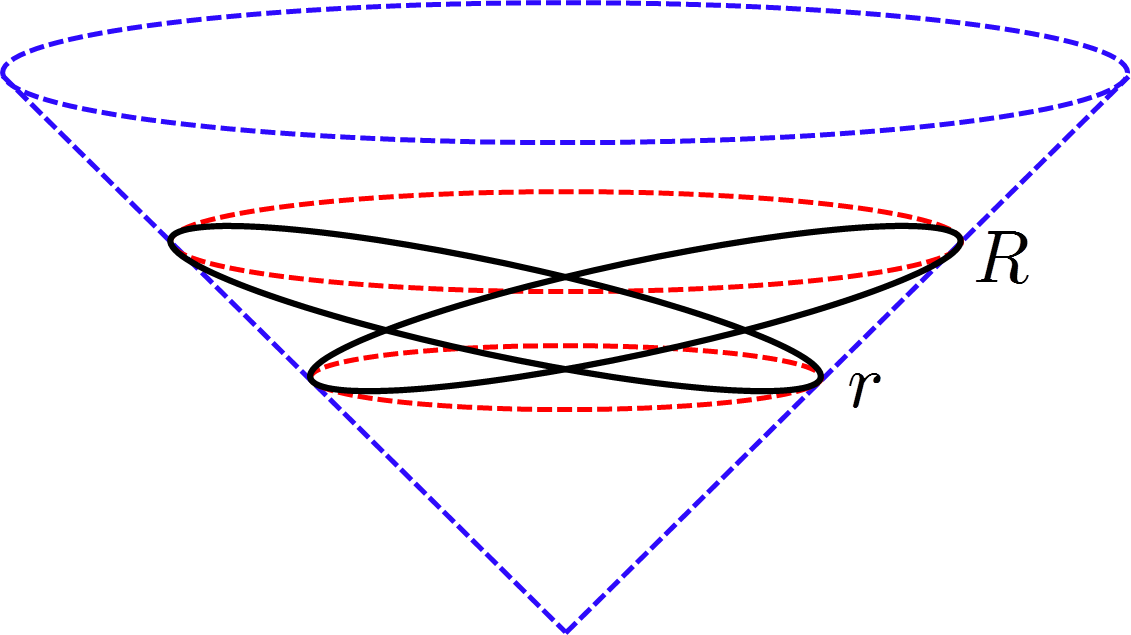}}
  \hspace{24pt}
   \subfloat[]{\includegraphics[width=.4\textwidth]{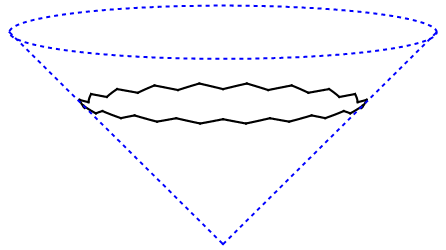}}
\bigskip
\caption{(Colour online) (a) Two (black) boosted circles lying on a common light cone cut across each other. The (red) constant time circles
on the cone, of radius $R$ and $r$, are tangent to the boosted circles. (b) An example of a wiggly circle appearing in the right-hand side of the SSA
relation between $N$ boosted circles rotated by angles $(2\pi/N) i$ with $i=0,1,\cdots,N-1$, around the axis of the null cone. The resulting wiggly circle is composed of $2N$ arcs.}
\labell{cone}
\end{figure}

These assumptions all refer to an essential new geometric feature that appears in the proof,  \ie the entangling surfaces contain corners lying on a null plane which arise where the circles described above cross each other -- for simplicity, we refer to these corners as `null cusps'. In a more conventional situation, where the entangling surface includes corners lie in a spatial plane, the entanglement entropy acquires additional logarithmic contributions, \eg see \cite{angle,HiraTak,pablo}. The appearance of such terms in the entanglement entropy of the wiggly circles, illustrated in figure \ref{cone}(b), would eliminate the power of SSA in proving the \F-theorem. The second and third assumptions above are based on the idea that there is no intrinsic local measure of the `angle' for a null cusp \cite{proof}. Hence, no local contribution from these features is expected to appear in the corresponding entanglement entropy, and in particular, no logarithmically divergent terms should appear.\footnote{In particular, if $\ut_1$ and $\ut_2$ are the two unit tangent vectors to the entangling surface at (\ie on either side of) the corner point, we have $\ut_1\cdot\ut_2=1$ when the corner lies in a null plane, \ie $\ut_1-\ut_2$ is a null vector. If the corner is close to being a null cusp, we have either $\ut_1\cdot\ut_2=\cos(\pi-\theta)<1$ when they lie in a common spatial plane with $\theta$ the deviation angle, or $\ut_1\cdot\ut_2=\cosh\beta>1$ when they lie in a common time-like plane with $\beta$ the relative boost parameter. The coefficient $s(\theta)$ of the logarithmic term in the entanglement entropy for a spatial corner  vanishes as $-k\, \theta^2$ for small $\theta$ \cite{angle,pablo}. For the time-like case, the coefficient of the logarithmic term can be found with the analytic continuation to imaginary angles $\theta=i \beta$ of the previous result, and we have $s(\beta)\sim k\, \beta^2$ for small $\beta$. Note that, SSA requires the coefficient of the log contributions to be negative for a spatial corner, but in contrast then, the coefficient of logarithmic term for a time-like corner has a positive sign.}

With these assumptions in mind, let $R$ and $r$ be the radius of the circles at
constant time which are tangent to the boosted circles  --- see figure
\ref{cone}. Then the SSA inequality for $N$ rotated boosted circles can be written as (see
\cite{proof} for details)
\begin{equation}
S(\sqrt{R r})\geq \frac{1}{N}\sum_{i=1}^{N}\tilde{S}\left( \frac{2 r R}{R+r-(R-r)\cos (
\frac{\pi i}{N})}\right)\,.
\end{equation}
Here $S(x)$ is the entanglement entropy of a circle of radius $x$ and $\tilde{S}(x)$ denotes the entropy of a wiggly circle with the same perimeter as a
circle of radius $x$. By assumption (c), taking the limit $N\rightarrow \infty$ converts $\tilde{S}$ into $S$ and this inequality becomes one involving
circle entropies alone.  The latter turns out to be equivalent to
\begin{equation}
S^{^{\prime \prime }}(R)\leq 0 \,.\labell{infinit}
\end{equation}
Then, using the \c-function
\begin{equation}
C(R)=\frac1{2\pi}\Big[R \,S^{\prime}(R)-S(R)\Big]\,,\labell{funi}
\end{equation}
we have
\begin{equation}
C^{\prime}(R)=\frac{R}{2\pi} \, S^{\prime\prime}(R)\le 0\,.\labell{ghgh}
\end{equation}
Now that according to eq.~(\ref{enti}), $C(R)$ coincides with the constant $c_0$ at conformal fixed points. Hence the fact that $C(R)$
decreases with increasing $R$ establishes the desired monotonicity property: $c_0^{\ssc{UV}}>c_0^{\ssc{IR}}$.

Now, we have shown mutual information can be used to provide a covariant regularization of
entanglement entropy  and so it is natural to revisit the assumptions of the
above proof to see how they may be accommodated in this regularization. There is a complication,
however, since mutual information gives a regularization introducing a cut-off
$\veps$, which is actually a distance in space. Hence we can regard it as a regularization
depending on a new element, the `framing' of the separation between the two
entangling surfaces. We might note that analogous framing regularizations have been used in the
literature studying Wilson loops \cite{framing}. In spite of the complications
introduced in the geometry, we find it valuable to go through the following
arguments, since it will help to clarify the correctness of the assumptions
(a)$-$(c) of the \F-theorem.

\subsubsection*{Eliminating frame dependence}

Suppose we have a generic region $A$ bounded by a smooth entangling curve ${\Gamma}(s)$,
parametrized by distance $s$ along the curve from some reference point. We would like to regulate the entanglement entropy across $\Gamma$
using mutual information, however, in order to obtain a finite result depending only on region $A$, we must first eliminate the frame dependence noted above. 
We will then also demonstrate that the resulting entropy also satisfies assumptions (a)$-$(c) above.  We begin
regulating the entanglement entropy of $\Gamma$ by evaluating the mutual information between regions defined by 
\beq
\Gamma_\pm\equiv {\Gamma}(s)\pm\frac12\, \veps(s) \,\un(s)
\labell{boundaries}
\eeq
where $\Gamma_+$ corresponds to the boundary of exterior region $A^+$ and $\Gamma_-$
corresponding to the boundary of interior region $A^-$, as illustrated in figure \ref{fi}(a). Further $\un(s)$ is a spatial unit
vector normal to $\Gamma$ and as in section \ref{newdef}, $\veps(s)$ is a macroscopic (\ie $\veps(s)\gg\delta$) but small distance which serves as the cut-off in the mutual information. In our general discussion here, we allow $\veps(s)$ to vary smoothly along the curve.\footnote{We also ask $\veps^\prime(s)$ to be of the same
order of $\veps$, such that $\veps^\prime(s)/\veps(s)$ converges to a finite number as the limit $\veps\rightarrow 0$ is taken.} 
Then, we expect
\begin{equation}
I(A^+,A^-)= \tilde{a}\, \int_\Gamma  \frac{ds}{\veps(s)}+ I_0(A)+ {\cal O}(\veps)\,,\labell{gamma}
\end{equation}
where $I_0(A)$ is a physical quantity which does not depend on the framing
$\veps(s)$.\footnote{In principle, there may be additional extensive contributions involving unusual (negative) powers of $\veps(s)$, as discussed in appendix \ref{kuprate}. However, the following arguments still carry through essentially unchanged in such a situtation.}  In the following, it is this quantity $I_0(A)$ which will play the role of the regulated entropy in the original proof of the \F-theorem discussed above. Comparing the above with eq.~\reef{mi1} for a circular entangling curve (in a fixed point and with constant $\veps$), we would have $I_0(A)= 2\pi R\,\tb -4\pi \tc_0$. Further let us note that, as in eq.~\reef{mi1}, $\tilde{a}$ is again the coefficient of the universal area term
for generic strips in the limit of small width -- see the discussion around eq.~\reef{strip}.
\begin{figure}[t]
  \centering
  \subfloat[]{\includegraphics[width=.4\textwidth]{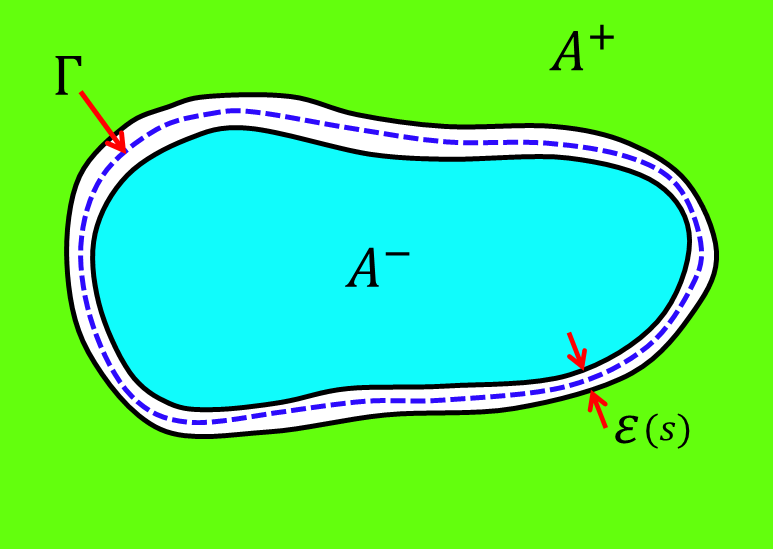}}
  \hspace{24pt}
   \subfloat[]{\includegraphics[width=.4\textwidth]{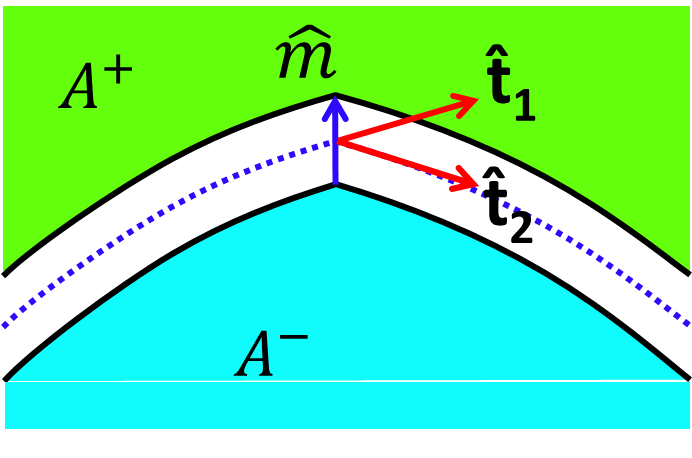}}
\bigskip
\caption{(Colour online) (a) Regularizing the entropy of $A$ with boundary $\Gamma$ (blue dashed line)
using the mutual information $I(A^{+},A^{-})$. (b) Detail of a null cusp, \ie a corner on a null
plane, with the size of $\veps $ the same on both sides of the corner point.}
\labell{fi}
\end{figure}

The reason for the structure in eq.~\reef{gamma} is the locality of the UV terms which depend on the
framing, similar to the discussion in section \ref{critb}. A local correction to eq.~(\ref{gamma}) depending on the derivatives of
$\veps(s)$ such as $\int ds\, \veps^\prime(s)/\veps(s)$ is a total derivative
and hence it does not contribute. Other local contributions involving higher derivatives of $\veps(s)$, or curvature corrections, must
be accompanied by extra powers of $\veps$ to compensate for the dimension of the derivatives and hence such terms would vanish with some power of $\veps$
as $\veps\to0$.\footnote{In making this argument, one should recall that the width is set such that $\veps\ll 1/m_i$ where $m_i$ may be any scale appearing in defining the RG flow between the corresponding fixed points. Hence we can only expect the appearance of coupling constants for relevant operators in these higher order contributions. But again such couplings must be accompanied by extra (positive) powers of $\veps$ which would cause these terms to vanish in the limit $\veps\to0$.} With eq.~\reef{boundaries}, we focused on symmetric framings or in the language of sections \ref{newdef} and \ref{critb}, we chose $\alpha=0$.\footnote{This choice further rules out a term of the form $\int ds\, \veps^\prime(s)/\veps(s)$  since locally the integrand is odd
under reflections in the plane orthogonal to $\Gamma$.} However, let us consider nonsymmetric framings for a moment, \eg $\Gamma_+= {\Gamma}(s)+ \veps_1(s)\un(s)$ and $\Gamma_-= {\Gamma}(s)-\veps_2(s)\un(s)$ with $\veps_1(s)\ne\veps_2(s)$ in general. Then we might find additional finite frame-dependent terms in eq.~\reef{gamma} involving the difference of the two framings, \eg 
\beq
\int ds\ \un\cdot\partial_s\ut\ 
\frac{\veps_1(s)-\veps_2(s)}{\veps_1(s)+\veps_2(s)}
\labell{badd}
\eeq 
--- compare to eq.~\reef{inte}. The $\alpha$-dependent term
in eq.~(\ref{expec}) implicitly has its origin in a contribution of this form. Of course, such a contribution vanishes with the symmetric framing in eq.~\reef{boundaries}. The symmetric choice also means that $I_{0}(A)$ is the same when $A^{+}$ and $A^{-}$ interchanged in eq.~(\ref{gamma}) and hence gives rise
to the symmetry $I_0(A)=I_0(\bar A)$ for the finite part. 

The proof of the \F-theorem requires that we consider boundaries $\Gamma(s)$ with null cusps, \ie corners on null planes. Hence we must determine if such local features introduce additional frame-dependent terms in the mutual information and if so,
how to eliminate them.  Let us decompose the mutual information as
 \begin{equation}
 I(A^+,A^-)=S(A^+)+S(A^-)-S(A^s)\,, \labell{decomp}
 \end{equation}
where $A^s$ is the strip, \ie $A^s=\overline{A^+\cup A^-}$.\footnote{Implicitly, eq.~\reef{decomp} assumes that the regulated entanglement entropies respect $S(V)=S(\bar V)$ for general regions $V$ in the ground state of the QFT. However, we do not really need to make this assumption here or in the following discussion. Rather it is a notational convenience to introduce $S(A^s)$ rather than using $S(A^+\cup A^-)$. For example, in eq.~\reef{tido}, we would need to replace $S(A_1^s\cup B^s_1)$ with $S((A^+\cap B^+)\cup A^-\cup B^-)$.}  Recall that while the entropies appearing in this decomposition are regulator dependent, the combination appearing in mutual information is universal. Further, we expect that any frame dependence from the null cusp will appear in a purely geometric contribution, whose form will depend on the details of the UV fixed point.  As noted above, since the corner lies in a null plane, there is no intrinsic local measure of the `angle' for a null cusp \cite{proof}. Hence the corresponding null cusps in $\Gamma_+$ and $\Gamma_-$ will not produce any extra local contributions in $S(A^+)$ or $S(A^-)$, respectively. 

However, the situation is different for the strip $A^s$ since we can now identify three distinct vectors at the cusps, as illustrated in figure \ref{fi}(b).
First, for the two lines crossing each other on the null plane, we have the unit tangent vectors $\ut_1$ and $\ut_2$ (taken with
the same orientation such that $\ut_1-\ut_2$ is null\footnote{Again, we note that $\ut_1\cdot\ut_2=1$ follows from the latter. Hence this inner product carries no information about the `angle' at the null cusp.}). Lastly, there is a unit vector $\um$ directed from the cusp on the inner boundary $\Gamma_-$ to that on the outer boundary $\Gamma_+$.\footnote{We emphasize that this unit vector is not the same as $\un$ in eq.~\reef{boundaries}, \eg see appendix \ref{broken}.} 
 The local geometry of the null cusp in the strip yields two new dimensionless quantities, which are both local and
Lorentz invariant, \ie
\begin{equation}
q_1\equiv \um\cdot  \ut_1\,,\hspace{1cm} q_2= \um\cdot \ut_2\,. \labell{qqq}
\end{equation} 
From the requirement that the boundaries be spatially separated to each other we have $q_1^2\le 1$, $q_2^2\le 1$. 
Giving these two variables the geometry of the cusp is determined except for an overall scaling. In particular, the ratio of the framing sizes on each side of the cusp, $\veps_1$ and $\veps_2$, is given in terms of these variables,
\be
\frac{\veps_1}{\veps_2}=\sqrt{\frac{1-q_1^2}{1-q_2^2}}\,.
\labell{gonzo2}
\ee  
 
Now the appearance of $\um$ in the above definition \reef{qqq} implicitly means that $q_1$ and $q_2$ depend on the framing at the null cusp. 
However, we argue that only a restricted amount of frame dependence can appear in $S(A^s)$ and hence in $I(A^+,A^-)$. In particular, the discussion in section \ref{entropy} indicated that the strip entropy $S(A^s)$ is a purely local UV contribution. However, the only scales appearing in this entanglement entropy which are local to the cusp would be $\veps_{1,2}$ and the cut-off $\delta$ and hence the contribution of the corner may be a function of the ratios $\veps_{1,2}/\delta$. However the mutual information will not contain contributions depending on the UV cut-off.\footnote{Of course, the situation where $\Gamma$ contains a corner lying in a spatial plane would be different. In this case, we expect a local contribution proportional to $\log(\veps/\delta)$ for the strip. The $\delta$ dependence of this term
would be compensated by terms with $\log(\delta/R)$, where $R$ some macroscopic scale, in the
entropies $S(A^+)$ and $S(A^-)$. The mutual information will then have an additional finite
($\delta$ independent) term proportional to $\log(\veps/R )$.} Hence only the ratio $\veps_{1}/\veps_{2}$ is relevant but the latter is completely determined by $q_1$ and $q_2$ in eq.~\reef{gonzo2}. Hence, the frame dependence of the null cusp can be expressed entirely in terms of  $q_1$ and $q_2$ 
and hence the mutual information can be written as
\begin{equation}
I(A^+,A^-)=\tilde{a} \int_\Gamma \frac{ds}{\veps(s)}+I_0(A)+f(q_1,q_2)\,,\labell{449}
\end{equation}
where $f(q_1,q_2)$ is a local frame-dependent term, depending exclusively on $q_1$ and $q_2$, and $I_0(A)$ is a function of $A$, independent of framing.

Since $f(q_1,q_2)$ is the cusp dependent term in the mutual information, it must be a universal quantity. However, eq.~\reef{449} still leaves the definition of $f(q_1,q_2)$, and the frame dependence of $I(A^+,A^-)$, open to certain ambiguities. We remove these uncertainties with the following choices:
First, we need to be careful with the integral in the first term on the right-hand side. We define it by simply integrating
all along the curve $\Gamma$ up to either side of the null cusp --- see figure
\ref{fi}(b) --- where the value of $\veps(s)$ is defined there as the
limit of the function $\veps(s)$ from either side of the corner. This choice is simply made for
convenience, however, other choices would produce finite redefinitions of the function $f(q_1,q_2)$. Finally, we fix $f(q_1,q_2)$ by demanding that $f(q_1,q_1)=0$, \ie this contribution vanishes when $\ut_1=\ut_2$ (and there is no null cusp).  This requirement eliminates the possible ambiguity in $f(q_1,q_2)$ (and hence on $I_0$) of redefining this contribution by the addition of a constant. 
\begin{figure}
\centering
\includegraphics[width=.65\textwidth]{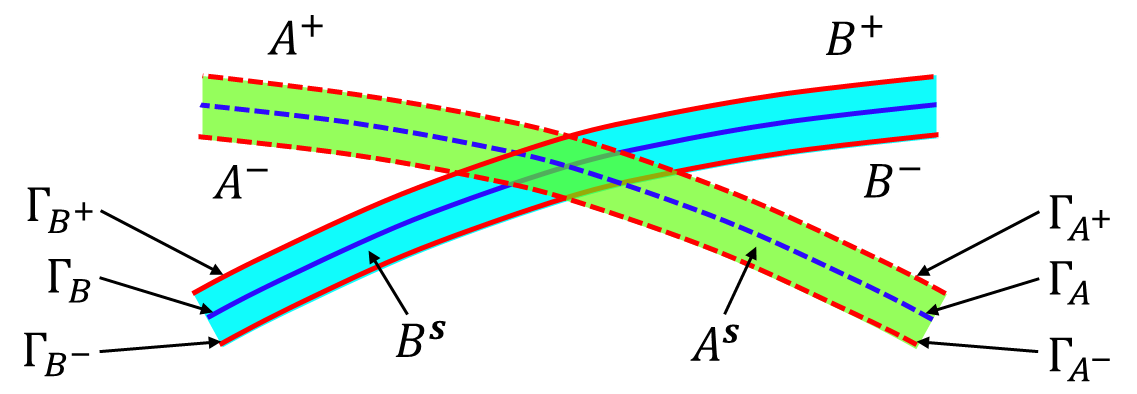}
\bigskip
\caption{(Colour online) Two entangling surfaces, $\Gamma_A$ and $\Gamma_B$, which intersect to form a null cusp. These surfaces are `expanded' according to eq.~\reef{boundaries} so that $\Gamma_{A^+}$ and $\Gamma_{B^+}$ still cross on a null plane and $\Gamma_{A^-}$ and $\Gamma_{B^-}$ cross on a nearby parallel null plane. The thin strips, $A^{s}$ (light blue) and $B^{s}$ (green), are complementary to the regions $A^+\cup A^-$ and $B^+\cup B^-$, respectively.}
\labell{cruce}
\end{figure}

Our expectations for the frame dependence of null cusps can be checked in the model of extensive mutual information, discussed in appendix \ref{extensive}.  The result for this model is evaluated in section \ref{broken} as
\begin{equation}
f(q_1,q_2)= \frac{\kappa_{\ssc UV}}{2} \left(2+\frac{(q_1(q_1+q_2)-2)\arcsin(q_1)}{(q_1-q_2) \sqrt{1-q_1^2}}+\frac{(q_2(q_1+q_2)-2)\arcsin(q_2)}{(q_2-q_1) \sqrt{1-q_2^2}}\right)\,,\labell{quo}
\end{equation}
where $\kappa_{\ssc UV}$ is a dimensionless constant characterizing the UV fixed point. 
As expected, $f(q_1,q_2)$ is a symmetric function of $q_1$ and $q_2$ since eq.~\reef{449} is invariant when $A^{+}$ and $A^{-}$ are exchanged while $q_1\leftrightarrow q_2$ under this interchange.
%rob
Further it is straightforward to show that with $q_2=q_1+\epsilon$, $f(q_1,q_2)$ is vanishes as $\epsilon^2$ in the limit $\epsilon\to0$. Again, this limit, \ie $q_2\to q_1$, corresponds to the vanishing of the null cusp, \ie the entangling curves become smooth with $\ut_1=\ut_2$ in figure \ref{fi}(b).
On the other hand, $f(q_1,q_2)$ diverges in the limit $q_1\rightarrow \pm 1$ or $q_2\rightarrow \pm 1$ because in this limit, the entangling curves $\Gamma_+$ and $\Gamma_-$ cease to be spatially displaced with respect to each other.

For general curves, we eliminate all frame-dependent terms from our definition of renormalized mutual information with
\begin{equation}
I_0(A)=I(A^+,A^-)-\tilde{a} \int_\Gamma  \frac{ds}{\veps(s)}-\sum_i f(q_1^i,q_2^i)\,,\labell{tuty}
\end{equation}
where the last sum is over all of the null cusps.
The $I_0(A)$ defined by eq.~(\ref{tuty}) is both universal (\ie independent of the UV regulator) and frame
independent. Hence the result is a finite Lorentz invariant function of $A$ alone.
Hence we can use $I_0(A)$ as the Lorentz invariant regularization of the entanglement entropy required for the proof of the \F-theorem, \ie this
quantity will satisfy the first of the three assumptions outlined at the beginning of this section. 

\subsubsection*{Strong subadditivity}

Let us now turn to the second of the assumptions required to prove the \F-theorem. In particular, we ask: is there a sense in which the mutual information satisfies SSA? Consider the geometry illustrated in figure (\ref{cruce}). We have two regions $A$ and $B$ with boundaries $\Gamma_A$ and $\Gamma_B$, respectively. These boundaries intersect at some point to form a corner in a null plane, as occurs in the proof of the \F-theorem.\footnote{In the following, we only consider a single null cusp but the discussion is easily extended to account for the contributions from any number of null cusps, as appear in eq.~\reef{tuty}.} Now following our prescription above, a strip is constructed around each of these boundaries so that we can regulate the entanglement entropy using the mutual information. The construction leads to two exterior regions, $A^+$ and $B^+$, and two interior regions, $A^-$ and $B^-$, as well as the two strips $A^s=\overline{A^+\cup A^-}$ and $B^s=\overline{B^+\cup B^-}$. The goal is to obtain an inequality for $I_0$ which is framing independent. Hence we are free to choose the framing of $\Gamma_A$ and $\Gamma_B$ such that $\Gamma_{A^+}$ intersects $\Gamma_{B^+}$ on a null plane and that these curves are spatial to one another, and similarly, the same for $\Gamma_{A^-}$ and $\Gamma_{B^-}$.

\begin{figure}[t]
\captionsetup[subfigure]{labelformat=empty}
  \centering
   \subfloat[]{\includegraphics[width=.25\textwidth]{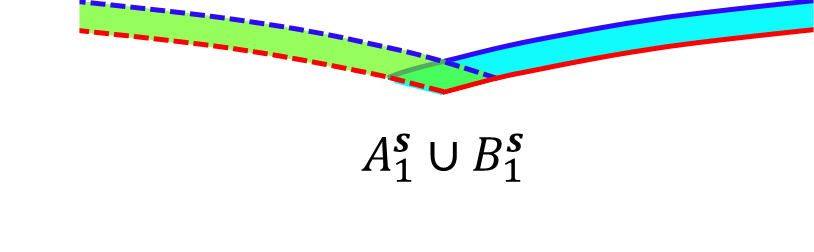}}
   \subfloat[]{\includegraphics[width=.25\textwidth]{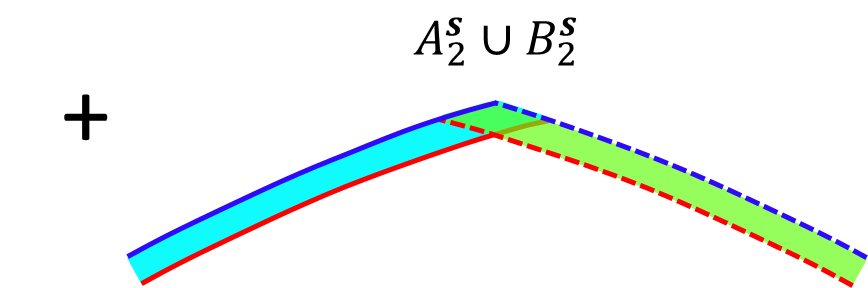}}
   \subfloat[]{\includegraphics[width=.25\textwidth]{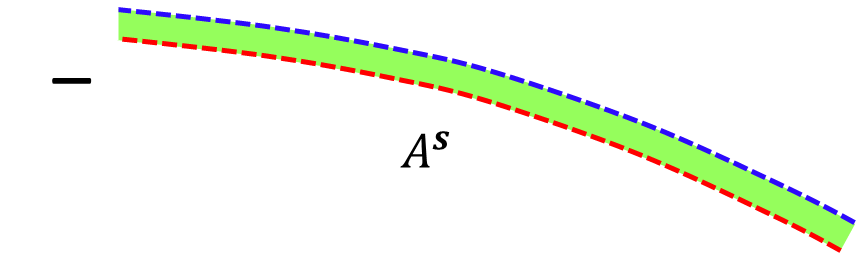}}
   \subfloat[]{\includegraphics[width=.25\textwidth]{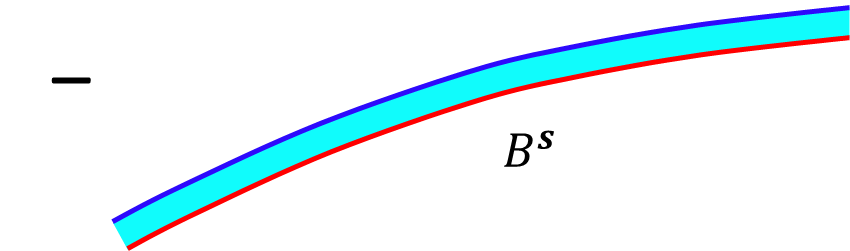}}
\bigskip
\caption{(Colour online) The quantity $J(A^s,B^s)$ is defined in eq.~\reef{tido} by the combination of entanglement entropies for the thin regions shown above.}
\labell{cruce1}
\end{figure}

Now, using the definitions of mutual information \reef{decomp} and strong subaddivity \reef{ssaa}, we find\footnote{Note that the two spatial regions $A^+$ and $B^+$ will generally lie on different Cauchy slices and hence the strict definition of $A^+\cap B^+$ would yield a linear region, \ie a codimension two surface. However, our intention here is that $A^+\cap B^+$ is a two-dimensional spatial region that lies within the intersection of the causal domains of $A^+$ and $B^+$. Similar definitions apply for the other intersections and unions of regions appearing in eq.~\reef{ine} and in the subsequent discussion.} 
\begin{eqnarray}
&&I(A^+,A^-)+I(B^+,B^-)=S(A^+)+S(A^-)-S(A^s)+S(B^+)+S(B^-)-S(B^s)\nonumber \\
&&\ge S(A^+\cup B^+)+S(A^+\cap B^+)+S(A^-\cup B^-)+S(A^-\cap B^-)- S(A^s)-S(B^s)    \nonumber \\
&&= I(A^+\cap B^+,A^-\cup B^-)+I(A^+\cup B^+,A^- \cap B^-)+ J(A^s,B^s)\,.\labell{ine}
\end{eqnarray}
In the last line above, we have introduced
\begin{equation}
J(A^s,B^s)=S(A_1^s\cup B^s_1)+S(A_2^s\cup B^s_2)-S(A^s)-S(B^s)\,,\labell{tido}
\end{equation}
where the strips have been decomposed as $A^s=A_1^s\cup A_2^s$ and $B^s=B_1^s\cup B_2^s$, according to the intersections at the null cusp  shown in figure
\ref{cruce1}.\footnote{The definition of the regions can be made more precise by writing expressions such as $A_1^s=A^s\cap \bar{B}^-$.}  
The inequality (\ref{ine}) has (almost) the same form of the SSA inequality \reef{ssaa} but now written in terms of the mutual information. However, there is one additional term, \ie $J(A^s,B^s)$, on the right-hand side of the inequality. The goal, however, is to write a SSA-like inequality for the regulated entropies, \eg $I_0(A)$, rather than the mutual information. Hence we proceed by noting that $I(A^+\cap B^+,A^-\cup B^-)$ and
$I(A^+\cup B^+,A^- \cap B^-)$ have the right form to provide the mutual
information regularization of the entanglement entropies for $A\cap B$ and $A\cup B$,
respectively. That is, if we take out the area terms and the frame-dependent terms coming from the null cusp, as in eq.~\reef{tuty}, the previous inequality becomes
\begin{equation}
I_0(A)+I_0(B)\ge I_0(A\cup B)+I_0(A\cap B)+2 f(q_1,q_2)+ J(A^s,B^s)\,.\labell{59}
\end{equation}
Note that the area terms can be removed since they are precisely the same on the right- and left-hand sides of eq.~\reef{ine}.

Let us now turn our attention on the extra term $J(A^s,B^s)$ appearing in the inequalities above. Here it is important that the corners in the strips $A_1^s\cup B^s_1$ and  $A_2^s\cup B^s_2$ are actually null cusps. If instead the corners in these strips were to lie in a spatial plane, the nontrivial angle would induce logarithmically divergent terms proportional to $\log\delta$ in the corresponding entanglement entropies appearing in $J(A^s,B^s)$. These log terms would not be compensated by the rest of terms in eq.~(\ref{ine}), which are $\delta$ independent,
and therefore this inequality would be insufficient for proving the \F-theorem. However, since 
the entangling curves in the proof are all placed on a common null cone, the resulting corners are null cusps and hence we do not have to worry
about $\log(\delta)$ terms here. Further, the area law terms proportional to $1/\delta$ also cancel in eq.~(\ref{tido}) and hence $J$ is a finite ($\delta$ independent) quantity, as much as the mutual information is. Therefore we must find the same universal result when $J$ is computed with any regularization for the entropies in eq.~\reef{tido}. But further in section \ref{entropy}, we have argued the entanglement entropy of thin regions must be extensive (in any
regularization). Thus, in the combination appearing in eq.~\reef{tido}, most of the contributions in the various strip entropies will cancel out and the only remaining contribution which could survive will be a strictly local (and UV) term associated with the corners in the first two strips. However, $J$ can not depend on $\veps$ since there are no other dimensionful parameters in the geometry. Again the strip entropy is probing correlations beyond the scales of the UV CFT and so there are no dimensionful couplings available in the field theory to appear in the characterization of this local feature. For example, no finite terms involving $\log\veps$ can arise for this reason.\footnote{Again, the situation would be different if the boundary corners lay in a spatial plane. In this case, the contribution from $J$ would still be localized on the cusp but we expect a logarithmic contribution $\log(\veps/\delta)$ where the UV cut-off appears as the second scale.}

Hence, we conclude that  $J$ must be a universal function of $q_1,q_2$, the only geometric parameter describing the null cusp, which is local and dimensionless. From the definition in eq.~\reef{tido}, it is also clear that source of this $q$ dependence must come from $S(A_1^s\cup B^s_1)$ and $S(A_2^s\cup B^s_2)$, the entanglement entropies of the two `bent' strips of figure \ref{cruce1}. However, it was precisely these two entropies which are added and subtracted in going from the second to the third line of eq.~\reef{ine}. Hence, the same two entanglement entropies appear in the mutual
informations $I(A^+\cap B^+,A^-\cup B^-)$ and $I(A^+\cup B^+,A^- \cap B^-)$ but with a negative sign. Therefore it must be that\footnote{It is important for this step that we have defined the additive constant on the function $f(q_1,q_2)$ by $f(q_1,q_1)=0$, since it is evident that for the limit of no cusp we also have $J=0$. } $J(q_1,q_2)=-2 f(q_1,q_2)$ and the two $q$-dependent terms cancel on the right-hand side of eq.~\reef{59}. With this cancellation, our regulated entropies satisfy an inequality like strong subadditivity \reef{ssaa}, \ie
\begin{equation}
I_0(A)+I_0(B)\ge I_0(A\cup B)+I_0(A\cap B)\,,
\end{equation}
for $A$, $B$, cutting each other on a null plane.  Hence we have shown that these regulated entropies satisfy assumption (b).

\subsubsection*{Limit of wiggly circles}
It remains to see if the regulated entropies \reef{tuty} satisfy the third assumption set out at the beginning of the section. That is, whether the regulated entropy of the wiggly circles $I_{0,w}(R)$ converges in the limit $N\rightarrow \infty$ to that
of a smooth circle $I_{0,s}(R)$ with the same perimeter.

Here again, we will invoke our favorite arguments involving locality. The difference between these two quantities in the limit of large $N$ must be given by the integration of some local term around the circumference. Further, the local term only depends on the UV CFT which, in particular, does not have any dimensionful couplings to introduce into this integral. Neither will the frame-dependent parameters $\veps$ or $q$ appear since $I_{0}$ does not depend on these parameters.
Thus, in order to eliminate the dimensions of the integral, we must divide by some local geometric quantity with dimensions of length. There are only two such quantities available: The first is the curvature of the circle, \ie $k\equiv\un\cdot\partial_s\ut\propto 1/R$ as described in section \ref{critb}. However, the construction of $I_0$ was chosen to respect the reflection symmetry discussed there and so by the same arguments as given in that section, we cannot have a contribution involving an integral of the curvature. The second possible geometric quantity would be the length of the arcs between the null cusps, \ie $\ell\equiv 2\pi R/2N$. Hence let us consider whether $I_{0,w}(R)$ might contain a contribution proportional to $\int ds/\ell = 2N$.\footnote{Let us note that other contributions with the right dimensions would have to take the form $\int ds\, k^{2n}\ell^{2n-1}\sim 1/N^{2n-1}$ which vanish in the limit $N\to\infty$. Contributions proportional to
higher powers of $N$, \eg $\int ds/ (k^2 \ell^3)\sim N^3$, can not arise
because they are singular for vanishing curvature. Note that in both of the examples here, we are demanding that $k$ only appear with even powers in order to respect the reflection symmetry.} This `area' term for the wiggly circle would also be present in the case where the boundary of the wiggly region lies on a null plane rather than on a null cone. But in considering $I_0$, we have regained Lorentz covariance. Thus by making a boost which keeps the null
plane fixed, we can vary arbitrarily the amount by which the arcs extend in the null direction. In particular, a sufficiently large boost will bring the wiggly `line' arbitrarily close to a straight line. Hence, this area term cannot be present since it is not present for the linear case. That is, the nonvanishing
contribution to $I_{0,w}(R)$ must be independent of $N$ in the limit of large $N$. Hence we must have
\beq
\lim_{N\to\infty}I_{0,w}(R)=I_{0,s}(R)\,,
\labell{flight}
\eeq
which establishes the third and final key ingredient required in the proof of the \F-theorem.\\

We can summarize the above arguments by saying that the mutual information,
stripped of the frame dependent terms in a universal way, can be defined given any
region $A$ with a spatial piecewise smooth boundary and where any corners
lie in a null plane. This universal part of the mutual information provides a regulated entanglement entropy which satisfiesl the three assumptions (a)--(c) which are required for the proof \cite{proof} of
the \F-theorem. For a smooth circular entangling curve, the construction here coincides with that introduced in section \ref{newdef} and hence an immediate consequence is that the constant term in $I_0(A)$ coincides with the one in $I(R+\veps/2,R-\veps/2)$, which we discussed in previous sections. Hence, combining our construction of the regulated entanglement entropy using the mutual information with the proof of \cite{proof}, we have established the three-dimensional \F-theorem as a proper \c-theorem.
In particular then, the new \c-function becomes
\begin{equation}
C(R)=\frac1{4\pi}\Big[R \,\partial_RI_0(R)-I_0(R)\Big]\,,\labell{funix}
\end{equation}
and the proof of \cite{proof} establishes $C^{\prime}(R)\le 0$.

\section{Discussion}
\labell{discuss}

Our investigation began with the issue that entanglement entropy in QFT cannot be defined without explicit reference to a UV regulator. As we discussed, this introduces various ambiguities in trying to define the universal contribution to entanglement entropy for odd dimensions, in particular, for $d=3$ --- unless one restricts calculations to the class of allowed regulators. Implicitly the problem is that the divergences in entanglement entropy can not be renormalized away. For example, while we do may add a counterterm for the area term divergence in $d=3$, we do not have a physical experiment which would allow us to fix the renormalized entanglement entropy against some physical result at a certain renormalization scale. 

We were able to resolve the above problem by we showing that the mutual information effectively provides a
`universal' geometric regularization scheme for the entanglement entropy, which
can be applied with any QFT. In this sense, mutual information is analogous to
the framing regularization used in the literature of Wilson loops. Further,
mutual information allows us to identify the physical renormalized `central charge' 
$c_0$. In particular this means that, as for any other
renormalized quantity, it is possible to compute this charge using any regularization that defines the continuum theory. In particular, our approach allows $c_0$ to be computed with a lattice regulator. We will also showed the proof of the \F-theorem \cite{proof} can be
written in terms of universal quantities derived from the mutual information.

Expressing $c_0$ in terms of the mutual information, instead of
the entanglement entropy, has both practical and theoretical advantages.
The main theoretical advantages are that it shows this quantity is universal,
well defined and intrinsic to the CFT at the fixed points. Let us add that the mutual information is also a gauge invariant quantity \cite{gauge} and so evades the subtleties arising in defining entanglement entropy for theories involving gauge fields. Our approach also
clarifies the correctness of the assumptions in the proof of the \F-theorem \cite{proof}.

From a more practical perspective, it was not clear from previous discussions in
the literature how a sufficiently geometric cut-off could be defined for general theories in order to allow the extraction of a universal quantity
from the entanglement entropy of a circle in odd dimensions. Zeta function or heat
kernel regularizations can be defined only for free fields, where the relevant
partition functions are given in terms of a differential operator. In practice,
the same can be said of the radial lattice. We have shown a universal way to
introduce a covariant universal short-distance cut-off is to consider the
mutual information as a regulated version of the entanglement entropy.

The mutual information is a renormalized quantity in the sense that it can be
computed under any regularization, specially for models defined in arbitrary
type of lattices. For these, it is not even necessary to know the expression of
the theory of the continuum. A lattice cut-off serves simultaneously to define
the theory in terms of the microscopic degrees of freedom and to produce a
finite entanglement entropy. However, the error in $R$, which is of the order of the lattice spacing, will
spoil the definition of the constant term in the expansion, as discussed in section \ref{prob}.
Alternatively, an unambiguous definition of the
constant term in the entanglement entropy can be achieved if we introduce a second cut-off with the sole
purpose of regulating the entropy, when the continuum theory is already known, as in section \ref{entropy}.

\bigskip

However, in its present form, the \F-theorem in $d=3$ cannot be considered as having the
same degree of mathematical rigor as the \c-theorem in $d=2$ or the $a$-theorem in $d=4$. The latter
are based on well established mathematical properties of correlation functions or the
S-matrix in QFT. Our understanding of entanglement entropy and related entanglement measures are not so well developed for
QFT's. However, our general and simple arguments establish a solid case. The natural
context for a proof in the mathematical sense seems to be the algebraic
approach to QFT, where the mutual information is well defined in terms of
states acting on local operator algebras.

\bigskip 

\subsection*{\F-theorem and topological theories}

%rob: some rewriting here

An example requiring further discussion is the case of topologically ordered phases or topological quantum field theories.
In this case, the constant term in the entanglement entropy of a circle coincides with
the topological entanglement entropy \cite{topo,topo2}: $s_{\textrm{topo}}=-\gamma$. This result is somewhat surprising since one finds
a nonzero (and possibly very large) central charge for a theory with no
local degrees of freedom. On the other hand, the mutual information for a topological theory {\it vanishes}, 
at least if the distances in the relevant geometry are all longer than the scale of the onset of topological order. 
The latter may be some macroscopic scale in the continuum theory, but in many instances, it will correspond to the scale of the microscopic cut-off.
A key question will be how $\veps$ compares to this `topological scale', which we denote $\vepst$. Specifically, if we can set $\veps\ll\vepst$ then the
constant contribution of the annular strip will vanish and mutual information
would yield the same charge as the circle entropy, \ie the constant term in the mutual information would simply be twice the topological entropy (when we set $R\gg\vepst$). 
However, if $\veps\gg\vepst$ then the entanglement 
entropy of the annulus will have a constant topological contribution and there will be a discrepancy between the charges defined by the mutual information and the circle entropy. 
%When $\veps$ is decreased from scales larger than the correlation length to smaller scales, both an area term an a constant term should appear together.

It may appear that the latter discrepancy has the potential to produce contradictions with \F-theorem, however, as we will now explain, it simply produces a reorganization of the fixed point contributions.\footnote{We are grateful to Tarun Grover for discussions on this point.} Let us consider an RG flow between two fixed points, both of which have a topological sector. We imagine that the topological contribution to the entanglement entropy can be separated from that of the local or gapless field degrees of freedom, \ie we can write $c_0=c_{0,\textrm{local}}+c_{0,\textrm{topo}}$. Then evaluating the circle entropy with a geometric regulator and comparing the results for the two fixed points, we would find
\bea
{\rm UV\ fixed\ point:}\qquad c^{\ssc UV}_0&=&c^{\ssc UV}_{0,\textrm{local}}+c^{\ssc UV}_{0,\textrm{topo}} \vphantom{\frac{1}{\Pi}}
\nonumber\\
{\rm IR\ fixed\ point:}\qquad c^{\ssc IR}_0&=&c^{\ssc IR}_{0,\textrm{local}}+c^{\ssc IR}_{0,\textrm{topo}} \vphantom{\frac{\Pi}{\Pi}}
\nonumber\\
c^{\ssc UV}_0-c^{\ssc IR}_0
&=&\big(c^{\ssc UV}_{0,\textrm{local}}-c^{\ssc IR}_{0,\textrm{local}}\big)+\big(c^{\ssc UV}_{0,\textrm{topo}}-c^{\ssc IR}_{0,\textrm{topo}}\big)
\vphantom{\frac{\Pi}{1}}
\labell{FF11}
\eea
Now let us make the same comparison using the mutual information. Further we imagine that the topological scale $\vepst$ for the UV fixed point corresponds to the cut-off scale, \ie the topological order is an inherent property of the continuum CFT describing the UV fixed point. Hence even though the annulus entropy probes the theory at some very short distance (but still macroscopic) scale $\veps$, it still acquires a topological contribution characteristic of the UV theory. The RG flow is then described by the mutual information as
\bea
{\rm UV\ fixed\ point:}\qquad \tc^{\ssc UV}_0&=&c^{\ssc UV}_{0,\textrm{local}} \vphantom{\frac{1}{\Pi}}
\nonumber\\
{\rm IR\ fixed\ point:}\qquad \tc^{\ssc IR}_0&=&c^{\ssc IR}_{0,\textrm{local}}+c^{\ssc IR}_{0,\textrm{topo}}-c^{\ssc UV}_{0,\textrm{topo}} \vphantom{\frac{\Pi}{\Pi}}
\nonumber\\
\tc^{\ssc UV}_0-\tc^{\ssc IR}_0
&=&\big(c^{\ssc UV}_{0,\textrm{local}}-c^{\ssc IR}_{0,\textrm{local}}\big)+\big(c^{\ssc UV}_{0,\textrm{topo}}-c^{\ssc IR}_{0,\textrm{topo}}\big)
\vphantom{\frac{\Pi}{1}}
\labell{FF22}
\eea
Hence both approaches agree on the change in the universal coefficient resulting from the RG flow. The key point is that the topological contribution in the entropy of the annular strip is proportional to $c^{\ssc UV}_{0,\textrm{topo}}$ at both fixed points. That is, the scale determining this contribution is the strip width $\veps$, which is held fixed while the radius $R$ varies between the UV and the IR scales.

We can extend the above discussion to consider the case where $\vepst^{\ssc UV}$ corresponds to some macroscopic scale,  which is still much shorter than the scales where the physics is described by the UV fixed point. As alluded to above, the mutual information description of the RG flow would correspond to that in eq.~\reef{FF22} as long as
$\veps\gg \vepst^{\ssc UV}$.  On the other hand, if we chose $\veps\ll \vepst^{\ssc UV}$, the mutual information results would coincide to those in eq.~\reef{FF11}
because the entanglement entropy of the annulus would no longer include a topological contribution. Again, both choices would agree on the running of the universal coefficient relevant for the \F-theorem. 

Hence the above discussion seems to clear away the potential tension between the \F-theorem and the discrepancy between the universal charges defined by the mutual information and the entanglement entropy. However, it calls into question whether or not our construction strictly satisfies the second requirement which we set out for a legitimate \c-theorem. That is, does the mutual information actually provide a physical quantity that is intrinsic to the fixed point of interest, \ie in eq.~\reef{FF22}, we see that $\tc^{\ssc IR}_0$ appears to depend on $c^{\ssc UV}_{0,\textrm{topo}}$, a property of the UV fixed point.\footnote{A possible resolution of this tension would be if the topological sector on the UV does not contribute to the nontrivial running of $\tc_0$ in the RG flow. That is, if the topological terms are the same at the UV  and IR ends of the RG flow, \ie $c^{\ssc IR}_{0,\textrm{topo}}= c^{\ssc UV}_{0,\textrm{topo}}$, then the topological contributions would be eliminated in $\tc^{\ssc IR}_0$. An interesting example of such a flow was discussed in \cite{tarum}. In that case, the UV fixed point corresponds to a Dirac fermion coupled to a Z$_2$ gauge field and while in the IR, the theory flows to an purely topological gapped phase. At both of these fixed points, $c_{0,\textrm{topo}}=\gamma/2\pi=\log(2)/2\pi$. Note, however, that the RG flow does affect the topological part of the theory in a nontrivial way. In particular, the topological sectors in the UV and the IR are distinct, \eg with Abelian and non-Abelian anyonic excitations, respectively. Unfortunately, it is not likely that the cancellation found in this example will be a generic property of general flows.}  However, we are not particularly concerned with this issue and instead suggest that the mutual information presents a perspective where the topological contribution to the entanglement should be thought as a `topological remnant' of the RG flow.  

For example, in the above discussion where $\vepst$ was considered a macroscopic scale, the essential difference between $\veps\gg\vepst$ and $\veps\ll \vepst$ is that in the latter situation, the high energy contribution to the mutual information, \eg eq.~\reef{stinkpot}, probes the physical short-distance correlations that build up the long-range topological entanglement. We observe that the same is true of standard probes of the topological order: For example, in order to identify the topological phase through its anyonic excitations, we need to probe the theory at the energy scale of these excitations which corresponds to the scale of the gap, \ie $1/\vepst$ in the current discussion. Similarly, we might consider the standard approaches for evaluating the topological entanglement entropy \cite{topo,topo2}.\footnote{To simplify the discussion here, we consider the case of a purely topological theory. With additional local degrees of freedom, the result of the standard constructions will be the sum of the topological entropy plus a contribution from the additional degrees of freedom. Further, the latter will depend on the precise details of the geometry.} In order for these constructions to yield unambiguous results, the cut-off scale and the distances between regions must be much smaller than the correlation length of the model, \ie $\vepst$ \cite{gauge}. For example, the Kitaev-Preskill construction \cite{topo} gives the topological entanglement entropy as 
\bea
-\gamma&=&S(A)+S(B)+S(C)-S(AB)-S(AC)-S(BC)+S(ABC)\nonumber\\
&=&I(A,B)+I(A,C)-I(A,BC)\labell{kp}
\eea  
for three adjacent regions $A$, $B$ and $C$ with a common vertex. Hence, the topological entropy can also be expressed in terms of the mutual information for regions which are close enough, \ie their separation is less than the correlation length. Again this restriction is natural since the only physical correlations capable of building up entanglement are over distances shorter than this scale. 

In our construction of the mutual information of circles, this restriction corresponds to having $\veps\ll\vepst$, the scale for the onset of topological order. Hence detecting the topological constant term with the mutual information requires that $\vepst$ is a macroscopic scale, different than the UV cut-off scale. In other words, there should be actual local degrees of freedom at high energies giving rise to the topological gapped sector at lower energies.  If there is a topological sector at the UV fixed point of the RG flow, this scale $\vepst$ could in principle be arbitrarily small and unrelated to the RG flow under study, which only occurs at much longer scales. While the $\veps\rightarrow 0$ prescription should always restore the correct value of the topological entropy in the mutual information, the peculiarity here is made clear by contrasting with the case of a regular CFT with gapless degrees of freedom, \eg a massless free scalar. In this case, producing the correct constant term only requires choosing $\veps\ll R$. That is, $\veps$ can be set independently of the possible physics at much shorter scales.    

Of course, if we do not restrict ourselves to working in Minkowski space, the topological sector at the fixed point can be identified by other means, \eg putting the theory on a topologically nontrivial manifold. This could in principle be used to set the correct topological entropy of the fixed point without going beyond the physics of the fixed point itself.

\bigskip

\subsection*{Alternate \c-functions}

As well as decreasing monotonically along RG flows, Zamolodchikov's \c-function in two dimensions is also stationary at fixed points \cite{Zamo}. It was pointed out in \cite{Klebanov2012} that the entropic \c-function \reef{funi} in three dimensions fails to be stationary in the particular case of a free scalar field. In fact, stationarity is also lost for the free scalar when using the entropic \c-function in two dimensions. This lack of stationarity is typically seen as a deficiency but we would like to advocate that, in fact, it is an advantage of using the entropic \c-function. In particular, Zamolodchikov's  original \c-function treats the scalar theories with a tachyonic $m^2$ on an equal footing as those with positive $m^2$.  In contrast, the entropic \c-function seems to recognize that the theories with a tachyonic mass are unphysical and only allows for flows towards positive $m^2$.

In any event, the above discussion highlights the fact that the \c-function is not uniquely defined. Rather given a particular \c-function, we can in principle construct infinitely many other \c-functions which are also monotonically decreasing and interpolate between the same fixed point values. Hence one may wonder whether a new version of the entropic \c-function can be found to ameliorate the situation with respect to stationarity. 

To begin, one might first consider the flow of our new \c-function defined using the mutual information. In fact, we argue that these flows are unchanged in comparison to those for eq.~\reef{funi} as follows: The entanglement entropy of the annular strip is a new ingredient in defining eq.~\reef{funix} and so one might hope that this is enough to change the behaviour near the free scalar fixed point. However, the arguments in section \ref{entropy} indicated that the entropy of this annular region is purely a UV quantity. Hence it will be unaffected by the introduction of a mass scale (with $\veps\ll 1/m$) and so it will not modify the flow of the \c-function near the fixed point of the massless free scalar  (or any other fixed point).

At RG fixed points, the conformal symmetry of the underlying field theory was used to show that the charge $c_0$ can also be identified with the universal contribution in the free energy of the three-sphere \cite{CHM}.\footnote{Recall that that in principle, the free energy has regularization ambiguities similar to the entanglement entropy. It would be interesting to investigate if a conformal mapping of the mutual information geometry might produce a practical new scheme in which the resulting `free energy' which is independent of the UV regulator.} 
We can also think of applying the conformal symmetry to map the geometry introduced in section \ref{newdef} for the mutual information to various new geometries (either in $R^3$ or $S^3$). While the mutual information of these new configurations would still identify the same charge $c_0$ at the fixed points, it would define new \c-functions on the RG flows that interpolate between fixed points since the conformal symmetry is lost along the flows. While one can easily test if some new \c-function defined in this way is stationary at the free scalar fixed point, the challenge would be to prove that in general, it decreases monotonically along any RG flow.

The geometry introduced in section \ref{newdef} for the mutual information is natural to consider because it implements a `point-splitting' of the circular entangling surface in the original entanglement entropy. However, another interesting direction to investigate would be to replace the mutual information with some measure of quantum entanglement $E(A^+,A^-)$ (\eg entanglement of formation, distillable entanglement or relative entropy of entanglement \cite{entang}) as the regulator of entanglement entropy. It can be expected that in the limit of coinciding entangling curves (\ie $\veps\to0$), the result should reproduce the behavior of the entropy to some extent. This is because
entanglement measures for complementary regions in a global pure state coincide
with twice the entanglement entropy, exactly as for the mutual information. However, this is likely a rather impractical approach due to the notorious difficulties in computing entanglement measures. However, we naturally wonder
how much the expansion in $\veps/R$ for entanglement measures differ from the one
of the mutual information. In particular, it would be interesting to see if the terms in the $\veps$ expansion coincide all
the way down to the same constant term (in odd dimensions or the logarithmic term in even
dimensions).

\bigskip

\subsection*{What is $c_0$?}

The original \c-theorem in two dimensions points to the central charge $c$ at the RG fixed points. However, $c$ plays many different roles in the physics of the underlying conformal field theories. Cardy's conjecture \cite{cardy0} suggests that a key role was as the coefficient of the trace anomaly. This conjecture gives a unified perspective for \c-theorems in any even number of spacetime dimensions, by suggesting that they all refer to the coefficient of the A-type anomaly. Of course, the more recent perspective is to relate \c-theorems in general dimensions to the universal contribution in the entanglement entropy of a sphere \cite{myers1}. Our arguments in section \ref{entropy} can be extended to higher dimensions and so our approach using the mutual information also produces the same `central charge' in general dimensions. While the new entanglement proposal chooses the same  charge as Cardy's conjecture for even dimensions, it also provides a unified framework to consider \c-theorems in both even and odd numbers of dimensions. 

This unified view can be carried further for free fields and we will show that the central charges defined by the entanglement entropy in odd dimensions
can actually be obtained by analytic continuation of those in even dimensions. Similar ideas were considered in \cite{igor1,dimreg}. The idea is that in order to compare
between different dimensions, in each dimension we  normalize the central charges by that
in a particular `reference' CFT, for example, the free massless scalar field. The central charges can be identified as the universal contribution to the free energy on a $d$-dimensional sphere, \ie the coefficient
of the logarithmic term for even dimensions and
the constant term for odd dimensions \cite{CHM}. For free field theories, we may use zeta-function regularization, which completely removes all of the power law divergences leaving only the desired universal contribution. In this case, the free energy is given by \cite{birrel}:
\begin{equation}
F=-\log (Z(S^d))=\frac{1}{2} \lim_{s\rightarrow 0}\Big[\mu^{ 2
s}\zeta^\prime(s,d)+\zeta(s,d)\, \log\mu^2\Big]\,,\labell{wq}
\end{equation}
where $\mu$ is a cut-off energy scale and $\zeta(s,d)$ is the appropriate zeta-function on the
sphere $S^d$. Then the ratios of the charges for two free models will be
$\zeta_1(0,d)/\zeta_2(0,d)$ for even dimensions and
$\zeta_1^\prime(0,d)/\zeta_2^\prime(0,d)$ for odd dimensions, since $\zeta(0,d)=0$ for odd dimensions. However, the latter also implies that
$\zeta(s,d)=s\, \zeta^\prime(0,d)+...$ for small $s$ in odd dimensions and so, in fact, in both cases we can write the ratios of central charges as
\begin{equation}
\lim_{s\rightarrow 0}\frac{\zeta_1(s,d)}{\zeta_2(s,d)}\,.
\labell{gulp}
\end{equation}
One might say that the anomaly coefficients (\ie the coefficients of the logarithmic terms) vanish in odd dimensions but the ratios of
anomalies do not. As an example, let us consider the ratio of the central charges for a Dirac fermion and a scalar in even dimensions. We also divide the ratio by the number of degrees of freedom of the fermion, \ie $2^{d/2}$ and expand for large $d$ \cite{cap}
 \begin{equation}
\frac{C_{\textrm{Dirac}}}{2^{d/2}\,
C_{\textrm{scalar}}}=(d-2)+k_0+\frac{k_1}{d}+\frac{k_2}{d^2}+...
 \labell{cape}\end{equation}
Now the exercise is to fit the coefficients $k_i$ using the known values of the even-dimensional central charges \cite{cap} and then continue the resulting formula to odd dimensions. Surprisingly, this analytic continuation does a good job at predicting the corresponding ratios for the odd-dimensional charges (also divided by the number of degree of freedom of the Dirac field), even for $d=3$. In particular, fitting the first 20 $k$'s over 
the first $100$ even dimensions, we find for $d=3$
\begin{equation}
\frac{C_{\textrm{Dirac}}}{2 C_{\textrm{scalar}}}=1.7157936606\cdots\,,
\end{equation}
which is extremely close the exact value: $1.7157936649\cdots$ \cite{Klebfree}.

Eq. (\ref{cape}) shows how different the entropic central charge can be from the standard
counting of numbers of  degrees of freedom \cite{cap}. The contribution of
topological entanglement entropy to the charge in $d=3$ is an extreme case
of mismatch between $c_0$ and number of degrees of freedom. This also calls into question the standard idea
that the \c-theorem describes the loss of degrees of freedom along
the RG flow. The relation of charge with entanglement
entropy (or mutual information) suggests that a more fruitful approach might be to attribute its
physical meaning to some measure of entanglement --- see for example
\cite{tarum,latorre}. 

One complication that obscures the physical meaning of this entropic central charge is that 
the $c_0$ term in eq.~\reef{unis} is overwhelmed by the power-law divergent contributions. Even in our mutual information approach, the universal contribution is still a small subdominant contribution compared to, \eg the area law term in eq.~\reef{mi1}. From this perspective, it is interesting to observe that this same coefficient appears as the coefficient of the entropy density of a CFT on a curved background $R\times H^{d-1}$ where the curvature and temperature are related by ${\cal R}=-4\pi^2 (d-1)(d-2)T^2$ \cite{myers1,CHM}.\footnote{We would like to thank Kris Jensen for discussing this point.} It is natural to think that by varying the temperature that the thermal entropy is probing the theory at different scales and in its role of specifying the entropy density, the central charge appears in the leading contribution. Formulating a \c-theorem in terms of the thermal entropy density (or free energy) in flat space was examined before but it was found that corresponding charge does not decrease in certain RG flows \cite{flaat1,flaat2}. Here the conformal mapping of the entanglement entropy calculation suggests that we should instead consider the thermal entropy density in a particular curved space background.  
%%rob edited here
Certainly, this suggestion faces a number of challenges. First of all, the entropy density at fixed temperature may not be very well defined in a hyperbolic geometry, where the volume and the area grow in the same way at large radii. Hence it would be difficult to isolate a volume contribution to the total thermal entropy from possible boundary terms. In particular, it is insufficient to define the entropy density as the total entropy divided by the volume for large regions in the hyperbolic geometry. Of course, a second challenge is to show that the (properly defined) entropy density defines a proper \c-function which decreases monotonically with decreasing temperature (and curvature). An additional complication in this regard is the following: We know how to uniquely define the theory on the curved background only at the conformal fixed points. When mass
scales are introduced, however, we do not know how to fix the nonminimal couplings with the curvature. Hence without a specific prescription, there will be many possible interpolating theories for a given RG flow in flat space.

The \F-theorem was originally discovered in a holographic framework \cite{myers1}.
In this context, one is considering a strongly coupled three-dimensional boundary theory with a large number of degrees of freedom, which is dual to some classical gravity theory in a four-dimensional bulk. In the gravity description, the \F-theorem corresponds to the monotonic behaviour of the corresponding entropy functional evaluated on some extremal surfaces (provided that the null energy condition holds for the matter fields in the bulk). One might note that this has some resemblance to the black hole area theorem. However, since the \F-theorem is valid for general quantum field theories in three dimensions, an interesting question is what the dual gravitational interpretation of the \F-theorem would be for theories which are not strongly coupled and/or do not contain a large number of degrees of freedom. It seems that some interesting monotonic behaviour continues to survive even in the realm of quantum gravity in the bulk.

\acknowledgments{We thank Nima Doroud, Eduardo Fradkin,
Tarun Grover, Kris Jensen, Igor Klebanov, Martin Kruczenski, Hong Liu, Roger Melko, Anton
van Niekerk, Ben Safdi, Erik Tonni, Stefan Theisen, Jonathan Toledo and Xiao-Gang Wen for useful discussions. 
HC is grateful to the Perimeter Institute for hospitality
during the initial stages of this work. Research at
Perimeter Institute is supported by the Government of Canada through Industry
Canada and by the Province of Ontario through the Ministry of Research \&
Innovation. RCM also acknowledges support from an NSERC Discovery grant and
funding from the Canadian Institute for Advanced Research. AY was supported by
a fellowship from the Natural Sciences and Engineering Research Council of
Canada. HC and MH acknowledge support from CONICET and Universidad de Cuyo,
Argentina.}

\appendix

\section{Holographic calculation of mutual information}
\labell{holographic}

In the framework of gauge/gravity duality, the entanglement entropy in the
boundary QFT can be evaluated with a geometric calculation in the bulk gravity
theory \cite{Ryu1}. In particular, the entanglement entropy between a (spatial)
region $V$ and its complement in the boundary is computed with the following prescription \cite{Ryu1}:
 \be
S(V) =\ \mathrel{\mathop {\rm ext}_{\scriptscriptstyle{v\sim V}} {}\!\!}
\frac{\cA(v)}{4\,G_\mt{N}}\,.
 \labell{define}
 \ee
That is, one evaluates the Bekenstein-Hawking formula on all surfaces $v$ in the
bulk which are homologous to the boundary region $V$ and the entanglement
entropy is given by the extremal result. This prescription was found to pass
wide range of consistency tests, \eg see \cite{CHM,Ryu1,head7,head9,highc},
however, quite recently a derivation of eq.~\reef{define} was also presented in
\cite{aitor}. We might add that eq.~\reef{define} assumes that the bulk theory
corresponds to Einstein gravity and this expression has been extended to
certain higher curvature theories as well \cite{highc,Jan2}
--- see also \cite{highc2}. Let us also note here that mutual information was
discussed in a holographic context by \cite{head9,holop,erik,erik2}. 

Now we wish apply this holographic formalism to evaluate the constant $\tc_0$ for the three-dimensional
boundary CFT, following the approach given in section \ref{mutual}. That is, we apply the holographic techniques to calculate the mutual information between a disc of radius $R_-=R - \frac12 \veps$ and the
exterior of a circle of radius $R_+=R+\frac12\,\veps$, as pictured in figure
\ref{fig:Mutual0}. To compute this mutual information from
eq.~(\ref{mutualdef}), we must evaluate eq.~\reef{define} for two kinds of
minimal surface in four-dimensional anti-de Sitter (AdS) space: first, minimal surfaces ending on a circle in the boundary
and second, an extremal bulk surface homologous to a two-dimensional annulus, as illustrated in
figure \ref{fig:Mutual2}. The first has already been studied in detail
\cite{CHM,Ryu1} and we will review the main ideas of this calculation here, as
it serves as a simple introduction to the more complicated calculations for the
second class of surfaces.
\begin{figure}
\centering
\includegraphics[width=.7\textwidth]{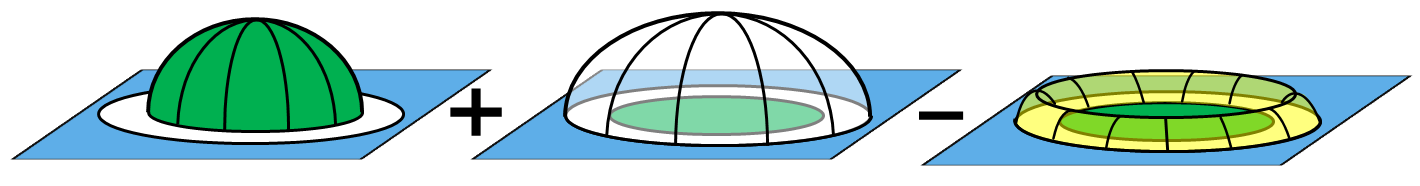}
\bigskip
\caption{(Colour online) The holographic mutual information, for the boundary
configuration illustrated in figure \ref{fig:Mutual0},
requires evaluating the area of three minimal surfaces in the bulk. First, there are two
hemi-spheres on the left which yield $S(A^+)$ and $S(A^-)$. Second, there is the annular
surface on the right yielding $S(A^+\cup A^-)$.} \labell{fig:Mutual2}
\end{figure}

Hence to begin, we determine the holographic entanglement entropy for a
circular entangling surface in the three-dimensional boundary CFT. The vacuum
state is represented by four-dimensional AdS space in the dual gravity theory
and the bulk metric may be written as
 \beq
ds^2 = \frac{{L}^2}{z^2} \left[ dz^2 + d\te^2+dr^2 + r^2 d\theta^2 \right]~,
 \labell{bulkmetric}
 \eeq
where $L$ is the AdS curvature scale. We are working in Euclidean signature with
$\te$ being the Euclidean time direction. The two remaining boundary directions
are written in polar coordinates, with $r$ and $\theta$. We choose the
entangling surface in the boundary to be the circle $r=R$ in the constant time
slice $\te=0$. If we describe the bulk surface with a profile $z(r)$, the
induced metric on this surface is
 \beq
ds^2 = \frac{{L}^2}{z^2} \left[ (1 + z'^2)dr^2 + r^2 d\theta^2 \right]~,
 \labell{inducedM}
 \eeq
where the prime denotes differentiation with respect to $r$. The corresponding
area is given by
 \myeq{ \cA =
2\pi {L}^{2} \int dr \frac{r}{z^{2}} \sqrt{1+z'^2}\,.
 \labell{areaS}}

According to eq.~\reef{define}, we must identify the surface with extremal area
and so extremizing the `action' \reef{areaS} yields the following equation:
 \myeq{ r z z'' +  z (z')^3 +  z z' + 2 r (z')^2
+ 2 r = 0~.
 \labell{eomS} }
Now we must find a solution satisfying the boundary conditions that it reaches
the AdS boundary at $r=R$, \ie $z(r=R)=0$ and that the profile is smooth at
$r=0$, \ie $z'(r=0)=0$. It turns out there is a simple solution, namely the
half-sphere \cite{Ryu1}
 \be
 z^2+r^2=R^2\,,
 \labell{halfsph}
 \ee
shown on the left of figure \ref{fig:Mutual2}. The minimal area may then be
written as
 \myeq{  \cA = 2\pi{L}^{2}\,R
\int_{\delta}^R \frac{dz}{z^{2}} ~,\labell{spherearea} }
where we have rewritten the result in terms of an integral over the bulk radius
$z$. We have also introduced a UV regulator surface $z=\delta$, which cuts off
this integral near the AdS boundary. In the boundary CFT, $\delta$ plays the
role of short distance cut-off. Hence the holographic entanglement entropy
becomes
\begin{equation}
S=\frac{\cA}{4 G_\mt{N}} =  \frac{\pi}2\, \frac{{L}^2}{G_\mt{N}}
  \left(\frac{R}{\delta}-1\right) \,.
\labell{spherearea2}
\end{equation}

Comparing this result to eq.~\reef{enti}, we would identify the corresponding
\c-function as \cite{myers1}
 \be
 c_0=\frac{{L}^2}{4\,G_\mt{N}}\,.
 \labell{holoc0}
 \ee
That is (up to a factor of $-2\pi$), this is the constant contribution in
eq.~\reef{spherearea2} if we simply ignore the divergent term proportional to
$R/\delta$. Of course, the holographic regulator is naturally a geometric one,
\eg \cite{calc2} and so this same $c_0$ also appears if we apply the formula of
\cite{liu} to eq.~\reef{spherearea2}, \ie $2\pi c_0=R\,\partial_R S(R) -
S(R)$. The advantage of the mutual information, as discussed in
section \ref{mutual}, is that it yields a finite result from which we may determine
$c_0$. Hence in the present holographic framework, all dependence on $\delta$
will vanish after we subtract the area for the annular surface from that of the
spheres.

Hence we move to remaining ingredient in our holographic calculation of the
mutual information \reef{mutualdef}, \ie $S(A^+ \cup A^-)$. Of course, this also
corresponds to the entanglement entropy of the annular region between the two
circles, \ie $\overline{A^+\cup A^-}$. Given this boundary region, there are two possible 
extremal surfaces in the bulk. The first would simply be the combination of the two hemispheres
appearing in the previous calculation of $S(A^\pm)$. The second possible extremal surface has the shape of a two-dimensional annulus in the bulk, as illustrated in the right-most diagram of figure \ref{fig:Mutual2}. Since we are interested in the limit $\veps/R\ll1$, this second one is the minimal area surface and hence it yields $S(A^+ \cup A^-)$ in eq.~\reef{define}. 
Now we can again describe this new annular surface with a
profile $z(r)$, as in the previous calculation, and the area and equation of
motion would be identical to the expressions in eqs.~\reef{areaS} and \reef{eomS}, respectively. The
only change is that the boundary conditions are now chosen as $z(r=R_A)=0$ and
$z(r=R_B)=0$. Here we note that finding the extremal bulk surface with these
boundary conditions has already been studied in the context of studying
correlation functions of circular Wilson loops
\cite{Zarembo,Drukker}.\footnote{While the Wilson loops were studied in these reference for a
four-dimensional boundary theory, the extremal surface lies within an AdS$_3$
slice of the five-dimensional bulk geometry. In the present case, the extremal surface is restricted to the AdS$_3$ slice defined by setting $\te=0$ in eq.~\reef{bulkmetric}. Hence precisely, the same extremal surface appears with the same boundary conditions in both calculations.} These surfaces have also been
considered in the context of holographic entanglement entropy by \cite{HiraTak,bounce,erik}. Hence we only need to review the
salient steps of these calculations here.

As a first step, we introduce `improved' coordinates $u$ and $v$ with
\cite{Drukker}
 \beq
\frac{r}L = \frac{e^v}{\sqrt{1+u^2}}\,, \qquad \frac{z}L =
\frac{u\,e^{v}}{\sqrt{1+u^2}}\,.
 \labell{params}
 \eeq
We note that both $u$ and $v$ are dimensionless and further that $u=z/r$. We
can think of $u$ as an `improved' radial coordinate in the bulk AdS$_4$
spacetime. In particular, we have $u=0$ at $z=0$. Similarly, $v$ plays the role
of an `improved' radial coordinate on the boundary. We will describe the
extremal surface with a profile $v(u)$ and then the corresponding area becomes
 \myeq{ \cA = 2\pi L^{2} \int
\frac{du}{u^2}\, \sqrt{ \frac{1+(1+u^2)^2\, (\partial_u v)^2}{1+u^2}}\,.
 \labell{area}  }
The advantage of the new coordinate choice becomes apparent here since the
integrand above does not contain any explicit dependence on $v$. Hence we have
a constant of the motion, which we denote as
 \be
\frac{\delta \cA}{\delta\,\partial_u v}=\frac{2\pi L^2}{q^2}\,.
 \labell{const8}
 \ee

In the following, it turns out that  we will be primarily interested in the
regime $q\ll1$, which will correspond to $\veps\ll R$. It will be convenient to
define a rescaled coordinate $u=q\,y$. With this choice, the area becomes
 \myeq{ \cA = \frac{2\pi L^{2}}{q^{2}} \int \frac{dy}{y^2}\,
\sqrt{ \frac{q^2+(1+q^2\, y^2)^2\,  (\partial_y v)^2}{1+q^2\, y^2}} \,.
 \labell{area2} }
Now using eq.~\reef{const8}, we find a first
order equation to determine $v(y)$
 \myeq{  \partial_y v =  \frac{q\,y^2}{1+ q^2 y^2}\,\frac{1}{\sqrt{1+q^2 y^2-y^4}} ~.
 \labell{vprime} }
where we have implicitly assumed that $\partial_y v>0$ here. At this point, we note that
our description of the minimal surface will have two branches, as shown in
figure \ref{fig:shape}. The parameter $y$ increases from $y=0$ (and
$v=\log(R_-/L)$) to the turning point at $y=y_+$ on the left branch, then
decreases from $y=y_+$ to $y=0$ (and $v=\log(R_+/L)$) on the right branch. At
the turning point, $\partial_v y =0$ or $\partial_y v$ diverges and so examining eq.~\reef{vprime}, we identify $y_+$ as the
positive root of $1 + q^2 y_+^2 -y_+^4=0$, \ie
 \beq
y_+^2=\sqrt{1+\frac{q^4}4}+\frac{q^2}2\,.
 \labell{root}
 \eeq

The profile of the extremal surface is illustrated in Figure \ref{fig:shape} in
terms of the original $z$ and $r$ coordinates. Note that this profile is
asymmetric, as is natural since the outer circle on the boundary has a larger
radius than the inner circle. Further note that the profile $z(r)$ is
multi-valued. That is, the left branch extends in to region $r<R_-$ before
bending back to meet the right branch at $y=y_+$. This inward `sagging' of the
the minimal surface arises because of the negative extrinsic curvature of the
inner component of the entangling surface on the boundary \cite{relative}. The
same properties will also true for the corresponding extremal surfaces in an
analogous construction in higher dimensions. In contrast, in terms of the
improved coordinates, the extremal profile $v(y)$ is symmetric about $y=y_+$,
as well as being single-valued.
\begin{figure}[t]
  \centering
    \includegraphics[width=.3\textwidth]{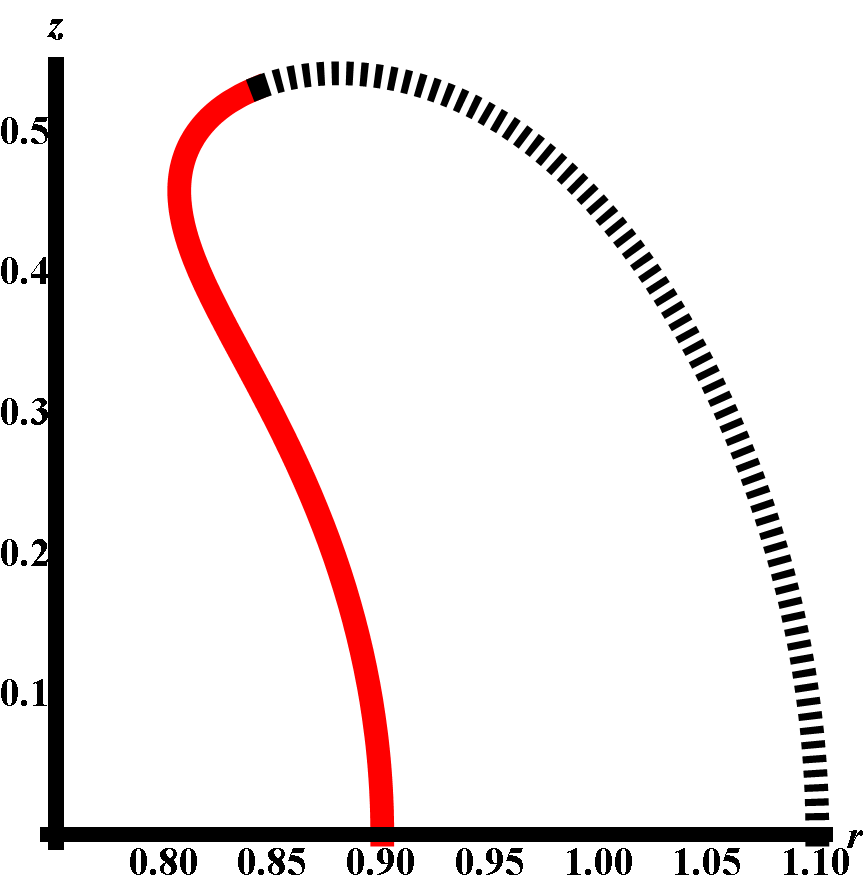}
  \caption{(Colour online) Shape of the minimal surface, with $R=1$ and $\veps=0.2$
($\alpha=0.1$), which ends on the annulus.  The red continuous curve
corresponds to the left branch, which sees $u$ start at $u=0$ at the
bottom before increasing to $u=u_+$.  The black dashed curve
corresponds to the right branch, with $u=u_+$ at the top decreasing to
$u=0$ at the bottom.  Notice that the minimal surface sags inward: this
is because the outer ring has a larger circumference than the inner
one.}
 \labell{fig:shape}
\end{figure}

Now substituting eq.~\reef{vprime} into the integral \reef{area2} yields the
extremal value of the area and the holographic entanglement entropy becomes
 \myeq{ S = \frac{\cA}{4 G_\mt{N}}=\pi\frac{L^{2}}{G_\mt{N}}\,\frac1{{q}}
\int_0^{y_+} \frac{dy}{y^2}\, \frac{1}{\sqrt{1+q^2 y^2-y^4}} ~,
 \labell{area3} }
where an additional factor of 2 was included here to account for integrating
over both branches, as described above. We also account for the boundary
conditions by integrating eq.~\reef{vprime} over the extremal surface to find
 \beq
 \log\frac{R_+}{R_-}=2\,{q}\,\int_0^{y_+} dy\,\frac{y^2}{1+ q^2 y^2}
 \,\frac{1}{\sqrt{1+q^2 y^2-y^4}} ~.
 \labell{range}
 \eeq
Now the above integral can be given in terms of elliptic integrals, however, the resulting expression is not very enlightening.
% \beq
%\log\frac{R_B}{R_A}=\frac{2y_+}{{q}}\, \left[\,K\!\left(iy^2_+\right) -
%\Pi\!\left(-q^2y_+^2,iy_+^2\right)\,\right]\,.
% \labell{range2}
% \eeq

It is clear that the integral in eq.~\reef{area3} diverges at the lower limit
$y=0$ but, of course, we expected such a UV divergence to appear in the
entanglement entropy. By subtracting the divergent behavior, \ie $\int dy/y^2$,
we can isolate a finite contribution to the entanglement entropy as
 \beq
S_{fin} = \pi\frac{L^{2}}{G_\mt{N}}\,\frac1{{q}}\left[ \int_0^{y_+}
\frac{dy}{y^2}\, \left(\frac{1}{\sqrt{1+q^2 y^2-y^4}}-1\right)\ -\
\frac1{y_+}\right] ~,
 \labell{area4}
 \eeq
which can be expressed as
 \beq
S_{fin}=\pi\frac{L^{2}}{G_\mt{N}}\,\frac1{{q}\,y_+}\left[ \, K(i y_+^2) - E(i
y_+^2)\,\right] \,,
 \labell{area9}
 \eeq
where $K(k)$ and $E(k)$ are complete elliptic integrals of the first and second kind, respectively.
Note that this contribution is completely independent of the short distance
cut-off. The remaining cut-off dependent term takes the simple form
 \beq
S_{div} =
\frac\pi2\,\frac{L^{2}}{G_\mt{N}}\,\left[\frac1{u_{+,min}}+\frac1{u_{-,min}}\right]
 \labell{dive0}
 \eeq
where $u_{\pm,min}={q}\,y_{\pm,min}$ is the minimum bulk `radius' reached at the
position of the regulator surface $z=\delta$ as $v$ approaches $\log(R_\pm/L)$.
Above, we have separated the two boundary contributions as they are not
identical. Using eqs.~\reef{params} and \reef{vprime}, we find
 \beq
\frac1{u_{\pm,min}}=\frac{R_\pm}\delta\,\left(1-\frac12\,\frac{\delta^2}{R_\pm^2}
+\frac13\,\frac{\delta^3}{R_\pm^3}+\cdots\right)\,.
 \labell{boundary}
 \eeq
An important point for the discussion below is that from this expansion, we can
see that eq.~\reef{dive0} does not contain a cut-off independent contribution,
\ie it will not contribute to $c_0$ in our calculation of the mutual information.
Rather we can rewrite $S_{div}$ as
 \beq
S_{div} = \frac\pi2\,\frac{L^{2}}{G_\mt{N}}\,\left[\frac{
R_+}{\delta}+\frac{R_-}{\delta}\right] \ + \ O(\delta)\,.
 \labell{dive}
 \eeq
Hence we see that $S_{div}$ contains the sum of the area law divergences appearing in
eq.~\reef{spherearea2} for the two entangling circles. Therefore, as expected, these
divergences will cancel in calculating the mutual information. Of course,
eq.~\reef{dive} also contains higher order contributions, which vanish in the
limit $\delta\to0$, but again no constant term appears.

Before proceeding with a detailed analysis of the finite contribution
$S_{fin}$, let us make an important (qualitative) observation. First from
eq.~\reef{range}, it is clear that the limit $q\to0$ coincides with the limit
$R_+\to R_-$. Now working in this regime, let us write the radii as
eq.~\reef{eps0}, \ie
 \beq
R_-=R-\left(\frac12-\alpha\right)\veps\,, \qquad
 R_+=R+\left(\frac12+\alpha\right)\veps\,.
 \labell{eps}
 \eeq
That is, as in eq.~\reef{radiusR}, we are allowing $R$ to be shifted slightly
from the average of the two circle radii. Substituting eq.~\reef{eps} on the
left-hand side of eq.~\reef{range}, we have
 \beq
\log\frac{R_+}{R_-}\simeq \frac\veps R -\alpha \left(\frac\veps R\right)^2+
\left(\frac1{12}+\alpha^2\right)\,\left(\frac\veps R\right)^3 -\alpha
\left(\frac1{4}+\alpha^2\right)\,\left(\frac\veps R\right)^4+\cdots\,.
 \labell{keyx}
 \eeq
Hence we see that introducing a nonvanishing $\alpha$ changes the nature of the
above expansion, in that it includes both odd and even powers of $\veps/R$. It
is straightforward to show that if $\alpha=0$ the expansion only contains odd
powers.\footnote{With $\alpha=0$ in eq.~\reef{eps}, $\veps\to-\veps$ is
equivalent to interchanging $R_+\leftrightarrow R_-$, which changes the sign of
the logarithm on the right-hand side of eq.~\reef{keyx}.} These observations
make clear that $\alpha=0$ is a distinguished case in our analysis but we now
examine the implications of this behavior for the mutual information.

In the regime of small $q$, we may rewrite eq.~\reef{range} using
eq.~\reef{keyx} and making a Taylor series expansion\footnote{The specific
values of the coefficients in the small-$q$ expansion here and in
eq.~\reef{expand98} are not important for the present argument. However, in the
following, we will explicitly calculate the leading terms and find that
$\beta_0=2\sqrt{\pi}\,\frac{\Gamma(3/4)}{\Gamma(1/4)}$ and
$\gamma_0=\pi^{1/2}\frac{\Gamma(3/4)}{\Gamma(1/4)}$.} on the right-hand side in
terms of $q$
 \beq
\frac\veps R -\alpha \left(\frac\veps R\right)^2+\cdots= {q}\,(\beta_0 +
\beta_1\,q^2 +\cdots)\,,
 \labell{expand99}
 \eeq
which yields
 \beq
{q}\simeq \frac{1}\beta_0\,\frac\veps R -\frac\alpha\beta_0\,
\left(\frac\veps R\right)^2+O\!\left[\left(\frac\veps R\right)^3\right]\,.
 \labell{expand98}
 \eeq
We can make a similar small-$q$ expansion in eq.~\reef{area4} and it is
straightforward to see that this finite contribution to the entanglement
entropy takes the form
 \beqa
S_{fin}&=& -\pi\frac{L^2}{G_\mt{N}}\,\frac1{{q}}\,\left(\gamma_0 +\gamma_1 q^2
+\cdots\right)
 \nonumber\\
&=& -\pi\frac{L^2}{G_\mt{N}}\,\beta_0\gamma_0\left( \frac R{\veps}+ \alpha +
O(\veps/ R) \right)\,.
 \labell{expand97}
 \eeqa
Hence we see that a constant term (\ie independent of $\veps/R$) appears
in $S_{fin}$ only when $\alpha\ne0$ or rather this contribution can be eliminated by
setting $\alpha=0$. Combining this with the observation that no constant term
appears in eq.~\reef{dive} for $S_{div}$, we see that in our holographic
calculation, $S(A^+\cup A^-)$ will not contribute to $\tc_0$ in eq.~\reef{mi1} as long as we choose $\alpha=0$ or rather
$R=(R_++R_-)/2$. Of course, $\alpha=0$ is precisely the choice which our general arguments in section \ref{critb} indicated was needed to eliminate the constant contribution coming from $S(A^+\cup A^-)$ and so here we have a confirmation of this general reasoning in our holographic calculations. Hence we complete the calculation of the mutual information in the following with $\alpha=0$.

We now evaluate the above expressions in the limit $q\to0$, which we saw above corresponds to taking
$\veps\ll R$. From eq.~\reef{root}, we have $y_+\to1$ in this limit 
and so eq.~\reef{area9} reduces to
 \beqa
S_{fin}&=&\pi\frac{L^{2}}{G_\mt{N}}\,\frac1{{q}}\left[ \, K(i) - E(i )\,\right]
+ O({q})
 \nonumber\\
&=&
%
%-2\pi^2\frac{L^{2}}{\lp^2}\,\frac1{\sqrt{q}}\, \frac{\Gamma(1/2)\,
%\Gamma(3/4)}{\Gamma(5/4)}=
%
-\pi^{3/2} \frac{\Gamma(3/4)}{\Gamma(1/4)}\,\frac{L^{2}}{G_\mt{N}}\,\frac1{{q}}
+ O({q})\,.
 \labell{area9a}
 \eeqa
Similarly, evaluating eq.~\reef{range} in this limit, we find
 \beqa
%
%\log\frac{R_B}{R_A}=
%
\frac{\veps}{R}+O\!\left[\left(\frac\veps R\right)^3\right]
 &=&2\,{q}\,\int_0^{1} dy\,\frac{y^2}{\sqrt{1-y^4}}+O(q^3)
 \nonumber\\
 &=&
%
%\frac{{q}}{2}\,\frac{\Gamma(1/2)\,\Gamma(3/4)}{\Gamma(5/4)}=
%
2\pi^{1/2} q\,\frac{\Gamma(3/4)}{\Gamma(1/4)}+O(q^3)\,.
 \labell{range3}
 \eeqa
Combining these expressions then yields
 \beq
S_{fin}=-\pi\, \frac{\Gamma(3/4)^2}{\Gamma(1/4)^2}\,\frac{L^{2}}{G_\mt{N}}
\,\frac{2\pi R}{\veps}+O(\veps/R)\,.
 \labell{area9b}
 \eeq

Note with $R=(R_++R_-)/2$, eq.~\reef{dive} can be written as
 \beq
S_{div} = \frac{L^{2}}{2\,G_\mt{N}}\,\frac{2\pi R}{\delta} \ + \ O(\delta)\,.
 \labell{dive5}
 \eeq
Now let us compare the leading results in eqs.~\reef{area9b} and \reef{dive} to
the holographic entanglement entropy of a strip of width $\veps$ and length $H$
\cite{Ryu1}\footnote{The reader may find it easier to compare with the strip entropy in \cite{m1}, which uses the same notation and conventions as in the present calculation.}
 \be
S_\mt{strip}\; =\;\frac{L^2}{2\,G_\mt{N}} \, \frac{H}{\delta} - \pi\,
\frac{ \Gamma(3/4)^2}{\Gamma(1/4)^2} \,\frac{L^2}{G_\mt{N}}\,
\frac{H}{\veps}\,.
 \labell{eqn12}
 \ee
Hence to leading order in the limit $\veps\ll R$, the present entanglement entropy
matches that for the strip above. In particular, with $H=2\pi R$, $S_{div}$ in
eq.~\reef{dive5} precisely matches the area law term while
$S_{fin}$ in eq.~\reef{area9b} reproduces the finite contribution proportional to $1/\varepsilon$.

Combining our holographic results for the individual entanglement entropies,
the mutual information \reef{mutualdef} becomes
 \be
I(A^+,A^-)= 2\pi^2\, \frac{\Gamma(3/4)^2}{\Gamma(1/4)^2}\,
\frac{L^{2}}{G_\mt{N}} \,\frac{R}{\veps} -\pi\,
\frac{{L}^2}{G_\mt{N}}+O(\veps/R)\,.
 \labell{hend6}
 \ee
We see that this result takes precisely the expected form, given in
eq.~\reef{mi1}. In particular from the constant term, \ie the term independent of $\veps/R$, we identify the value of the
\c-function in our holographic CFT as
 \be
 \tilde{c}_0=\frac{{L}^2}{4\,G_\mt{N}}\,.
 \labell{holoc00}
 \ee
Further we have a precise match $\tc_0=c_0$, ie, the above result agrees with the value identified in eq.~\reef{holoc0} from the entanglement entropy of a circle. We should not be surprised by this agreement here since holography naturally provides covariant regulator in our calculation of the entanglement entropy. Of course, this match occurred because only the circle entropies contributed to the constant term in
eq.~\reef{hend6}. That is, as was stressed above, with the choice $\alpha=0$, the entanglement entropy
of the annular region did not have a constant term. It would interesting to
verify the analogous result in a holographic framework where the bulk theory
involved a higher curvature interactions, \eg \cite{highc,Jan2} or where the bulk
solution describes a holographic RG flow, \eg \cite{liu,m1,klop}. 

%%rob
\subsection{Strip entanglement entropy in RG flows}

\labell{kuprate}

In this section, we wish to illustrate with a holographic calculation that if the RG flow is driven by an operator which is close to being marginal, one may find additional `area law' terms in the mutual information \reef{mi1} which diverge in the limit $\veps\to0$. Our calculations here are a simple extension of the holographic analysis of \cite{calc2}. There the focus was on identifying situations where the RG flow would generate additional universal terms accompanied by logarithmic divergences $\log(\mu\delta)$. However, as we will see below additional `unconventional' power law divergences may also be generated and in the appropriate limit, the mutual information is also modified. For example, with $d=3$, if the RG flow is initiated by a relevant operator with conformal dimension in the range $3>\Delta>5/2$, then the entanglement entropy \reef{enti} will contain an additional contribution of the form $2\pi R\, a'\, \lambda^2/\delta^{2\Delta-5}$ where $\lambda$ is the coupling constant for the new operator and $a'$ is a dimensionless coefficient. An analogous term of the form $2\pi R\, \ta'\, \lambda^2/\veps^{2\Delta-5}$ also appears in the mutual information \reef{mi1} if $\lambda\,\veps^{3-\Delta}\ll 1$. We note that similar contributions are also generated in appropriate circumstances with the extensive model studied below in appendix \ref{extensive} --- see footnote \ref{foot99}.

The following calculations follow quite closely the analysis presented in \cite{calc2}: For simplicity, we focus on the case of Einstein gravity coupled to a scalar field in four bulk dimensions, with an action of the form
 \beq
I = \frac{1}{16\pi G_\mt{N}} \int \mathrm{d}^{4}x \, \sqrt{-g}\, \left[ \frac{6}{L^2}+ R
-\frac12(\partial\Phi)^2-\frac{m^2}2\, \Phi^2 - U(\Phi)\right]\,,
 \labell{action99}
 \eeq
where the potential $U(\Phi)$ contains interactions which are cubic or higher order in $\Phi$.
According to the standard AdS/CFT dictionary \cite{revue} (in the vacuum with $\Phi=0$), the scalar field is dual to a scalar operator with conformal dimension
 \be
 \Delta=\frac{3}2+\sqrt{\frac{9}4+m^2L^2}
 \labell{dimension}
 \ee
in the boundary theory.  We write the asymptotically AdS$_4$ bulk metric with a flat boundary as
 \be
ds^2=\frac{L^2}{z^2}\left(dz^2+f(z)\, \eta_{ij} dx^i dx^j\right)\,,
 \labell{back2}
 \ee
where $f(z)\to1$ as $z\to0$. Now in a holographic RG flow, the leading asymptotic behaviour of the fields will be
 \beqa
  f(z)&=&1-\frac18\, \lambda^2\,{z}^{2(3-\Delta)}+\cdots\,,
\labell{series}\\
\Phi(z)&=&\lambda\,z^{3-\Delta}+\cdots\,,
 \nonumber
 \eeqa
where $\lambda$ is the (dimensionful) coupling constant in the boundary theory which initiates the RG flow.
Both of the ellipses above represent terms involving higher powers of the combination $\lambda\,{z}^{3-\Delta}$.

Now let us consider a strip where the entangling surface consists of two straight parallel lines separated by a distance $\veps$. With some work (see \cite{calc2} for details), the holographic entanglement entropy becomes\footnote{Note there is a typo in the corresponding eq.~(3.15) in \cite{calc2}!}
 \be
S_\mt{strip}=\frac{L^2}{2  G_\mt{N} }\,H\,
\int_{\delta}^{z_*}  dz\, {f(z)^{1/2}  \over
z^{2}}\left[ 1-{f(z_*)^2 \, z^4\over f(z)^2\,z_*^4 }\,  \right]^{-1/2}\,,
 \labell{EEbelt}
 \ee
where $z=\delta$
is the UV regulator surface in the bulk and $H$ is an IR scale introduced to regulate the length of the strip --- we note that the area of the entangling surface is then $2H$. Further $z_*$ is the maximum value of $z$ which the extremal surface reaches in the bulk, and it is determined in terms of the width of the strip by
\be
 \veps=2\int_0^{z_*}\!\!dz\,\, \frac{f(z_*)\,z^2}{f(z)^{3/2}\,z_*^2}\,
 \left[ 1-{f(z_*)^2 \, z^4\over f(z)^2\,z_*^4 }\,  \right]^{-1/2}
 \, . \labell{grog}
 \ee

Given the asymptotic behaviour of the metric function $f(z)$ in eq.~\reef{series} and that $z\le z_*$ in both of the previous integrals, we can evaluate eqs.~\reef{EEbelt} and \reef{grog} perturbatively in $\lambda$ if we have $\lambda\, z_*^{3-\Delta}\ll1$. It is straightforward to see from eq.~\reef{grog} that $\veps$ and $z_*$ are linearly related (to leading order) and therefore we are actually evaluating the entanglement entropy \reef{EEbelt} as a perturbative expansion in dimensionless combination $\lambda\,\veps^{3-\Delta}$.
Let us begin with eq.~\reef{grog}, which yields
\beqa
\veps&=&2z_*\int_0^1\!\! \frac{u^2\,du}{\sqrt{1-u^4}}\left[1+\frac{(\lambda z_*^{3-\Delta})^2}8\, \left(
\frac{u^{6-2\Delta}}{2} -\frac{1-u^{6-2\Delta}}{1-u^4}\right)+\cdots\right]
 \nonumber\\
 &=&\gamma\, z_*\left[1
+\frac{(\lambda z_*^{3-\Delta})^2}{16}\, \left( 
1-  \frac{3-\Delta}2\,\frac{\Gamma\!\left(\frac{1}4\right)}{\Gamma\!\left(\frac{3}4\right)}\frac{\Gamma\!\left(\frac{9-2\Delta}4\right)}{\Gamma\!\left(\frac{11-2\Delta}4\right)}
 \right)+\cdots\right]\,,
 \labell{gnome}
\eeqa
where
\beq
\gamma=2\sqrt{\pi}\,\frac{\Gamma\!\left(\frac34\right)}{\Gamma\!\left(\frac14\right)}\,.
\labell{troll}
\eeq
Alternatively, it will be convenient to write
\beq
 \frac{1}{z_*}=\frac{\gamma}\veps\left[1
+\frac{1}{16}\left(\frac{\lambda\, \veps^{3-\Delta}}{\gamma^{3-\Delta}}\right)^2
\left( 1- 2\,
 \frac{3-\Delta}{7-2\Delta}\,\frac{\Gamma\!\left(\frac{1}4\right)}{\Gamma\!\left(\frac{3}4\right)}\frac{\Gamma\!\left(\frac{9-2\Delta}4\right)}{\Gamma\!\left(\frac{7-2\Delta}4\right)}
 \right)+\cdots\right]\,.
\labell{gnome2}
\eeq
Similarly, we can write eq.~\reef{EEbelt} as
 \beqa
 S_\mt{strip}&=&\frac{L^2}{2  G_\mt{N} }\,\frac{H}{z_*}\int_{\delta/z_*}^1\! \frac{du}{u^2\sqrt{1-u^4}}\left[1
-\frac{(\lambda z_*^{3-\Delta})^2}{8}\,u^4\left(
\frac{u^{2-2\Delta}}{2} +\frac{1-u^{6-2\Delta}}{1-u^4}\right)+\cdots\right]\,.
 \labell{EEbelt2}
 \eeqa
We evaluate the above expression in the limit $\delta/z_*\to0$ but also with the assumption that we are considering a relevant operator with conformal dimension in the range $3>\Delta>5/2$. The final expression for the entanglement entropy then takes the form
 \beqa
S_\mt{strip}
&\simeq&c_1\, \frac{2H}\delta - c'_1\, \frac{2H\,\lambda^2}{\delta^{2\Delta-5}} - \ta\,\frac{H}\veps - \ta'\, \frac{H\,\lambda^2}{\veps^{2\Delta-5}}+\cdots\,,
 \labell{strip33}
 \eeqa
with
\beqa
c_1&=&\frac{L^2}{4\GN}\,,\qquad c'_1= \frac{L^2}{64\GN}\,\frac{1}{2\Delta-5}\,,
\qquad
\ta=\frac{L^2}{4\GN}\,\gamma^2\,,
\labell{strip44}\\
&&\quad
\ta'=-\frac{\sqrt{\pi}\,L^2}{32\,\GN}\,\gamma^{2\Delta-5}
 \frac{3-\Delta}{2\Delta-5}\,\frac{\Gamma\!\left(\frac{9-2\Delta}4\right)}{\Gamma\!\left(\frac{11-2\Delta}4\right)}
 \,.
\nonumber
\eeqa
Examining this result, we see that the two terms proportional to $c_1$ and $\ta$ reproduce the entanglement entropy \reef{eqn12} of the strip calculated in the AdS vacuum.
The term proportional to $c'_1$ is a new area law divergence which will generally appear in the entanglement entropy with any geometry for the entangling surface. Hence in calculating the mutual information, each of the separate entropies in eq.~\reef{mutualdef2} would contain such a term and so they would cancel amongst each other to yield a finite result --- see below. Finally, the above expression is calculated in the regime where the width of the strip was smaller than the scale set by the corresponding coupling, \ie $\lambda\,\veps^{3-\Delta}\ll 1$, and this limit yields the $\ta'$ term, which is also proportional to the area of the entangling surface. In our construction, this the would produced an extra contribution of the form $2\pi R\, \ta'\, \lambda^2/\veps^{2\Delta-5}$ in the mutual information \reef{mi1}. 

The appearance of these unusual area law terms in the mutual information does not effect the discussion in the main text in an essential way. For example, in section \ref{proof}, the definition of $I_0(A)$ in eq.~\reef{gamma} or eq.~\reef{449} would simply involve removing an additional extensive contribution, \ie $\ta'\,\lambda^2\,\int_\Gamma ds/\veps(s)^{2\Delta-5}$. In general, defining $I_0(A)$ would require removing all of contributions in $I(A^+,A^-)$ which are divergent in the limit $\veps\to 0$. However, the subsequent discussion of the properties of $I_0(A)$ would remain unchanged and hence the proof of the \F-theorem remains intact.

Again an essential assumption in the above calculation for $d=3$ was that the dimension of the relevant operator lay in the range $3>\Delta>5/2$. We note that in special case $\Delta=5/2$, the additional power law contributions above would be replaced by a logarithmic divergence, as discussed in \cite{calc2}. Further, the above analysis is easily extended to higher dimensions using the results of \cite{calc2} and one finds additional unconventional power law contributions for $d>\Delta>(d+2)/2$.

It is interesting to also apply the above analysis to evaluate the holographic entanglement entropy for a circle in $d=3$. Again, this is a straightforward extension of the calculations in \cite{calc2} and we only present the results here. In particular, in the regime where the conformal dimension satisfies $3>\Delta>5/2$ and the radius of the circle is small, \ie $\lambda\,R^{3-\Delta}\ll 1$, we find
 \beqa
S_\mt{circle}
&\simeq&c_1\, \frac{2\pi R}\delta - c'_1\, \frac{2\pi R\,\lambda^2}{\delta^{2\Delta-5}} - 2\pi\,c_0 -  2\pi\,b'\,R^{2(3-\Delta)}\,\lambda^2+\cdots\,,
 \labell{circle33}
 \eeqa
where $c_1$ and $c'_1$ are given by the same expressions as in eq.~\reef{strip44}, $c_0=L^2/(4\GN)$ as in eq.~\reef{holoc0} and 
\be
 b'=\frac{L^2}{16\GN}\,\frac{3-\Delta}{(7-2\Delta)(2\Delta-5)}\,.
\labell{circle44}
\ee
Hence the mutual information for these small circles becomes
\be
I(A^+,A^-)=\ta\,\frac{2\pi R}\veps + \ta'\, \frac{2\pi R\,\lambda^2}{\veps^{2\Delta-5}}- 4\pi\,c_0 -  4\pi\,b'\,R^{2(3-\Delta)}\,\lambda^2+\cdots\,,
\labell{mutu44}
\ee
where above we must also have $\veps\ll R$, and the finite part becomes
\be
I_0(A)=- 4\pi\,c_0 -  4\pi\,b'\,R^{2(3-\Delta)}\,\lambda^2+\cdots\,.
\labell{mutu445}
\ee
Note that the latter now includes `nonlocal' contribution that depends on the radius $R$ in a sublinear way.\footnote{We note that this nonlocal contribution persists for any relevant operator. However, in the regime $\Delta<5/2$, the behaviour becomes faster than linear. As usual, a logarithm appears at precisely $\Delta=5/2$, \ie the above power is replaced by $R\log(R\lambda^2)$.} As a result, this contribution is not eliminated in the c-function \reef{funix}. Rather we find
\be
C(R)=c_0 + (2\Delta-5)\,b'\,R^{2(3-\Delta)}\,\lambda^2+\cdots\,,
\labell{funix55}
\ee
however, we recover $C(R\to0)=c_0$, as expected.

%%%%%%%%%%%%%%%%%%%%%%%%%%%%%%%%%%%%%%
%%%%%%%%%%%%%%%%%%%%%%%%%%%%%%%%%%%%%%
%%%%%%%%%%%%%%%%%%%%%%%%%%%%%%%%%%%%%%

\section{The extensive mutual information model}
\labell{extensive}

In general, it is difficult to find explicit examples where the mutual
information can be computed exactly. The holographic model in the previous appendix provides one such example. In this appendix, we consider a model \cite{split} for the mutual information in relativistic QFT which follows by imposing an additional property of extensivity 
\begin{equation}
I(A,B)+I(A,C)=I(A,B\cup C)\,.\labell{ext}
\end{equation}
With this constraint, the general form of the mutual information satisfying the
combined requirements of strong subaddivity and Lorentz covariance is known. While the resulting expression
can be shown to provide the mutual information of a free fermion in two dimensions, it is not known at
present if this form corresponds to the actual mutual information of any other specific QFT's. We use this model here to investigate the new \c-function proposed in section \ref{mutual} since it has the advantage of having a simple analytical expression. We also note that by construction, this expression satisfies
all of the constraints imposed by strong subadditivity, which are rather nontrivial in the Minkowski space geometry and largely determines the essential
features of the entanglement entropy. In particular, the extensive mutual information model already involves a \c-theorem for a specific
\c-function, a fact which will help us to test our conjectures by relating this property with the putative \c-theorem for $\tilde{c}_0$.

Explicitly, for the extensive mutual information in $d=3$, we have \cite{split}
\begin{equation}
I(A^+,A^-)=-\int_{\partial A^-}ds_x\, \int_{\partial A^+}ds_y\,H(|x-y|) \ \ut_x\cdot\ut_y 
\,,\labell{yy}
\end{equation}
with $\ut_x$ the unit vector tangent to the boundary at the point $x$ and $H(\ell)$ a function of the distance
between the points --- see figure \ref{concentric}. The direction of tangent vectors $\ut$ is chosen such
that $\varepsilon_{abc}\, \uu^a\, \un^b\, \ut^c>0$ where $\uu$ is a future-directed time-like vector and $\un$ is the unit normal to the boundary
pointing into the region on the Cauchy surface on which $A^+$ and $A^-$ are defined.\footnote{Implicitly we are assuming that $\veps_{txy}=+1$ here. However, we also note the result for $I(A^+,A^-)$ would not change if another convention was chosen to determine the direction of $\ut$.} 
Further we have
\begin{equation}
H(\ell)=-\int^\ell dy\, \frac{\kappa(y)}{y^3}\,. \labell{ff}
\end{equation}
An integration constant is left undetermined here but it is irrelevant since adding a constant to $H(\ell)$ does not change the result for the mutual information in eq.~(\ref{yy}).
The function $\kappa(\ell)$ is a \c-function in the sense that it must be positive, dimensionless, decreasing with increasing $\ell$, and constant
at the conformal fixed points.
\begin{figure}[t]
\centering
\includegraphics[width=.6\textwidth]{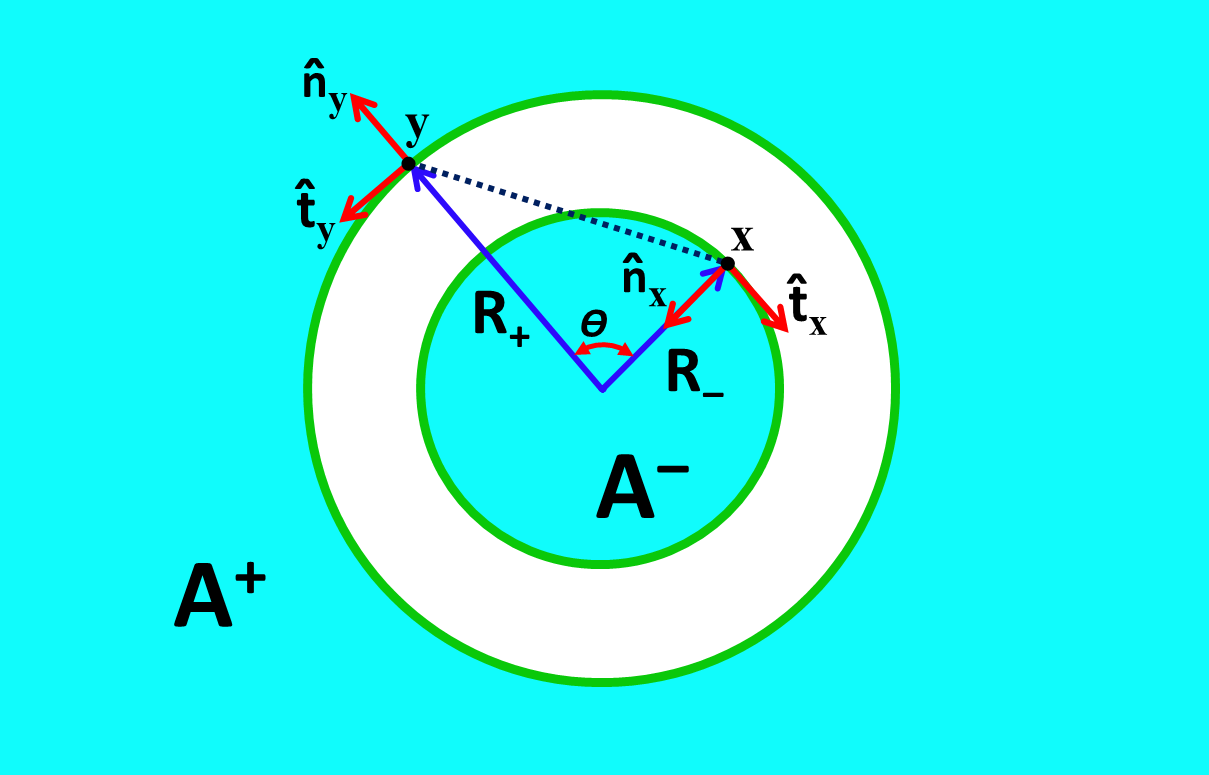}
\bigskip
\caption{(Colour online) Mutual information in the extensive model is given
directly by a double integral of a certain function over the points $x$ and $y$
on the boundaries of the regions $A+$ and $A^-$.} \labell{concentric}
\end{figure}

Let us first calculate the mutual information for the annular geometry in figure \ref{concentric} at a conformal point with $\kappa(\ell)=\kappa$ is some fixed constant. Eq.~\reef{ff} then yields
$H(\ell)=\kappa/(2 \ell^2)$ and we have
\begin{equation}
I(A^+,A^-)=\pi \kappa\,R_+\, R_- \int_0^{2 \pi} d\theta\, \frac{\cos\theta}{R_+^2+R_-^2-2 R_+ R_-\, \cos\theta}=\pi^2 \kappa \left(\frac{R_+^2+R_-^2}{R_+^2-R_-^2}-1\right)\,.
\end{equation}
Substituting for $R_+$ and $R_-$ with eq.~\reef{eps0}, \ie $R_-=R-(1/2-\alpha)\veps$ and $R_+=R+(1/2+\alpha)\veps$, and then expanding
for small $\veps/R$, we find
\begin{equation}
I(A^+,A^-)\equiv I(R,\veps,\alpha)= \pi^2 \kappa\, \frac{R}{\veps}+\pi^2 \kappa\, (\alpha-1)+{O}(\veps/R)\,. \labell{Mcft}
\end{equation}
Hence the mutual information has the expected form given in eq.~\reef{mi1} with $\ta=\pi \kappa/2$, $\tb=0$ and $\tc_0=\pi \kappa\, (1-\alpha)/4$.
We also see that if we change $\alpha$, then the constant $\tc_0$ is shifted as in eqs.~(\ref{expec}) and (\ref{changes}).

Our next task is to calculate the mutual information in the case $\kappa(\ell)$ describes a RG flow running between a UV fixed point with $\kappa(\ell\to0)=\kappa_{\ssc UV}$ and an infrared point where $\kappa(\ell\to\infty)=\kappa_{\ssc IR}$. For simplicity, we will assume the derivative of
this \c-function vanishes at the origin, \ie $\kappa^\prime(0)=0$.\footnote{Note that if the small$\,-\,\ell$ expansion of the \c-function %in the vicinity of $\ell=0$ 
takes the form $\kappa=\kappa_{\ssc UV}+\lambda\,\ell^a+\cdots$ with $0<a<1$, then the mutual information will contain subleading `area law' contributions of the form $R\,\lambda/\veps^{1-a}$. (A contribution proportional to $\log(\veps)$ appears with $a=1$.) These unusual `area law' contributions indicate that the RG flow is generated by perturbing the UV conformal point with a relevant, but close to marginal, operator, as discussed in appendix \ref{kuprate}. \labell{foot99}}
To simplify the calculations, we will fix $\alpha$ and specifically we choose $\alpha=0$ in accord with the prescription discussed in section \ref{critb}. The mutual information then becomes
\begin{equation}
I(R,\veps)=-2 \pi (R^2-\veps^2/4) \int_0^{2 \pi} d\theta\, \cos\theta\, \int^{q(\theta)}\! dy\, \frac{\kappa(y)}{y^3}\,, \labell{eli}
\end{equation}
with
\begin{equation}
q(\theta)=\sqrt{2 R^2+\veps^2/2-(2R^2-\veps^2/2)\,\cos\theta}\,.\labell{eli2}
\end{equation}
%We can eliminate the $\veps^2$ term in the prefactor $(R^2-\veps^2/4)$ in eq.~(\ref{eli}) since it will only give rise to contributions of $O(\veps/R)$ in $I(A,B)$
%and we are not interested in terms of that order here.
In order to analyze the running of the constant term, we calculate the \c-function in eq.~\reef{guff} here but we do not yet the limit $\veps\to0$, \ie
\begin{equation}
C_\veps(R)=\frac1{4\pi}\left(R\,\frac{\partial I(R,\veps)}{\partial\,R}-I(R,\veps)\right)\,.\labell{guffe}
\end{equation}
Now using eqs.~\reef{eli} and \reef{eli2} and integrating by parts once, the final expression can be written as
\beq
%C_\veps(R)&=&-\frac12\left(R^2+\veps^2/4\right) \int_0^{2 \pi} d\theta\, \cos\theta\, \int^{q(\theta)}\! dy\, \frac{\kappa(y)}{y^3}
%\nonumber\\
%&&\qquad\qquad- R^2\left(R^2-\veps^2/4\right) \int_0^{2 \pi} d\theta\, \cos\theta\,(1-\cos\theta)\, \frac{\kappa(q(\theta))}{q(\theta)^4} 
%\labell{gruff32}\\
C_\veps(R)=R^4 \int_0^{2 \pi} d\theta\,\frac{\kappa(q(\theta))}{q(\theta)^4}\left[ \frac12\left(1-\frac{\veps^4}{16R^4}\right)\sin^2\theta
-\left(1-\frac{\veps^2}{4R^2}\right)\cos\theta\,(1-\cos\theta)\right]  \,.
\labell{gruff32}
%\nonumber
\eeq
One can readily confirm that the integrand above is everywhere nonsingular in the limit $\veps\to0$ and hence in the model with extensive mutual information, our \c-function \reef{guff} becomes
\beq
C(R)=\lim_{\veps\to0}C_\veps(R)=\frac{1}{8}\int_0^{2 \pi} d\theta\    \kappa(2R \sin(\theta/2) )
=\frac{1}{2}\int_0^{2 R} \frac{dy\ \kappa(y )}{\sqrt{4R^2-y^2}}\,.
\labell{guffex}
\eeq
Either of the above expressions shows that $C(R)$ averages the \c-function appearing in the extensive model over all scales from zero out to the diameter of the annulus geometry used to define our new \c-function. In fact, if we observe that $2R \sin(\theta/2)$ is the chord length between two points separated by an angle $\theta$ on the circle of radius $R$, then the averaging in the angular integral is over the distance between all pairs of points on the circle.  But we also observe that the measure in the $y$ integration in the second expression gives the greatest weight to scales near $y\simeq 2R$, \ie near the  infrared scale defined by the circle. In the limits of large and small circles, we find that the \c-function interpolates between $C(R\to0)=\pi \kappa_{\ssc UV}/4$ and $C(R\to\infty)=\pi \kappa_{\ssc IR}/4$. Further, it is evident that $C(R)$ decreases monotonically with
increasing $R$ as a consequence of the decreasing character of $\kappa(y)$. Incidentally,
this simple example shows \c-functions are by no means unique, \ie given a \c-function, we can construct infinitely many other \c-functions which are also
monotonically decreasing and interpolate between the same fixed point values.

It is also interesting to examine the running of the area term in the mutual information, \ie the term proportional to $2\pi R$ in eq.~\reef{mi1}. First we can integrate by parts to re-express the mutual information \reef{eli} as
\begin{equation}
I(R,\veps)=2 \pi R^4\left(1-\frac{\veps^2}{4R^2}\right)^2 \int_0^{2\pi} d\theta\,\frac{\kappa(q(\theta))}{q(\theta)^4}\,\sin^2\theta\,. \labell{eli3}
\end{equation}
Now the limit $\veps\to0$ produces a singularity in the integrand at $\theta=0$ (and $2\pi$) since eq.~\reef{eli2} yields $q(\theta\simeq0)\simeq\sqrt{R^2\theta^2+\veps^2}$. This potential singularity is the source of the contribution $2\pi R \ta /\veps$ and we can approximate the result $\theta\simeq0$ portion of the integral by replacing $\kappa(q(\theta))\simeq\kappa(0)=\kappa_{\ssc UV}$. The result is then the same as found above in eq.~\reef{Mcft} for a conformal fixed point with $\ta=\pi\kappa_{\ssc UV}/2$. We note that as this result applies for all values of the radius $R$ and as expected, the coefficient depends only on information from the UV portion of the RG flow.

Now for small $R$, \ie smaller than any scales appearing in the flow defined by $\kappa(\ell)$, $\kappa(0)=\kappa_{\ssc UV}$ is a good approximation for the entire integration in eq.~\reef{eli3} and so the entire result will coincide with that in eq.~\reef{Mcft} (with $\alpha=0$). In particular, then $\tb=0$ for small $R$ but this could again be anticipated because we are only probing the UV fixed point and there are no dimensionful parameters from which $\tb$ could be constructed. In contrast, while large $R$ definitely probes the IR fixed point, eq.~\reef{eli3} actually samples the  interpolating function $\kappa(\ell)$ over all scales from $\ell=0$ to $2R$. Hence we can expect a nonvanishing $\tb$ will appear in the large radius limit. In fact, considering eqs.~\reef{mi1} and \reef{eli}, we find
\begin{equation}
\tilde{b}=\frac{1}{2\pi}\lim_{\veps\to0}\left[\frac{\partial I(R,\veps)}{\partial\,R}\bigg|_{R\to\infty}-\frac{\partial I(R,\veps)}{\partial\,R}\bigg|_{R\to0}\right]
=-2\int_0^{\infty} \frac{dy}{y^2}\, \left[\kappa_{\ssc UV}-\kappa(y)\right]\,.
\labell{runn2}
\end{equation}
Notice that our assumption that $\kappa'(0)=0$ ensures that the $y=0$ end of the integration yields a finite result. Further, because $\kappa(\ell)$ decreases with growing $\ell$, we note that the integrand is everywhere positive and hence $\tb<0$. That is, the area term in eq.~\reef{mi1} picks up a finite
negative contribution in the infrared. This result is in accordance with the results of \cite{proof,newcas} on the running of the area term for the entanglement entropy of a circle.\footnote{We can extend the comparison with \cite{proof,newcas} by considering
\beq
\frac{1}{2\pi}\lim_{\veps\to0}\left[\frac{\partial I(R,\veps)}{\partial\,R}\bigg|_{R}-\frac{\partial I(R,\veps)}{\partial\,R}\bigg|_{R\to0}\right]
=-2\int_0^{2R} \frac{dy}{y^2}\, \frac{\kappa_{\ssc UV}-\kappa(y)}{\sqrt{1-(y/2R)^2}}\,,
\labell{runn3}
\eeq
which decreases with increasing $R$, \ie from the UV to the IR.}

We conclude here by summarizing our results for the extensive mutual information model for small and large circles and a generic value of $\alpha$:
\begin{eqnarray}
I(R,\veps,\alpha)&=& \frac{\pi^2\kappa_{\ssc UV}}{\veps}\,R-\pi^2 \kappa_{\ssc UV} +\alpha \pi^2 \kappa_{\ssc UV} +O(\veps)\hspace{2.25cm} \textrm{for}\, R\rightarrow 0\,,\labell{bigandsmall}\\
I(R,\veps,\alpha)&=& \left(\frac{\pi^2\kappa_{\ssc UV}}{\veps}+2\pi\, \tilde{b}\right) R-\pi^2 \kappa_{\ssc IR} +\alpha \pi^2 \kappa_{\ssc UV}+O(\veps) \hspace{.5cm} \textrm{for}\, R\rightarrow \infty\,.\nonumber
\end{eqnarray}
where $\tb$ is given by eq.~\reef{runn2}.
We see the constant term, \ie the term independent of $R$, at the infrared fixed point is intrinsic to the infrared theory alone only with the choice $\alpha=0$, as argued in section \ref{critb}. With our analysis here, it may seem quite remarkable that any choice of $\alpha$ yields a $\tc_0$ which is only proportional to the IR quantity $\kappa_{\ssc IR}$, and is independent of the UV value $\kappa_{\ssc UV}$ or any of the details of RG flow incorporated in the interpolating function $\kappa(\ell)$.

\subsection{Contribution of a strip with a null cusp}\labell{broken}

For the discussion in section \ref{proof}, it is interesting to evaluate the contributions coming from null cusps within extensive mutual information model. In particular, we would like to evaluate the quantities $f(q_1,q_2)$ and $J(q_1,q_2)$ defined in eqs.~\reef{449} and (\ref{tido}). Since these
are purely local contributions, we can simplify the calculation by evaluating these quantities for null cusp formed by the intersection of two straight lines lying in a null plane. Using standard Minkowski coordinates, we can choose without generality the null plane such that its tangent space is spanned by the unit vector $\hat{\bf x}$ and the null vector $\hat{\bf v}=\hat{\bf t}+\hat{\bf y}$. Now a framing must be constructed for each of the boundary lines as in eq.~\reef{boundaries} but again for simplicity, we choose $\veps(s)$ to be a (different) fixed constant for each of the lines. The geometry under consideration will then consist of two straight strips with edges lying in two parallel null planes. However, there is still a certain freedom in choosing the precise framing. We fix the latter\footnote{One can verify that given any framing as described above, $\um$ can always be brought to the desired form by applying an $SO(1,2)$ transformation $\Lambda$, which satisfies $\hat{\bf v}^T\cdot \Lambda\cdot \hat{\bf v}=0$.} by choosing the unit vector directed between the two intersection points of the strip edges to be $\um=\hat{\bf y}=(0,0,1)$. At the same time, we may write the tangent vectors to the two lines as
\begin{equation}
 \ut_1=\hat{\bf x}+\sg\,\hat{\bf v}=(\sg,1,\sg)\,,\qquad{\rm and}\qquad
 \ut_2=\hat{\bf x}+\roo\,\hat{\bf v}=(\roo,1,\roo)\,.
 \labell{vecs}
\end{equation}
The tangent vectors are parametrized above according to the definitions in eq.~\reef{qqq},  \ie
\beq
q_1=\um\cdot\ut_1\qquad{\rm and}\qquad q_2=\um\cdot\ut_2\,.
\labell{pqs}
\eeq
Recall that the limit $q_2\to q_1$ corresponds the case where the tangent vectors become equal and hence the cusp vanishes. 

Our following discussion is adapted to null cusp illustrated in figure \ref{cruce}, with $\ut_1$ being the tangent to $\Gamma_{B}$ while $\ut_2$ is tangent to $\Gamma_{A}$. Further note that in the configuration shown there, we would have $q_1>0$ and $q_2<0$.  Now
let us take the first pair of edges, \ie the boundaries of $A^-$ and $B^-$, to lie in the null plane passing through the origin with\footnote{As before, we use standard Cartesian coordinates $(t,x,y)$ in describing the geometry here.}
\begin{equation}
\Gamma_{A^-}: \,\,\,(\roo\, x, x, \roo\, x)\qquad{\rm and}\qquad\Gamma_{B^-}: \,\,\,
(\sg\, x,x,\sg\, x)\,, \labell{plus}
\end{equation}
where the curves were chosen to intersect at $(0,0,0)$. The second pair of edges, \ie the boundaries of $A^+$ and $B^+$, are then placed in the second null plane with 
\begin{equation}
\Gamma_{A^+}:\,\,\,(\roo\, x, x, \roo\, x+\tilde\veps)\qquad{\rm and}\qquad\Gamma_{B^-}:\,\,\,
(\sg\, x,x,\sg\, x+\tilde\veps)\,, \labell{minus}
\end{equation}
which then intersect at $(0,0,\tilde\veps)$.
We note that with the above parameterization, it may appear that, \eg $\Gamma_{A^+}$ and $\Gamma_{A^-}$ are separated by the distance $\tilde\veps$. However, this is the displacement along the vector $\um$ rather than the normal vector between the two curves, as specified in eq.~\reef{boundaries}. It is straightforward to show that the appropriate unit normal vectors for the two strips are
\beq
\un_{\ssc A}=(-\roo^2,-\roo,1-\roo^2)/\sqrt{1-\roo^2}\quad{\rm and}\quad \un_{\ssc B}=(-\sg^2,-\sg,1-\sg^2)/\sqrt{1-\sg^2}\,.
\labell{normal}
\eeq
Then let us specify the widths across the $A$ and $B$ strips measured as the proper distance along these normals as $\veps_A$ and $\veps_B$. We find that these widths are related to $\tilde\veps$ by
\beq
\tilde\veps=\frac{\veps_A}{\sqrt{1-\roo^2}}=\frac{\veps_B}{\sqrt{1-\sg^2}}\,.
\labell{gonzo}
\eeq
In the main text, we used the notation $\veps_A=\veps_2$ and $\veps_B=\veps_1$, and hence eq.~\reef{gonzo2} follows directly from these relations. 

We can now determine $f(q_1,q_2)$ by evaluating the mutual information $I(A^+\cup B^+,A^-\cap B^-)$, however, we must introduce an infrared cut-off to produce finite integrals. On $\Gamma_{A^-}$ and $\Gamma_{B^-}$, we limit the range of $x$ to $-L/2<x<L/2$. The range of the $x$ integration along $\Gamma_{A^+}$ is chosen so that the ends of the corresponding strip $A^s$ lies along the normal $\un_{\ssc A}$, given in eq.~\reef{normal}. Hence we choose
$-L/2-\roo\tilde{\veps}<x<L/2-\roo\tilde{\veps}$ for $\Gamma_{A^+}$. Similarly, we choose $-L/2-\sg \tilde{\veps}<x<L/2-\sg\tilde{\veps}$ on $\Gamma_{B^+}$ so that the ends of $B^s$ lie parallel to $\un_{\ssc B}$.

Now the corresponding mutual information is given by 
\begin{eqnarray}
&&I(\roo,\sg,\veps,L)=\int_{0}^{L/2-\roo\tilde{\veps}} dx_1\int_0^{L/2}dx_2\,  \frac{\kappa/2}{(x_1-x_2)^2+(\roo x_1+\tilde{\veps}-\roo x_2)^2-(\roo x_1-\roo x_2)^2}\nonumber \\
&&\qquad + \int_{0}^{L/2-\roo\tilde{\veps}} dx_1\int_{-L/2}^0 dx_2\, \frac{\kappa/2}{(x_1-x_2)^2+(\roo x_1+\tilde{\veps}-\sg x_2)^2-(\roo x_1-\sg x_2)^2} 
\nonumber \\
&&\qquad + \int_{-L/2-\sg\tilde{\veps}}^{0} dx_1\int_{-L/2}^0 dx_2\,  \frac{\kappa/2}{(x_1-x_2)^2+(\sg x_1+\tilde{\veps}-\sg x_2)^2-(\sg x_1-\sg x_2)^2}
\labell{longm}\\
&&\qquad + \int_{-L/2-\sg\tilde{\veps}}^{0} dx_1\int_0^{L/2} dx_2\,  \frac{\kappa/2}{(x_1-x_2)^2+(\sg x_1+\tilde{\veps}-\roo x_2)^2-(\sg x_1-\roo x_2)^2}\,.
\nonumber
\end{eqnarray}
As in eq.~\reef{449}, we must subtract $\ta\,\int ds/\veps(s)$, the area term along the curve running along the midpoint line of the cusp. Recall that in eq.~\reef{Mcft}, we found $\ta=\pi\kappa/2$ for this model and hence we have
\beqa
\ta\,\int \frac{ds}{\veps(s)}&=&\frac{\pi\kappa}2\int_{-(L+\sg\tilde{\veps})/2}^0\frac{ds}{\veps_B}+\frac{\pi\kappa}2\int_0^{(L-\roo\tilde{\veps})/2}\frac{ds}{\veps_A}
\nonumber\\
&=&\frac{\pi\kappa}4\left(\frac{L}{\veps_A}+\frac{L}{\veps_B}+\frac{\sg}{\sqrt{1-\sg^2}}-\frac{\roo}{\sqrt{1-\roo^2}} \right)\,,
\labell{midline}
\eeqa
where we have simplified the final expression using eq.~\reef{gonzo}. Here we have ended the curves abruptly at the infrared cut-off and the corresponding end-points will introduce another frame dependent contribution. The latter can be independently evaluated for a single interval with no cusp and is given by 
%$-\frac{\kappa}{2}\, [\log(L/\veps_A)+\log(L/\veps_B)+2] $.  
$-\kappa\, [\log(L/\sqrt{\veps_A\veps_B})+1] $. 
Subtracting this contribution and the area term (\ref{midline}), we find
\bea
{f}(q_1,q_2)&=&\lim_{\tilde\veps\rightarrow 0}\Bigg(I(\roo,\sg,\veps,L)+\kappa\, \big[\log(L/\sqrt{\veps_A\veps_B})+1\big]
\labell{quo2}\\
&&\qquad\qquad\qquad\qquad\qquad
-\frac{\pi\kappa}4\left[\frac{L}{\veps_A}+\frac{L}{\veps_B}+\frac{\sg}{\sqrt{1-\sg^2}}-\frac{\roo}{\sqrt{1-\roo^2}} \right]\Bigg)
\nonumber\\
&=& \frac{\kappa}{2} \left(2+\frac{(q_1(q_1+q_2)-2)\arcsin(q_1)}{(q_1-q_2) \sqrt{1-q_1^2}}+\frac{(q_2(q_1+q_2)-2)\arcsin(q_2)}{(q_2-q_1) \sqrt{1-q_2^2}}\right)\,,
\nonumber
\eea
as was given in eq.~(\ref{quo}). 

Note that the infrared regulator $L$ is arbitrary and does not enter in the final result. As a result, the width $\tilde\veps$ also drops out of the final expression which must be dimensionless. Hence the result depends only on $q_1$ and $q_2$ and in fact, it is symmetric in exchanging $q_1\leftrightarrow q_2$. Further, it is straightforward to verify that the expression above for $f(q_1,q_2)$ vanishes in the limit $q_2\to q_1$.
 
Implicitly, we have also presented the calculation for a constant $\kappa$, however, one can readily verify that in the case of a nontrivial RG flow with $\kappa(\ell)$, the quantity that enters the final expression \reef{quo2} is $\kappa_{\ssc UV}=\kappa(0)$. One may also repeat the calculation choosing the bent strip in the positive $y$ region, \ie evaluating $I(A^+\cap B^+,A^-\cup B^-)$, and  exactly the same result is found for $f(q_1,q_2)$.

Finally we would like to determine $J(q_1,q_2)$ as defined in eq.~\reef{tido}, however, we need the entanglement entropy to evaluate this expression. The natural definition of $S(A)$ in this model is given modifying the double integral (\ref{yy}) so that $x$ and $y$ run along the boundary of the region $A$, \ie along the same boundary,\footnote{Essentially the same expression for the entanglement entropy emerges from the heuristic model in \cite{swingle}.}
\begin{equation}
S(A)=-\int_{\partial A}ds_x\, \int_{\partial A}ds_y\,H(|x-y|) \ \ut_x \cdot\ut_y\,.\labell{yy1}
\end{equation}
Substituting this expression into the definition of the mutual information \reef{mutualdef} then reproduces the expected expression in eq.~\reef{yy}. Of course, the above expression for $S(A)$ is divergent but a short-distance cut-off can
introduced by only integrating over regions where $|x-y|>\delta$. However, it is not difficult to see that the integrals in eq.~\reef{tido} involving $x$ and $y$ on the same null plane, which are the only ones which depend on $\delta$, cancel in the combination $J(q_1,q_2)$. Of course, this was to be expected since in the main text, we argued that $J(q_1,q_2)$ was a finite regulator-independent quantity. Hence it is straightforward to see that evaluating the expression (\ref{tido}) for $J(q_1,q_2)$ involves the same integrals as above and we find
\begin{equation}
J(q_1,q_2)=-2 f(q_1,q_2)\,,
\end{equation}
as argued in section \ref{proof}.

\section{Circle entropy for a scalar field}
\labell{scalar} 

In this appendix, we review two computations of the constant term in the entanglement entropy \reef{enti} of
a circle for a free massless scalar.  First, we use the conformal mapping from the
causal domain of the disk to $R\times H^2$, where $H^2$ is the two-dimensional
hyperbolic plane. The second calculation involves a similar mapping to (the static patch of) three-dimensional de Sitter space.
Our purpose here is to highlight the ambiguity in the definition of this constant
term, even when using ``covariant'' cut-offs. In the Section \ref{entropy}, we
show how this ambiguity can be eliminated with a prescription emerging from
what we learned from the mutual information.

\subsection{Mapping to $R\times H^2$}

We calculate the entanglement entropy of a circle for a conformally coupled massless scalar field in three space-time dimensions as
the thermodynamic entropy of thermal gas of scalar particles in hyperbolic space \cite{CHM,sphere1}.

The causal domain of dependence of a disk $r<R$ (and $t=0$) in flat space
\begin{equation}
ds^2=-dt^2+dr^2+r^2 d\theta^2 \labell{mink3}
\end{equation}
is simply given by $r<R-|t|$. This region is conformally mapped to $R \times H^{2}$ by the change of coordinates
\begin{eqnarray}
 t&=&R\frac{\sinh(\tau/R)}{\cosh u + \cosh(\tau /R)}\,,\\
r&=&R \frac{\sinh u}{\cosh u + \cosh(\tau /R)}\,. \labell{coo}
\end{eqnarray}
After eliminating a Weyl rescaling, the metric in these new coordinates is
\begin{equation}
ds^2=-d\tau^2+R^2 (du^2 + \sinh u^2 d\theta^2) \,, \labell{neuf}
\end{equation}
where we recognize the spatial metric describes a hyperbolic space with scalar curvature
\begin{equation}
{\cal R}=-\frac{2}{R^2}\,.
\end{equation}

As explained in \cite{CHM}, the density matrix transforms unitarily under conformal transformations,
and in the new hyperbolic space, corresponds  to the thermal equilibrium state with temperature $T=1/(2\pi R)$. Hence, the entanglement entropy of the circle becomes the
 thermodynamic entropy $S(T)$ in $R\times H^{2}$ and can be written as
\begin{equation}
S(T)= (1+T\, \partial_{T})\log Z(T)\,.
\end{equation}
In the present calculation,  $\log Z(T)$ is the effective action of a massless conformally coupled scalar field $\phi$,
\begin{equation}
\left(\partial_\tau^2 -\Delta_{H^{2}}+\frac{1}{8}{\cal R}\right)\phi=0\,.
\end{equation}
We refer the interested reader to \cite{CHM,sphere1} for further details.

We are interested in the constant term of the expansion of the entropy. The time coordinate is compactified to a circle of size $1/T$ after passing to Euclidean coordinates.
To calculate $\log Z$, we use the heat kernel method which gives
\begin{equation}
 \log Z=\frac{1}{2} \int_0^{\infty}\frac{dt}{t}\,K(t)\,,
\end{equation}
 where $K(t)$ factorizes as
\begin{equation}
 K(t)=K_{H^{2}}(t)\, K_{S^1}(t)\, e^{-t {\cal R}/8}\,.
\end{equation}
The explicit expressions for the heat kernels of the Laplace operator in hyperbolic space $K_{H^{2}}(t)$ and
the circle $K_{S^1}(t)$ are well known \cite{heatkernel},
\begin{eqnarray}
 K_{S^1}(t)&=&\frac{2 }{\sqrt{4 \pi t} \,T}\sum_{n=1}^{\infty}e^{-\frac{n^2 }{4T^2 t}}\,,\\
K_{H^{2}}(t)&=& f(t)  \int d^2x\, \sqrt{g}\,.
\end{eqnarray}
In the second expression, the integral is over the volume of the hyperbolic space and
\begin{equation}
 f(t)=\lim_{\rho\rightarrow 0}\, \frac{  R^3
 e^{-\frac{ t}{4 R^2}}}{2^{5/2}\pi^{3/2}t^{3/2}}\ \int_{\rho/R}^{\infty}ds\,\frac{s e^{-s^2 R^2/4t}}{(\cosh s-\cosh \rho/R)^{\frac{1}{2}}}\,.
\end{equation}

Then the calculation of $\log Z$ involves solving the integrals in $t$ and $s$, as well as an infinite sum in $n$. After some algebra, we obtain
\begin{equation}
S= \text{vol}(H^2)~ \frac{\sqrt{2}}{8\pi R^2} \int_{0}^{\infty}\!\! ds\,
\frac{2 \coth(s/2) + s~ \text{csch}(s/2)^2- s^2\coth(s/2) ~ \text{csch}(s/2)^2}{s^2 (\cosh(s)-1)^{1/2}}\,. 
\end{equation}
The result of this integral gives
\begin{equation}
 S= \frac{\text{vol}(H^2)}{16 \pi R^2}  \left[\log2-\frac{3\zeta(3)}{2\pi^2}\right]\,.\labell{35}
\end{equation}
Here, we can recognize the factor in the brackets as also appearing in the final result for $\coss$ in eq.~\reef{plat}.

Hence the constant term in the entropy comes from the product of a numerical constant times the volume of the hyperbolic space in eq.~\reef{neuf}. Of course, the latter is divergent and so we must introduce a cut-off $u_{\textrm{max}}$ in the radial integral
\begin{equation}
\textrm{vol}(H^{2})=  2 \pi R^{2} \int_0^{u_{\textrm{max}}}  du\, \sinh(u)\,.
\end{equation}
The desired universal contribution to the entropy is then determined by taking the constant term in an expansion of the regulated volume with respect to the short-distance cut-off in the entanglement entropy calculation. With this
aim, let us define the short-distance cut-off $\delta$ in the original radial
coordinates in flat space by summing the entropy contributions up to $r_{max}=R-\delta$.
Then, taking into account the mapping of coordinates (\ref{coo}), we find
\begin{equation}
 u_{max}=-\log\left(\frac{\delta/R}{2-\delta/R}\right)\,.\labell{umax}
\end{equation}
Thus expanding in $\delta/R$, we have
\begin{equation}
 \text{vol}(H^2) =  2 \pi R^2~ \left(\frac{R}{\delta}-\frac{3}{2}+O(\delta/R)\right)\,.\labell{voli}
\end{equation}
Substituting this result back in eq.~(\ref{35}), the constant term in the entropy \reef{enti} is
\begin{equation}
2\pi c_0=\frac{3}{2}\,2\pi \coss\,,
\end{equation}
where the coefficient $\coss$ is given in eq.~\reef{plat}. Of course, our result here differs from the expected constant term  by a factor $3/2$.

Evidently, the result above depends on the details of our cut-off prescription.
For example, let us consider slightly modifying the short-distance cut-off and integrating up
to $r_{\textrm{max}}=R-\hat\delta$ with
\begin{equation}
\hat\delta=\delta\,\left(1 + a_1\, \frac{\delta}{R} + a_2\,\frac{\delta^2}{R^2}+\cdots\right).\labell{expa}\,.
\end{equation}
This small change in the prescription makes a small change in the total regulated volume but it has a finite effect on the constant term in the $\delta$ expansion, with eq.~\reef{voli} becoming
\begin{equation}
\text{vol}(H^2) =  2 \pi R^2~ \left(\frac{R}{\delta}-\left(\frac{3}{2}+a_1\right)+{O}(\delta/R)\right)\,.\labell{voli2}
\end{equation}
Hence we may reproduce the desired coefficient \reef{plat} in this calculation by choosing $a_1=-1/2$. However, this seems a rather mysterious and arbitrary choice as presented here.

As an example, the correct coefficient can be obtained if one defines the radial cut-off with $\delta=2 R \exp[-u_{max}]$ rather than eq.~(\ref{umax}). This prescription, used in \cite{Klebfree}, is motivated by the corresponding holographic calculations \cite{myers1,CHM}. This choice yields to an expansion as in eq.~(\ref{expa}) for the effective cut-off $\hat\delta$ in the flat space integration in terms of $\delta/R$
\begin{equation}
\hat\delta= \delta\left(1-\frac{1}{2}\,\frac{\delta}{R}+\frac{1}{4}\,\frac{\delta^2}{R^2}+\cdots\right)\,.\labell{parti}
\end{equation}
As explained in section \ref{entropy}, the success of this prescription has its origin in respecting a certain inversion symmetry.  Of course, the correct coefficient will also be produced by any other prescription where the corresponding
expansion (\ref{expa}) yields $a_1=-1/2$. However, at this point, in this specific calculation, and without using the results
from zeta function regularization of the free energy of the sphere,
we do not have in principle any reason to prefer one expansion from the other. This is a manifestation of the ambiguities in the evaluating constant term in the entanglement entropy \reef{enti}.

\subsection{Mapping to de Sitter space}

This section is based on Dowker's calculation of the entanglement entropy for spheres in even dimensions \cite{Dowker:2010}, using a conformal mapping to de Sitter space.
Here we apply the same approach to calculate the universal contribution in $d=3$.

In this case, we apply the following coordinate transformation,
\begin{eqnarray}
t&=&R \frac{\sqrt{1-\hat{r}^2/R^2} ~\sinh(\tau /R)}{1+ \sqrt{1-\hat{r}^2/R^2} ~ \cosh(\tau/R)}\,, \\
r&=& \frac{\hat{r}}{1+\sqrt{1-\hat{r}^2/R^2} ~ \cosh(\tau/R)}\,,
\labell{desitter}
\end{eqnarray}
to the flat space metric \reef{mink3}. After eliminating of the Weyl scaling, the causal domain of the disk $r<R$ has been mapped to the static patch of de Sitter space with metric
\begin{equation}
ds^2=-\left(1-\frac{\hat{r}^2}{R^2}\right)d\tau^2+\frac{d\hat{r}^2}{1-\frac{\hat{r}^2}{R^2}}+\hat{r}^2 d\theta^2\,.
\labell{harmony}
\end{equation}

As happens with the mapping to the hyperbolic geometry, in this new conformal frame, the entanglement entropy of the circle can be thought as the thermal entropy in de Sitter space. Here, the temperature is associated with the cosmlogical horizon of the static patch \reef{harmony} and the static observer measures $T=1/(2 \pi R)$, where $R$ is now also the curvature scale of the de Sitter geometry. Hence the one route to evaluating the entropy is first to compute the energy density at a finite temperature $T$ and then to integrate this density over the static patch to obtain the total energy $E(T)$. The entropy is determined with the thermodynamic formula
\begin{equation}
 S(T)=\int_{0}^{T}\frac{dT}{T}\,\frac{d E(T)}{dT} \,.\labell{esebeta}
\end{equation}
The final step is to fix $T$ to be the equilibrium temperature of the horizon, \ie  $T = 1/(2\pi R)$.

The energy density for a conformally coupled free scalar at finite temperature in de Sitter space was calculated in \cite{Dowker:1987} by
conformally transforming the known expressions in Rindler space at finite temperature with respect to Rindler time. 
Following \cite{Dowker:2010}, for convenience we change radial coordinates in de Sitter metric as 
\begin{equation}
z^2=\frac{R-\hat{r}}{R+\hat{r}}\,,
\labell{z}
\end{equation}
with $z\in(0,1)$. With this radial coordinate, the horzion becomes $z=0$ while the origin sits at $z=1$. 
The de Sitter metric \reef{harmony} now becomes
\begin{equation}
 ds^2=\frac{4}{(1+z^2)^2}\left(-z^2 d\tau^2+R^2 dz^2\right)+R^2\left(\frac{1-z^2}{1+z^2}\right)^2d\theta^2\,.
\end{equation}
In these coordinates, the energy density in thermal de Sitter space is given by \cite{Dowker:1987}
\begin{equation}
\left\langle\, T_\tau{}^\tau\,\right\rangle=\frac{T(1+z^2)^3}{64\sqrt{2}\,\pi\, R^2\, z^{3}}\,\left[W_3-\frac{1}{4}W_{1}\right]
\labell{trindler}
\end{equation}
where
\begin{equation}
 W_n\equiv-2^{\frac{4-n}{2}}\int_0^{\infty}\frac{dy}{\cosh(y/2)^n}\,\frac{\sin(2\pi^2 R\,T)}{\cosh( 2 \pi R\,T\,y)-\cos(2 \pi^2  R\,T)}\,.
\end{equation}
The entropy is now obtained from eq.~(\ref{trindler}) by integrating the energy density over the volume of the static patch and using eq.~(\ref{esebeta}). Solving the integrals, we find
\begin{eqnarray}
 S&=&\frac{1}{2^{7/2}}\int_{0}^{1/(2 \pi R)} \frac{dT}{T}\ \frac{d (T\,[W_{3}-\frac{1}{4}W_{1}])}{dT}  \int_{z_{\textrm{min}}}^1\frac{(1-z^2)}{z^{2}}dz\nonumber\\
 %&&\hspace{4cm}
&=&\frac{1}{2}\left( z_{\textrm{min}}-2+\frac{1}{z_{\textrm{min}}}\right) (2 \pi\,\coss)\,.
\end{eqnarray}

Here the volume of the static patch is finite but the result is divergent due to the singular contribution of the energy
density for $T<1/(2 \pi R)$ as one approaches the horizon.
To give a meaning to the regularization on the coordinate $z>z_{\textrm{min}}$ used here, let us translate it to planar
coordinates by the transformations in eqs.~(\ref{desitter}) and (\ref{z}). Choosing $r_{\textrm{max}}=R-\hat\delta$
with $\hat\delta=\delta\,(1+a_1 \delta/R+\cdots)$ as in eq.~(\ref{expa}), we find the coefficient of the constant term is
\begin{equation}
c_0=\left(\frac{3}{2}-a_1 \right) \,\coss\,.
\end{equation}
Hence we recognize the same coefficients and ambiguities as in the previous calculations with the hyperbolic geometry.

\end{document}